\documentclass[pre,aps,floats,superscriptaddress,floatfix,twocolumn]{revtex4}
\usepackage{amssymb,amsmath}
\usepackage{amsmath,amssymb}
\usepackage{graphicx}
\usepackage{psfrag}
\usepackage{color}
\usepackage{soul}
\usepackage{dcolumn}
\usepackage[utf8]{inputenc}
\usepackage{hyperref}

\usepackage{bm}
\usepackage[normalem]{ulem}
 \hypersetup{
    colorlinks=true,
    citecolor=blue,
    }
 \usepackage{amsmath}

\begin{document}

 \title{Availability, storage capacity, and diffusion: Stationary states of an asymmetric exclusion process connected to two reservoirs}
 \author{Sourav Pal}\email{isourav81@gmail.com}
\affiliation{Theory Division, Saha Institute of
Nuclear Physics, A CI of Homi Bhabha National Institute, 1/AF Bidhannagar, Calcutta 700064, West Bengal, India}
\author{Parna Roy}\email{parna.roy14@gmail.com}
\affiliation{Shahid Matangini Hazra Government General Degree College for Women, Purba Medinipore 721649, West Bengal, India}
\author{Abhik Basu}\email{abhik.123@gmail.com, abhik.basu@saha.ac.in}
\affiliation{Theory Division, Saha Institute of
Nuclear Physics, A CI of Homi Bhabha National Institute, 1/AF Bidhannagar, Calcutta 700064, West Bengal, India}

\begin{abstract}
  We explore how the interplay of finite availability,  carrying capacity of particles at different parts of a spatially extended system and particle diffusion between them  control the steady state currents and density profiles in a one-dimensional current-carrying
channel connecting the different parts of the system. To study this, we construct a minimal model consisting of two particle reservoirs of finite carrying capacities connected by a totally asymmetric simple exclusion process (TASEP). In addition to particle transport via TASEP between the reservoirs, the latter can also directly exchange particles  via Langmuir kinetics-like processes, modeling particle diffusion between them that can maintain a steady current in the system. We investigate the steady state density profiles and the associated particle currents in the TASEP lane. The resulting phases and the phase diagrams are quite different from an open TASEP, and are characterised by the model parameters defining particle exchanges between the TASEP and the reservoirs,  direct particle exchanges between the reservoirs, and the filling fraction of the particles that determines the total resources available. These parameters can be tuned to make the density on the TASEP lane globally uniform or piecewise continuous, and can make the two reservoirs preferentially populated or depopulated.
\end{abstract}

\maketitle

 \section{INTRODUCTION}
 \label{introduction}

  Driven diffusive systems have emerged as fascinating subjects of research in nonequilibrium physics, providing fertile ground for investigations of the fundamental issues of nonequilibrium steady states (NESS) and phase transitions~\cite{driven-diff1,driven-diff2,driven-diff3,driven-diff4}. One of the most prominent examples is the totally asymmetric simple exclusion process (TASEP)~\cite{krug,krug1,krug2,derrida}, originally proposed as a simple model to describe protein synthesis in biological cells~\cite{macdonald}. Subsequently, it was reinvented as a paradigmatic nonequilibrium model when it was discovered to have boundary-induced phase transitions~\cite{krug,krug1,krug2}. A TASEP consists of a one-dimensional (1D) lattice with open boundaries. The dynamics is stochastic in nature, and involves unidirectional particle hopping, subject to exclusion at all the sites, i.e., any site can contain at most one particle at a time. In TASEP dynamics, a particle enters at one of the boundaries at a specified rate, unidirectionally hops to the following sites at rate unity subject to exclusion until it reaches the other end, from which it leaves at another specified rate. Parametrised by the entry and exit rates at the boundary, the TASEP exhibits three distinct phases: the steady state densities in bulk can be either less than half, giving the low-density (LD) phase, or more than half, giving the high-density (HD) phase, or just half, which is the maximal current (MC) phase.  Alongside these uniform phases, a coexistence line  separating the LD and HD phases can also be observed where nonuniform densities in the form of domain wall (DW) can be found. This DW is delocalised in space, attributed to the particle nonconservering dynamics governing the open TASEP.


   In the present study, our focus is on a class of TASEP models \textit{different from those with conventional open boundaries having infinite supply of particles}.  Rather, we consider a single TASEP that is assumed to be connected to {\em two different particle reservoirs} at its two boundaries. This allows it to access   a fixed ``restricted'' or ``finite'' access to resources  or particles. In order to ensure a finite steady current in the nonequilibrium steady states, the two reservoirs are allowed to exchange particles at given specified rates instantaneously. Clearly, this puts a constraint on the total number of particles in the combined system of the TASEP and the attached reservoirs, which is held constant by the dynamics. This is principally motivated by biological and physical processes that are often under the constraint of limited resources. For instance, TASEPs with finite resources provide a simple description of the basic mechanism for the biological processes of protein synthesis in cells~\cite{reser1}, due to the finite amount of ribosomes available in a cell, and also in the context of traffic~\cite{traffic1}, due to the finite number of automobiles in a network of roads; see also Ref.~\cite{limited} in this context. Some of the  examples of the applications of a TASEP connected to a reservoir are limited resources in driven diffusive systems~\cite{lim-driv}, different biological contexts like mRNA translations and motor protein dynamics in cells~\cite{lim-bio1} and traffic problems~\cite{lim-bio2}. Notable experiments relevant to these model systems include studies on spindles in eukaryotic cells~\cite{lim-exp1,lim-exp2}.


 At the basic level of modeling of a TASEP with finite resources, a single TASEP is connected at both ends to a particle reservoir without any spatial extensions (i.e., a point reservoir), with the entry rate to the TASEP from the reservoir being given by a function of the instantaneous reservoir occupation; the exit rate from the TASEP to the reservoir is taken to be constant, as in an open TASEP. This has been generalised to
 more than one TASEP connected to a reservoir as a model to study competition between TASEPs for finite resources; see Ref.~\cite{reser3}.
 Detailed numerical stochastic simulations and analytical mean-field theory studies reveal rich nonuniform steady state density profiles, including domain walls in these
 models~\cite{reser1,reser2,reser3}.
 In another interesting extension of a TASEP with finite resources, the effect of a limited number of two different fundamental transport resources, \textit{viz.} the hopping particles and the ``fuel carriers'', which supply the energy required to drive the system away from equilibrium has been considered in Ref.~\cite{brackley} that reports multiple phase transitions as the entry and exit rates are changed.
 In a recent study~\cite{astik-1tasep}, both the entry and exit rates are assumed to depend on the instantaneous reservoir population leading to dynamics-induced competition between the entry and exit rates, which reveals {\em reservoir crowding effects} on the steady states.

  How asymmetric exclusion process, a paradigmatic nonequilibrium process, couples with equilibrium diffusion or Langmuir kinetics (LK) to produce nontrivial steady states is a fundamentally important question. Diffusion is ubiquitous in cell biological contexts~\cite{diff-cell}. In a cell, molecular motors can undergo driven motion along the microtubules, which are effectively 1D channels, or diffusion in the cell cytoplasm. Competition between driven and diffusive dynamics has been studied by using various models, characterised by specific modeling of particle diffusion, its coupling to the driven part of the dynamics and the presence or absence of any overall particle conservation. Modeling the system as a filament (microtubule) executing TASEP confined in a three-dimensional particle reservoir representing the cell cytoplasm, where the motors diffuse around, the consequences of the effects of diffusion on the steady states of the filament (TASEP) has been explored in Refs.~\cite{klump-PRL,klump-JSTAT,klump-PRE,ciandrini-2019,dauloudet-2019}.


 In a complementary approach to these studies, a two-lane lattice-gas model has been considered, where diffusion has been modeled as a 1D symmetric exclusion process (SEP), coupled with a TASEP. For instance, in a ring geometry involving a TASEP segment connected at its two ends with a SEP segment of equal size, that strictly conserves total particle number, it was shown that in the limit where the diffusivity scales linearly with the system size, the TASEP, in addition to the usual LD, HD and MC phases of an open TASEP admits a localised domain wall (LDW) over a finite region of the phase space spanned by the diffusivity and mean particle number~\cite{hinsch}, with the LDW converting to a delocalised domain wall (DDW), akin to the one found in an open TASEP, when the diffusive segment is much larger than the TASEP segment~\cite{anjan-parna}. In another study on a coupled TASEP-SEP model in a half-open geometry for transport of molecular motors along protein filaments, where the TASEP is coupled to SEP in the bulk via a mechanism of mutual lane exchange of particles, how the interplay between the active and diffusive transport results into density localisation and tip localisation of the molecular motors is investigated~\cite{frey-graf-PRL}; see also Ref.~\cite{bojer-graf-frey-PRE}. More recently Ref.~\cite{sm} explored the complex space-dependence of the TASEP and SEP density profiles, when a TASEP and a SEP channels, both with open boundaries, are connected via a mechanism of mutual lane exchange of particles akin to Langmuir kinetics (LK) ~\cite{lk-tasep}. In LK, particles adsorb at an empty site or desorb from
an occupied one.  If interactions between the particles other
than the hard-core repulsion are neglected, the equilibrium
density is solely determined by the ratio of the two kinetic
rates~\cite{fowler}, as given by the Gibbs ensemble. This makes LK a simple paradigm for equilibrium processes.


 In spite of the significant development in researches on TASEP with finite resources, there are many questions or issues which remain largely unexplored. For instance, in more general situations, the resources may be ``distributed'', a simple reflection of closed but spatially extended systems. Assuming for simplicity the existence of several point reservoirs, such distributed resources may be connected by channels carrying steady currents, which can be tuned by both the ``supply side'' as well as the ``receiving side''.
While our motivation originates from general cell biological transport, our aim is not to describe a particular biological system in detail.
 The overarching goal of the present study is to theoretically study and understand the nonequilibrium steady states in systems with multiple point reservoirs representing distributed resources having finite capacities. We consider two such reservoirs connected by a single current carrying channel executing TASEP dynamics. Further, the reservoirs exchange particles at given specified rates in a manner reminiscent of LK. The model naturally ensures a steady current in the TASEP channel. Overall particle number conservation is imposed. This provides a simple conceptual model for a system with multiple distributed resources coupled by exchanging particles and connected by channels with steady currents. Since direct particle exchanges between two isolated point reservoirs is an equilibrium process, our model serves as a prototype to investigate the consequences of competition between an equilibrium (particle exchange) and a nonequilibrium (TASEP) processes.  We study two specific limits of the model, when the particles exchange process competes with the current in the TASEP channel, and a second case when the former dominates over the TASEP current. Our studies here generalise the work in Ref.~\cite{reser3} in the simplest possible way. Our principal qualitative results are:

 (i) For appropriate choice of the model parameters, the TASEP density profile can be spatially uniform, or display macroscopic nonuniformity in the form of a localised domain wall (LDW).

 (ii) Depending upon the availability of the total resources, the TASEP may or may not be able to access all the phases of an isolated open TASEP.

 (iii) The relative populations of the two reservoirs can also be controlled dynamically.

 (iv) The above properties are controlled by the combined effect of finite resources, particle exchange and reservoir crowding effect.

 (v) At a more technical level, (a) our model can describe steady states both with or without particle-hole symmetry, with the former holding at a special point in the parameter space. Further, (b) in the steady states, when the particle exchanges are infrequent or ``weak'', i.e., cannot dominate over the stationary TASEP current, the two reservoirs maintain their distinct identities with the fluctuations remain relevant even in the thermodynamic limit, leading to quantitative mismatches with mean-field theory predictions. In contrast,  when the particle exchange takes place rapidly, i.e., ``strong'', the two reservoirs maintain a strict ratio in their respective particle numbers and fluctuations are irrelevant in the thermodynamic limit.

 (vi) From the general standpoint of nonequilibrium statistical physics, our model shows how a simple, paradigmatic equilibrium process, {\em viz.} particle exchange, couples with TASEP, one of the simplest paradigm for nonequilibrium systems, to produce complex phase behaviour, with fluctuations remaining important even in the thermodynamic limit in some of cases.

 The rest of the article is organised as follows: In Section \ref{Model}, we provide a brief description of the model. Section \ref{steady-state density model L} presents a mean-field analysis of its nonequilibrium steady state properties. Next, in Section \ref{mean-field phase diagrams}, we present the phase diagrams when the particle exchange competes with the TASEP current, called the weak coupling limit. The density profiles and boundaries between different phases are determined within the framework of MFT in Section \ref{ssd and pb}. Particle-hole symmetry is discussed in Section \ref{ph sym and pd}, followed by an exploration of the nature of phase transitions in Section \ref{dw nature}. The discrepancy between MFT and MCS results in the weak coupling limit is discussed in Section \ref{Role of fluctuations}. Finally, we summarise our findings and provide an outlook in Section \ref{conclusion and outlook}. Additional details including results on the case when the particle exchange dominates, called the strong coupling limit, are given in Appendix.

 \section{MODEL}
 \label{Model}

\begin{figure}[!h]
 \centerline{
 \includegraphics[width=10.0cm, height=8cm, clip=true]{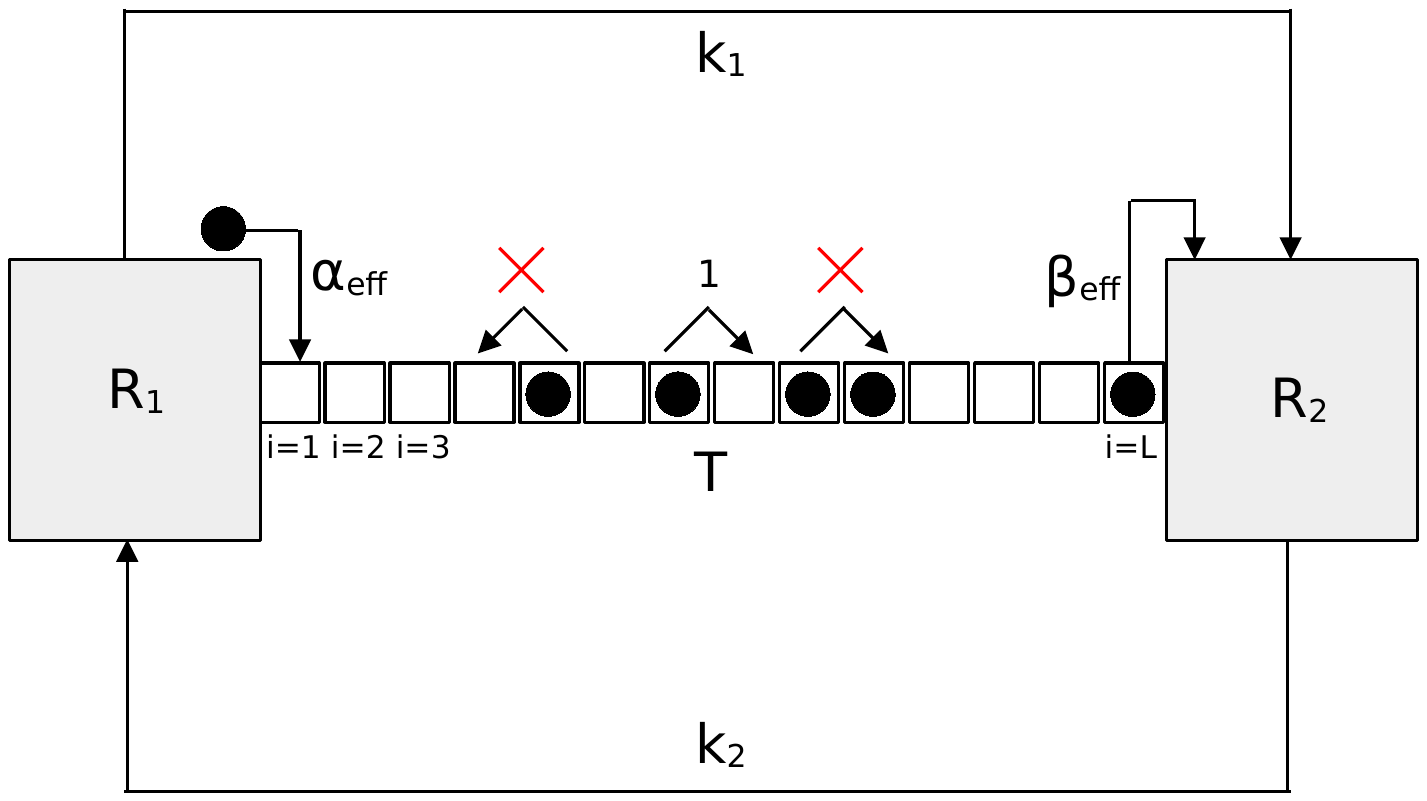}}
 \caption{\textbf{Schematic model diagram}: The model consists of two reservoirs $R_{1}$ and $R_{2}$, with $N_{1}$ and $N_{2}$ number of particles, respectively which are connected by a TASEP lane, denoted as $T$, having $L$ sites. Particles, which are denoted by black solid circles, enter the first site ($i=1$) of $T$ from $R_{1}$ with an effective rate $\alpha_\text{eff}$ and hop from left to right with rate unity subject to exclusion. Eventually, particles are removed from the last site ($i=L$) of $T$ into $R_{2}$ with an effective rate  of $\beta_\text{eff}$. Additionally, particle exchanges occur directly between $R_{1}$ and $R_{2}$ with rates $k_{1}$ and $k_{2}$ modeling diffusion in the system.}
 \label{Schematic diagram of the model.}
 \end{figure}

 The model consists of a 1D lattice $T$ with $L$ number of sites, which executes TASEP dynamics and is connected to two particle reservoirs $R_{1}$ and $R_{2}$, having finite particle storage capacities; see Fig.~\ref{Schematic diagram of the model.} for a schematic diagram of the model. The reservoirs are assumed to be point reservoirs without any spatial extent or internal dynamics. The sites of $T$ are labeled by an index $i$, $i \in [1, L]$ with $i=1(L)$ as the entry(exit) end, and can accommodate no more than one particle at a given time. Particles enter $T$ from the left reservoir $R_{1}$ provided the first site  of $T$ is empty, hop uni-directionally along $T$ subject to hard-core exclusion, i.e., only if the next site is vacant, and eventually leave $T$ to enter thr right reservoir $R_{2}$. We choose the hopping rate in the bulk of $T$ as unity, which sets the time scale of the model. Our model strictly follows \textit{particle number conservation} (PNC) globally, which reads

 \begin{equation}
  N_{0}=N_{1}+N_{2}+\sum_{i=1}^{L}n_{i},
 \end{equation}

 \noindent
 where $N_{0}$ is the total particle number of the system (reservoirs and TASEP lane combined), $N_1$ and $N_2$ are the particle numbers in $R_1$ and $R_2$, respectively, and $n_{i} \in 0$ or $1$, denotes the occupation number of site $i$ in the TASEP lane. The entry and exit rates are parametrised by $\alpha$ and $\beta$, which may take any positive value.  Actual entry and exit rates, which should also be positive, depend on the instantaneous reservoir population and are given by
 \begin{equation}
 \label{effective entry and exit rates}
 \alpha_\text{eff}=\alpha f(N_{1}), \hspace{5mm} \beta_\text{eff}=\beta g(N_{2})
 \end{equation}
 These rates model the coupling of the $T$ with its ``environments'', i.e., the two reservoirs. It is reasonable to expect that a rising population of $R_{1}$ ($R_{2}$) facilitates (obstructs) the inflow (outflow) of particles to (from) $T$. In keeping with this expectation, $f$ and $g$ are assumed to be monotonically increasing and decreasing functions of $N_{1}$ and $N_{2}$ respectively. We make the following simple choices:
 \begin{equation}
 \label{Model 1}
 f(N_{1})=\frac{N_{1}}{L}, \hspace{5mm} g(N_{2})=1-\frac{N_{2}}{L},
 \end{equation}
where these reservoir population-dependent rate functions $f$ and $g$ must be positive to make the effective rates in (\ref{effective entry and exit rates}) positive. The positivity of $g$ in (\ref{Model 1}) requires the restriction $N_{2} \le L$.  While more complex forms for monotonically rising $f$ and decreasing $g$ can be conceived, such simple choices as given in (\ref{Model 1}) suffice for the present purposes. Since the particle current through $T$ is unidirectional from $R_1$ to $R_2$, in order to maintain a steady non-zero current, there must be another mechanism for particle exchanges between $R_1$ and $R_2$. To that end, we allow the reservoirs to exchange of particles directly, modeling diffusion between them, at some fixed rates: $R_{1}(R_{2})$ releases particles at rate $k_{1}(k_{2})$ which $R_{2}(R_{1})$ receives instantaneously.  We further define filling factor by $\mu=N_{0}/L$. Taken together, our model has five control parameters --- $\alpha$, $\beta$, $\mu$, $k_{1}$, and $k_{2}$ --- which determine the steady states of the model.

 \section{MEAN-FIELD THEORY}
 \label{steady-state density model L}

 In this Section, we analyse the  model at the mean-field level following same line of reasoning as in earlier works~\cite{blythe}. MFT entails neglecting the correlation effects and replacing averages of products by the products of averages. In the mean-field description, the equations of motion for the density $\rho_i(t)$ at site $i$ are
 \begin{eqnarray}
  &&\frac{d\rho_i}{dt}=\rho_{i-1}(1-\rho_{i})-\rho_{i}(1-\rho_{i+1}),\,i=2,..,L-1, \label{mft1}\\
  &&\frac{d\rho_1}{dt}=\alpha_\text{eff}(1-\rho_{1})-\rho_{1}(1-\rho_{2}), \label{mft2}\\
  &&\frac{d\rho_L}{dt}=\rho_{L-1}(1-\rho_{L})-\beta_\text{eff}\rho_{L}. \label{mft3}
  \end{eqnarray}
  Similarly, the the reservoir populations $N_1$ and $N_2$ follow
  \begin{eqnarray}
  &&\frac{dN_1}{dt}=k_{2}N_{2}-k_{1}N_{1}-J_{T}, \label{mft4}\\
  &&\frac{dN_2}{dt}=k_{1}N_{1}-k_{2}N_{2}+J_{T}. \label{mft5}
  \end{eqnarray}
  In what follows below, for the convenience of presenting our results, we assume unit geometric length of the lattice, introduce  a quasi-continuous variable $x=i/L \in [0,1]$ where $\epsilon=1/L$ is the lattice constant. With this new coordinate $x$, the discrete density $\rho_{i}(t)$ is replaced by a continuous density $\rho(x,t)$. The equation of motion of $\rho(x,t)$ in the bulk of $T$ reads~\cite{blythe}
  \begin{eqnarray}
  &&\frac{\partial \rho(x)}{\partial t}=-\epsilon\partial_x(\rho(x)[1-\rho(x)])+ {\cal O}(\epsilon^2), \label{mft1-x}
  \end{eqnarray}
  supplemented by the boundary conditions $\rho(0)= \alpha_\text{eff},\,\rho(1)=1-\beta_\text{eff}$. Eq.~(\ref{mft1-x}) has a conservation law form, reflecting the conserving nature of the TASEP dynamics in the bulk, and thus allows us to extract a particle current $J_T$ that in the MFT has the form
  \begin{equation}
   J_T=\rho(1-\rho).\label{tasep-curr}
  \end{equation}
  This current $J_T$ is related to the reservoir occupation numbers, $N_1$ and $N_2$, through a flux balance equation in the steady state:
\begin{equation}
 \label{Flux balance equation}
 \frac{dN_{1}}{dt}=k_{2}N_{2}-k_{1}N_{1}-J_{T}=-\frac{dN_{2}}{dt}=0.
 \end{equation}
 Moreover, PNC imposes the constraint
  \begin{equation}
 \label{PNC}
 N_{0}=N_{1}+N_{2}+N_{T}.
 \end{equation}

Our model admits a special symmetry, called the {\em particle-hole symmetry} when $k_1=k_2$. The transformations
 \begin{eqnarray}
  &&\rho_{i} \leftrightarrow 1-\rho_{L-(i-1)}, \label{rho tr}\\
  &&\alpha_\text{eff} \leftrightarrow \beta_\text{eff}, \label{a b tr}
 \end{eqnarray}
  leave Eqs.~(\ref{mft1}-\ref{mft3}) invariant. Also, since $\alpha_\text{eff}=\alpha N_{1}/L$ and $\beta_\text{eff}=\beta(1-N_{2}/L)$, transformation (\ref{a b tr}) can be decomposed into two parts: $\alpha \leftrightarrow \beta$ together with $N_{1}/L \leftrightarrow (1-N_{2}/L)$. These properties allow us to obtain the phase diagrams of our model for $\mu>3/2$ (greater than the half-filled limit) from those at $\mu<3/2$ (less than the half-filled limit) when $k_1=k_2$ (discussed later). For $k_1\neq k_2$, there is no such symmetry.  Eqs.~(\ref{mft1}-\ref{a b tr}) formally define MFT for our model.

In the steady states, $\partial\rho/\partial t=0$, giving $J_T$ a constant.
 For a given constant current $J_{T}$, the possible steady state densities in TASEP are
  \begin{equation}
  \rho=\frac{1}{2}\left(1\pm \sqrt{1-4J_{T}}\right)=:\rho_\pm.\label{rho-sols}
 \end{equation}
  Thus the steady state TASEP density can be spatially uniform with $\rho=\rho_-\le 1/2$ (LD phase), or $\rho=\rho_+>1/2$ (HD phase). In the particular case when $J_{T}$ is at its maximum value $J_{T}=1/4$, one has $\rho=1/2$ (MC phase). There is also a second possibility in which case $\rho$ is piecewise discontinuous, with $\rho=\rho_-$ and $\rho=\rho+$ occurring at different regions of $T$. This gives rise to a domain wall (DW). Further, in the steady state, $dN_1/dt=0=dN_2/dt$, and hence flux balance condition (\ref{Flux balance equation})  gives
 \begin{equation}
 k_2N_2=k_1N_1 +J_T.\label{flux-bal-sol}
 \end{equation}
 Solving Eq.~(\ref{flux-bal-sol}) together with (\ref{PNC}), we get the following expressions for the reservoir populations:
 \begin{subequations}
 \begin{equation}
 \label{N1}
 N_{1}=\frac{k_{2}(N_{0}-N_{T})-J_{T}}{k_{1}+k_{2}},
 \end{equation}
 \begin{equation}
 \label{N2}
 N_{2}=\frac{k_{1}(N_{0}-N_{T})+J_{T}}{k_{1}+k_{2}}.
 \end{equation}
 \end{subequations}
 Thus $N_1$ and $N_2$ depend explicitly on $J_T$, the steady TASEP current. Depending on whether or not this dependence on $J_T$ is significant in the limit of the system size $L\rightarrow\infty$, we two distinct cases; see below. Unsurprisingly,
 the total reservoir population $N_R=N_1+N_2=N_0-N_T$ is independent of $J_T$.

\begin{figure}[!h]
 \centering
 \includegraphics[width=\columnwidth]{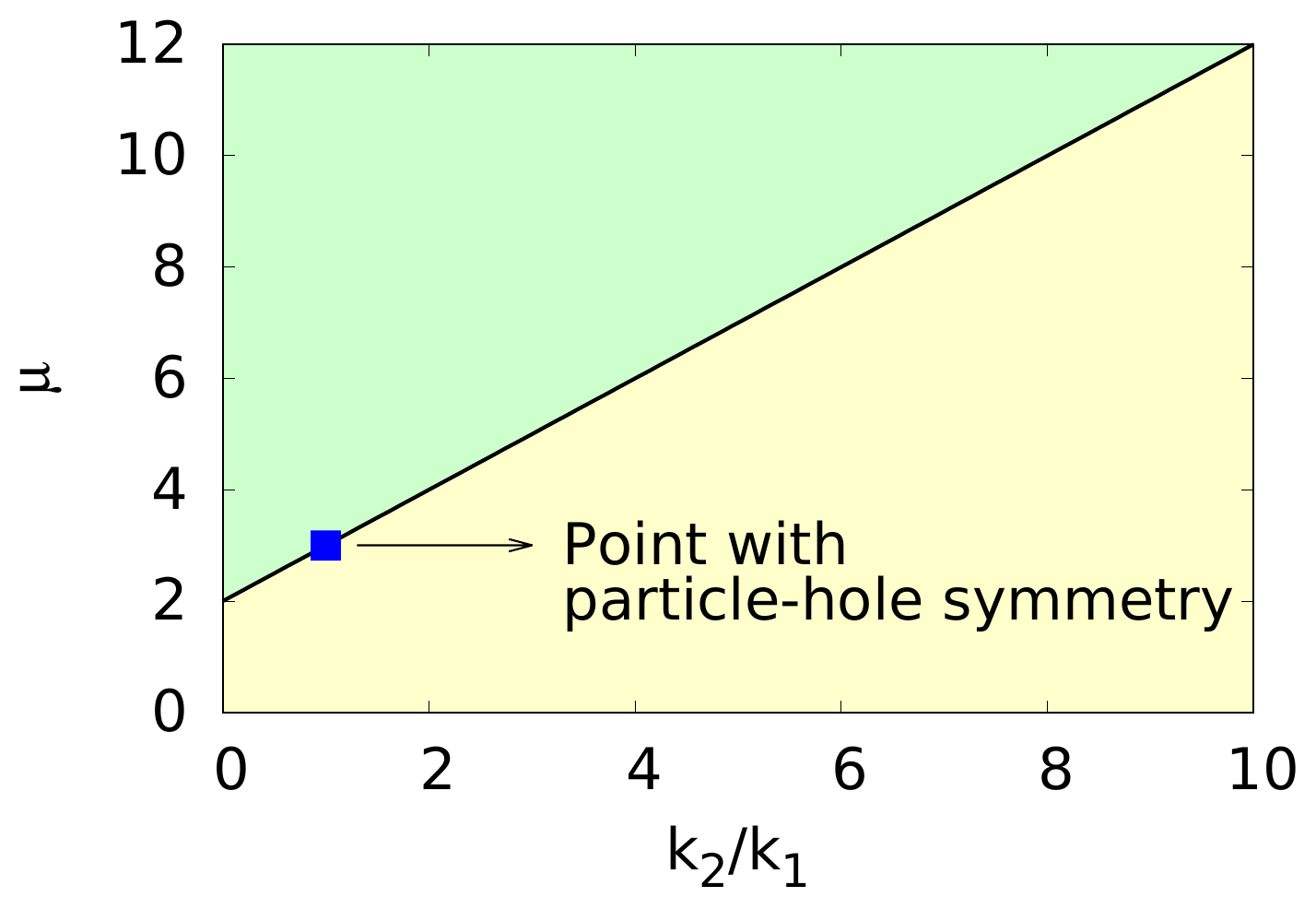}
 \caption{ State diagram of the model in the $k_2/k_1-\mu$ plane. The black inclined line given by $\mu=\mu_\text{max}=(2+k_2/k_1)$ gives the maximum possible filling fraction admissible for a given $k_2/k_1$. In the yellow region below the black line one has $\mu\leq \mu_\text{max}$, where the model is well-defined. In the greenish region above the line, $ \mu>\mu_\text{max}$, for which the model is ill-defined. The blue square on the line at $k_2/k_1=1$ is the point where the model admits the particle-hole symmetry, which is absent elsewhere in the yellow region below the black line.}
  \label{mumax}
 \end{figure}

 After determining the reservoir populations, our next task is to calculate the maximum of the filling factor $\mu\equiv N_0/L$. We have $\mu=0$, its minimum value, when there are no particles present in the system. Consequently, in the TASEP, this implies a steady state particle density of zero. When $\mu$ reaches its maximum value $\mu_{\text{max}}$, the particle density in the TASEP lane is equal to 1. Hence, the particle current becomes zero since there is no available space for particles to flow within the completely filled system. Eqs.~(\ref{N1}) and (\ref{N2}) imply
 \begin{equation}
 \frac{N_{1}}{N_{2}}=\frac{k_{2}}{k_{1}},
 \end{equation}
  which gives
  \begin{equation}
  N_{1_{\text{max}}}= \frac{k_{2}}{k_{1}}N_{2_\text{max}} = \frac{k_{2}L}{k_{1}}.
  \end{equation}
  Additionally, the number of particles in the TASEP lane is $N_{T}=L$ when $\mu=\mu_{\text{max}}$. By substituting these values into the PNC relation~(\ref{PNC}), we determine the upper limit of $\mu$ as
  \begin{equation}
  \mu_{\text{max}}=\bigg(2+\frac{k_{2}}{k_{1}}\bigg).\label{mumax-exp}
  \end{equation}
  Thus, $\mu_\text{max}$ is controlled by the exchange rate ratio $k_2/k_1$. Since $k_{2}/k_{1}$ itself varies between 0 (when $k_{2}=0$ or $k_{2}<<k_{1}$) and $\infty$ (when $k_{1}=0$ or $k_{1}<<k_{2}$), the upper limit of $\mu$ is not fixed and can be any number between 2 and $\infty$.  In the particular case, when the exchange rates are equal ($k_{1}=k_{2}$), the value of $\mu_{\text{max}}$ is 3. At this point, the transformation $\mu \leftrightarrow 3-\mu$ captures the particle-hole symmetry of the model. See Fig.~\ref{mumax} for a state diagram of the model in the $k_2/k_1-\mu$ plane.

 As we have explained above that the non-negativity constraint on function $g$ implies that the population of reservoir $R_{2}$ has an upper limit defined by the system length $L$, leading to $N_{2} \le L$. Consequently, the range of function $g$ is bounded between 0 and 1. In contrast, the upper bound of function $f$ is determined by the maximum value of $N_{1}$. As $N_{1_{\text{max}}}=k_{2}L/k_{1}$, it follows that $f_{\text{max}}=k_{2}/k_{1}$, which can be any (positive) number.

  Eqs.~(\ref{rho-sols}-\ref{mumax-exp}) can be used to solve for the steady state TASEP densities and the reservoir populations in MFT.

Having identified the possible phases in $T$, our next task is to identify the locations of these phases in the space of the control parameters $\alpha,\beta$ for fixed $\mu$, $k_{1}$, and $k_{2}$. Before proceeding with the actual calculations, we can identify the two distinct cases. From (\ref{flux-bal-sol}) we note that while $J_T< 1$ is independent of $L$, whereas $N_1,\,N_2$ should scale with $L$. Another way to see this is from (\ref{N1}) and (\ref{N2}), where $J_T<1$ but $N_0$ and $N_T$ should rise with $L$ indefinitely. Thus, {\em if} $k_1,\,k_2\sim {\cal O}(1)$, we can neglect $J_T$ in (\ref{flux-bal-sol}) or (\ref{N1}) and (\ref{N2}) in the thermodynamic limit (TL) $L\rightarrow \infty$, giving $k_1N_1=k_2N_2$ asymptotically exactly, independent of $J_T$. Thus in TL the relative population of the two reservoirs is independent of $J_T$, and is controlled only by the ratio $k_2/k_1$, independent of the TASEP parameters $\alpha$, $\beta$.  This effectively eliminates one of the reservoirs, which can allow us to describe the TASEP steady states in terms of the population of an effective single reservoir. This is the particle exchange or diffusion dominated regime. Since in this regime, the two reservoirs effectively become one reservoir, losing their separate identities, in so far as the TASEP steady states are concerned, we name it the {\em strong coupling} limit below. We further consider a case where the diffusion process competes with the hopping in $T$. This is ensured by introducing a mesoscopic scaling $k_{1(2)}=k_{10(20)}/L$ where $k_{10},\,k_{20} \sim \mathcal{O}(1)$. In this regime, the two reservoirs cannot be replaced by a single effective reservoir and they retain their separate identities, we name this the {\em weak coupling} limit. In the main text of the paper, we discuss the weak coupling limit case. Results for the strong coupling limit are given in Appendix.

\section{MEAN-FIELD PHASE DIAGRAMS IN THE WEAK COUPLING CASE}
\label{mean-field phase diagrams}

\begin{figure*}[htb]
 \includegraphics[width=\columnwidth]{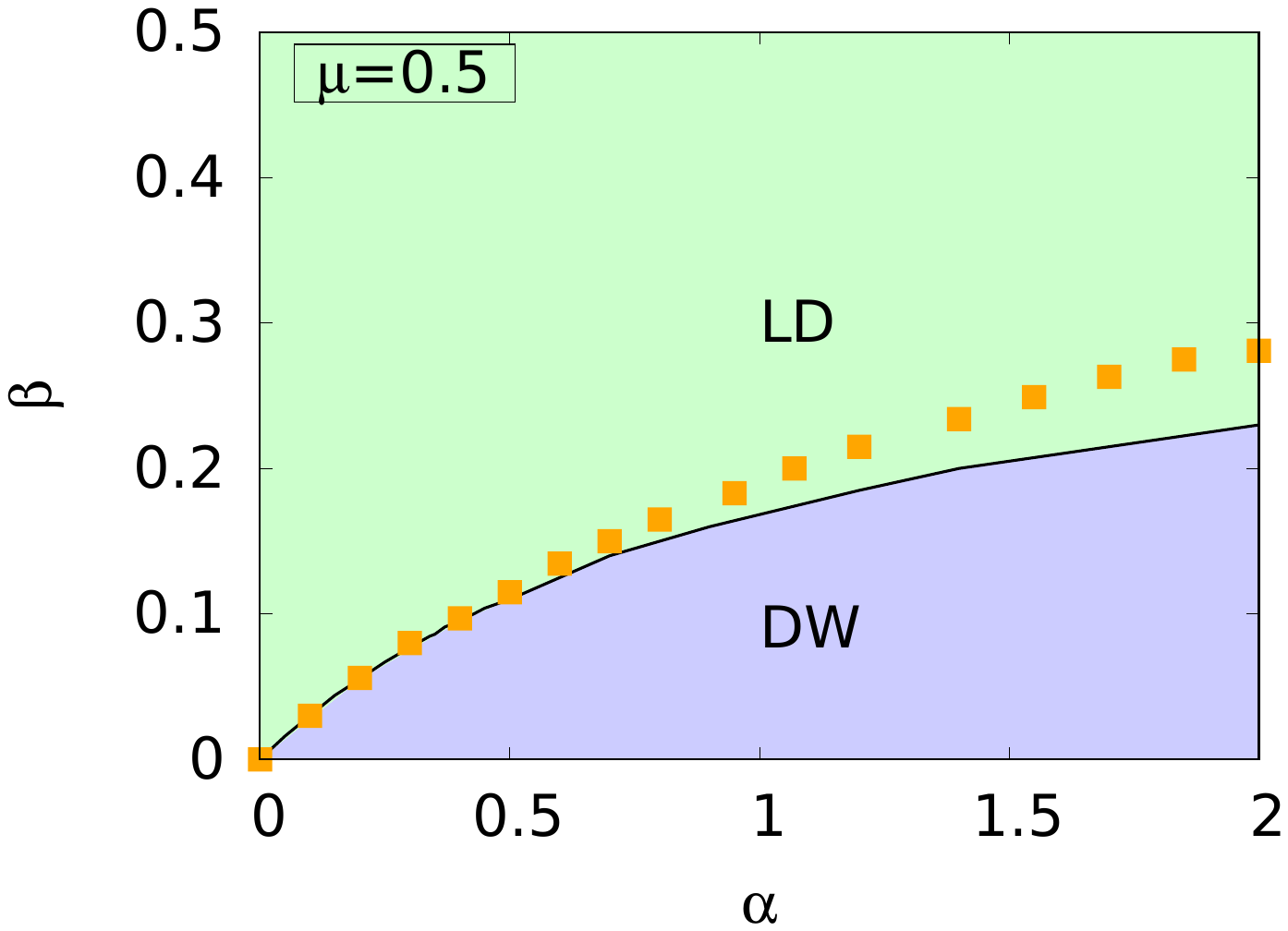}
 \hfill
 \includegraphics[width=\columnwidth]{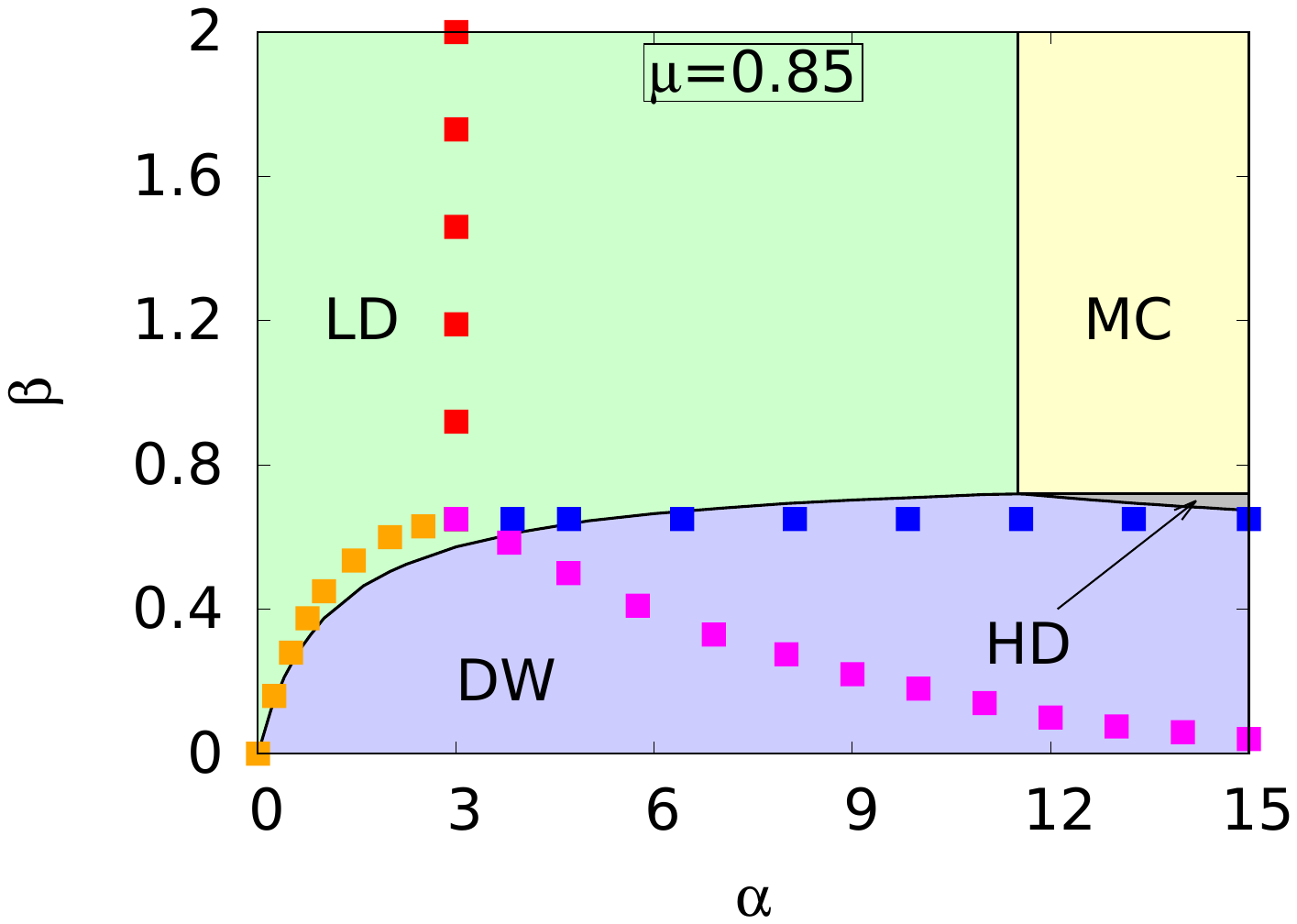}
 \\
 \includegraphics[width=\columnwidth]{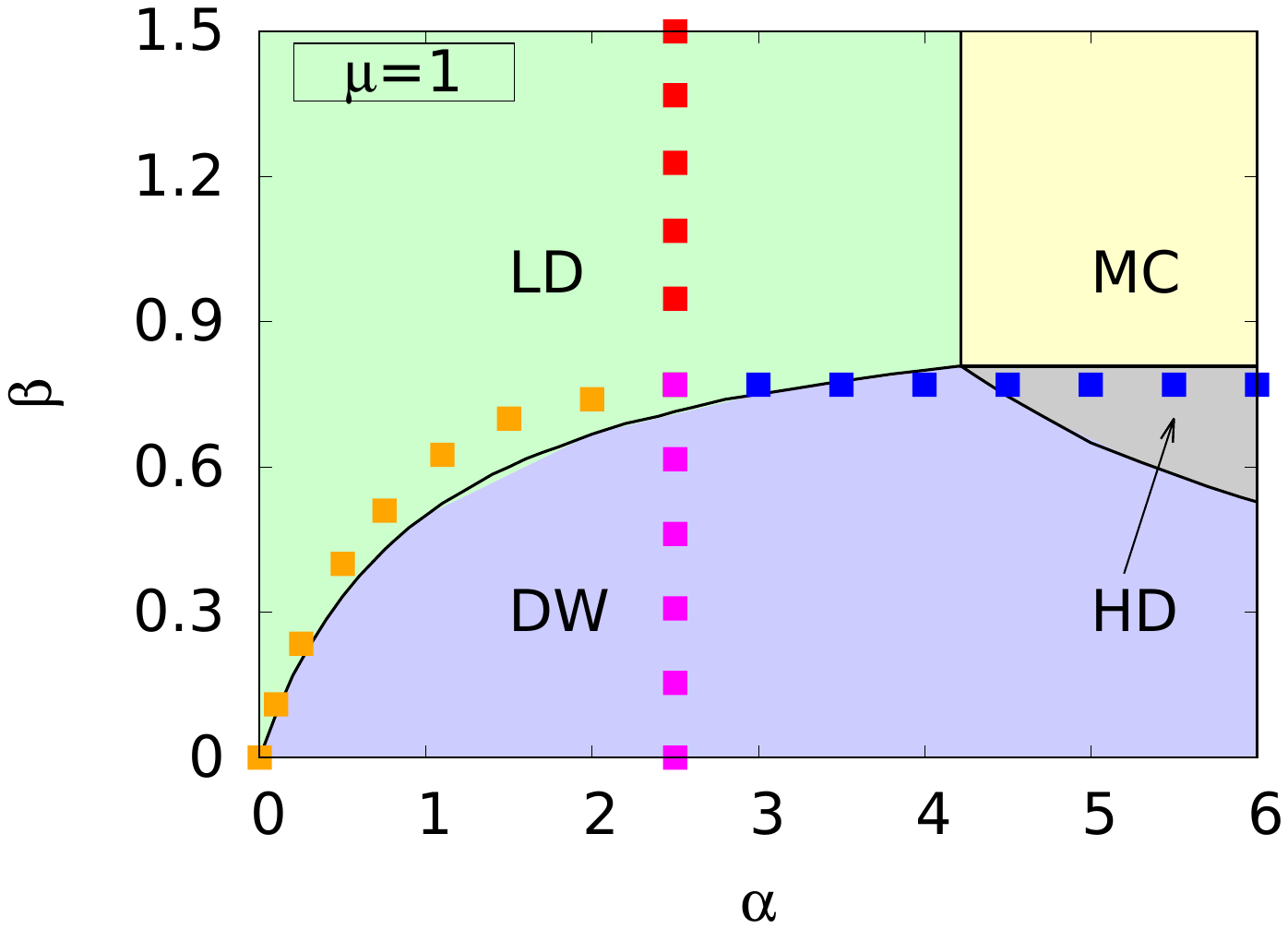}
 \hfill
 \includegraphics[width=\columnwidth]{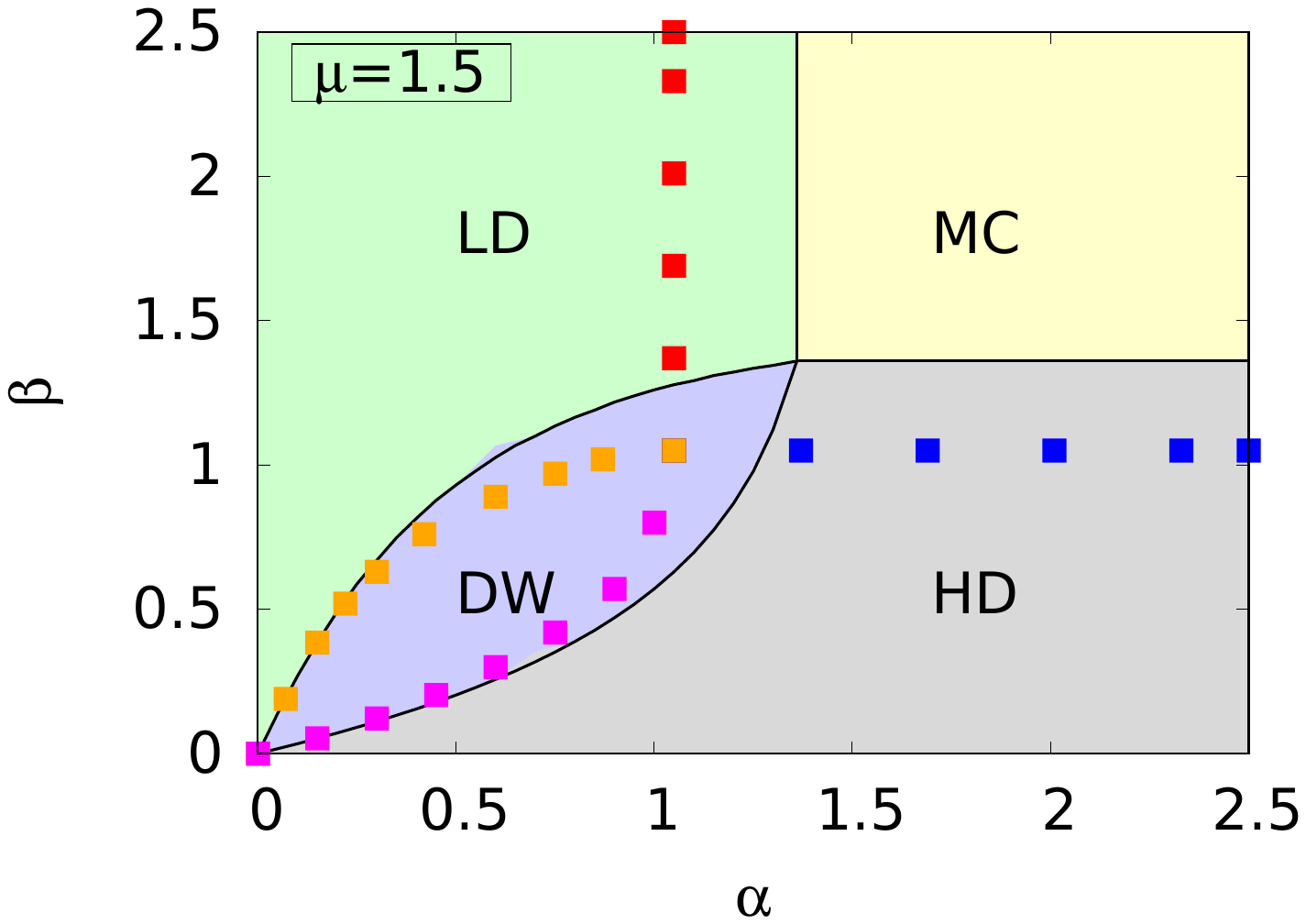}
 \\
\caption{Phase diagrams in the weak coupling limit of the model with particle hole symmetry are shown for different filling factors $\mu$ with equal exchange rates ($k_{10} = k_{20} = 0.95$). Depending on the value of $\mu$ with $0<\mu<\mu_\text{max}=(2+k_{20}/k_{10})=3$, two or four distinct phases are observed simultaneously, using both MFT and MCS. These phases are LD, HD, MC, and DW, represented by green, gray, yellow, and blue regions respectively with solid black lines separting the phases, according to MFT; while MCS results on the phase boundaries are shown with discrete colored points: red (LD-MC), blue (HD-MC), orange (LD-DW), and magenta (HD-DW). (top left) $\mu=0.5$, (top right) $\mu=0.85$, (bottom left) $\mu=1$ and (bottom right) $\mu=1.5$.  Significant discrepancies between MFT and MCS results on the phase diagrams are observed. With $k_{10} = k_{20}$, the phase diagrams for $\mu>3/2$ are related to those with $\mu<3/2$ by particle-hole symmetry, where $\mu=3/2$ is the half-filled limit; see Fig.~\ref{ph-wkl} below.}
\label{pd weak 3}
\end{figure*}

\begin{figure*}[htb]
 \includegraphics[width=\columnwidth]{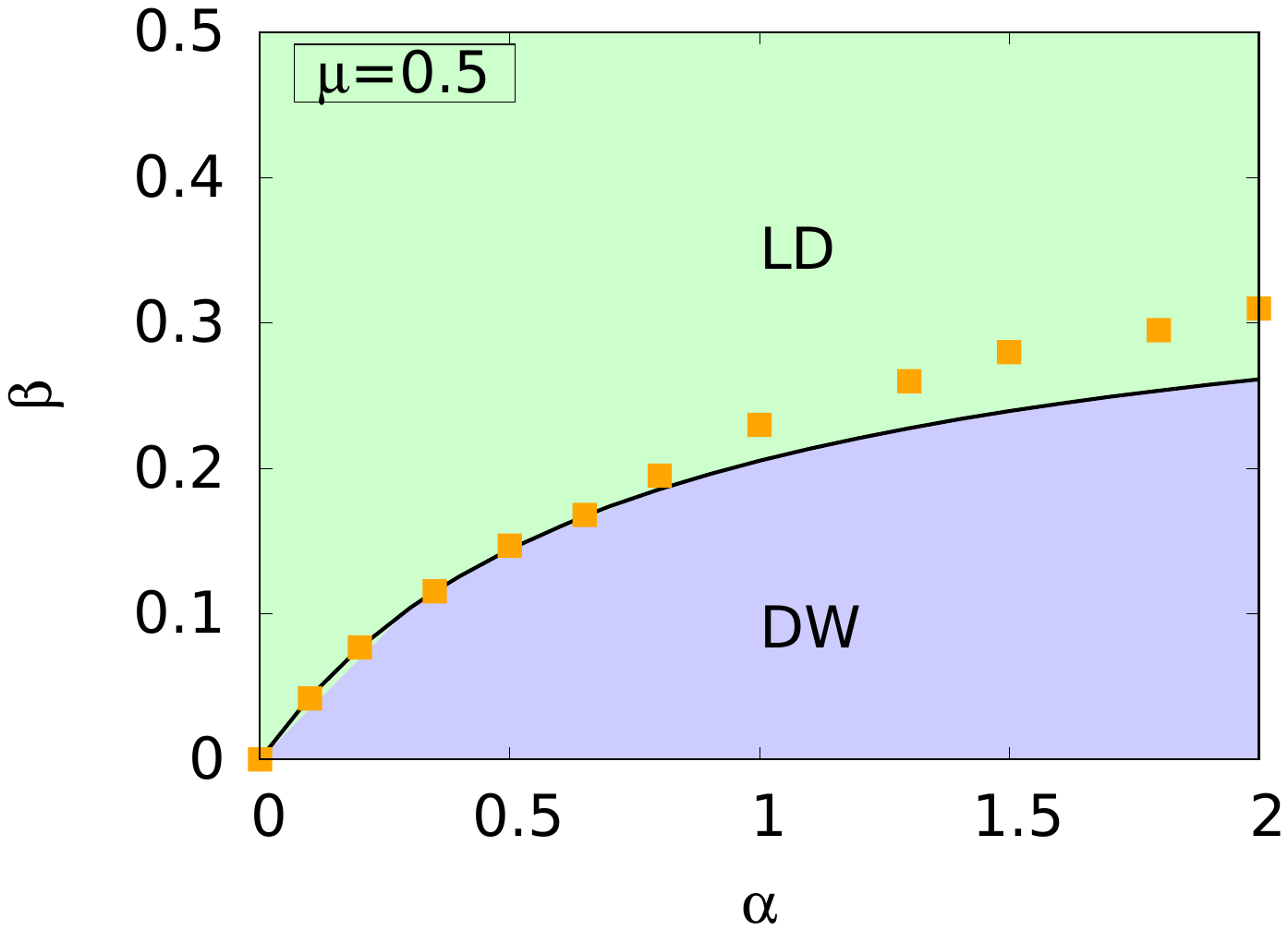}
 \hfill
 \includegraphics[width=\columnwidth]{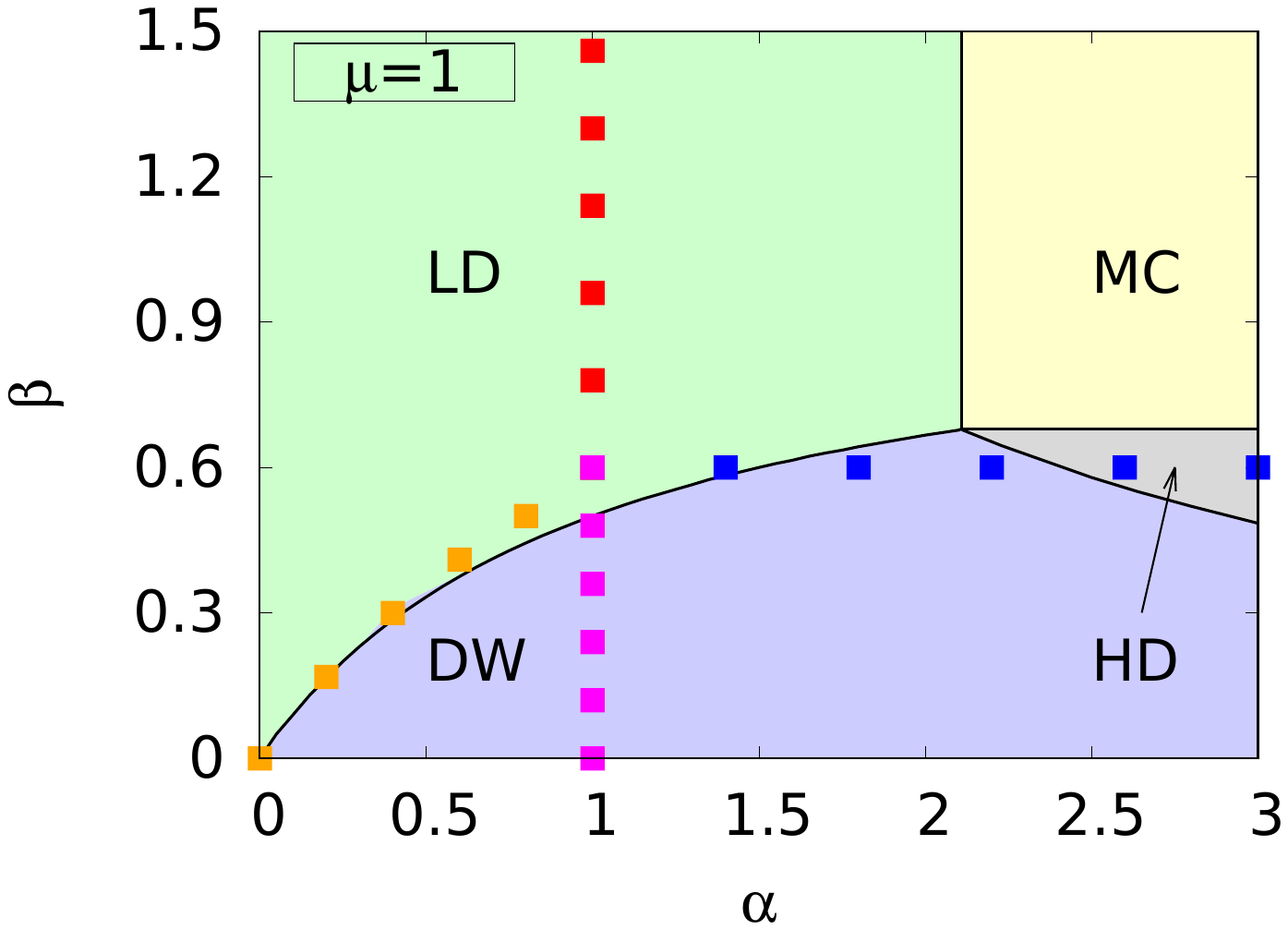}
 \\
 \includegraphics[width=\columnwidth]{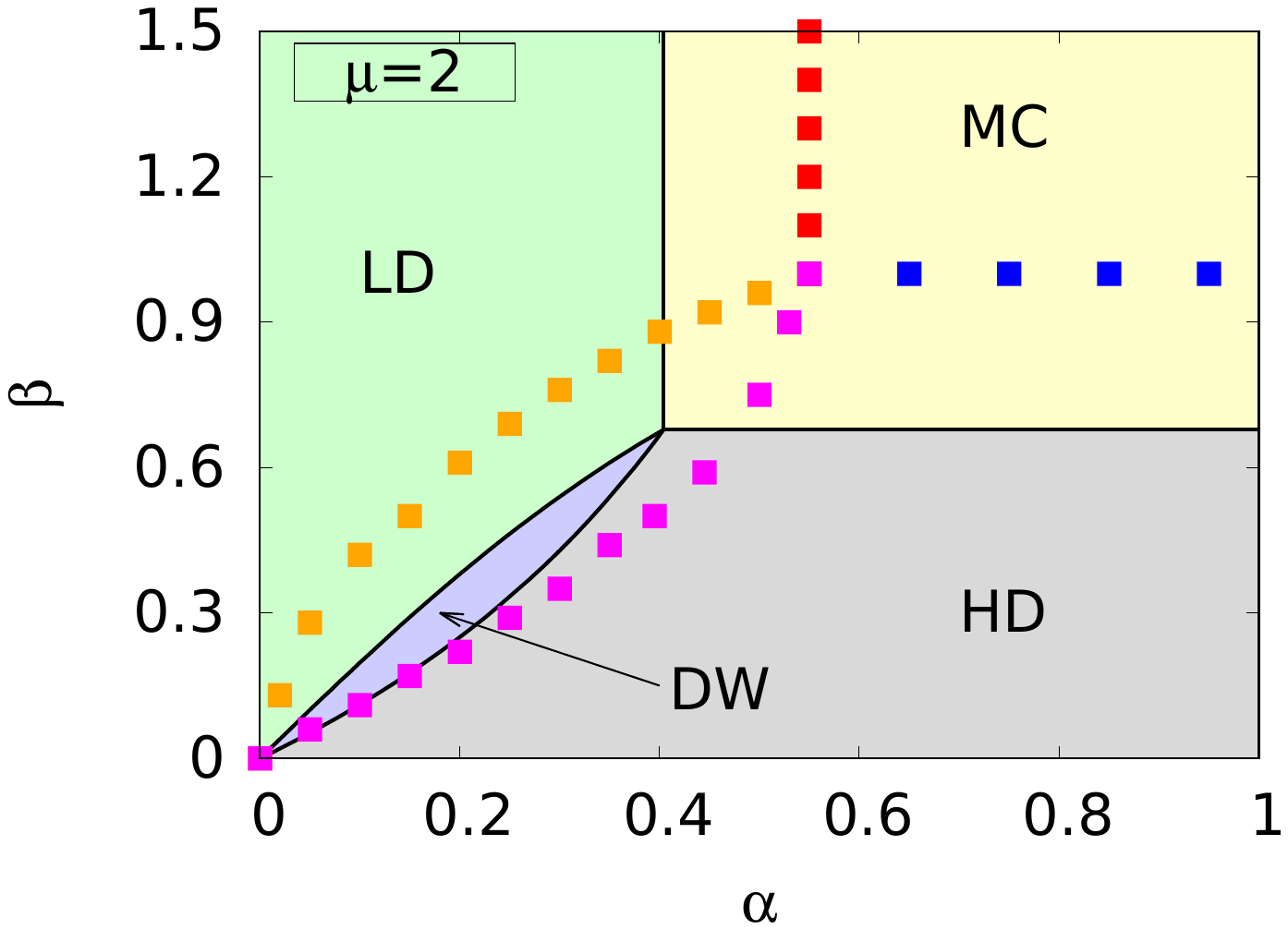}
 \hfill
 \includegraphics[width=\columnwidth]{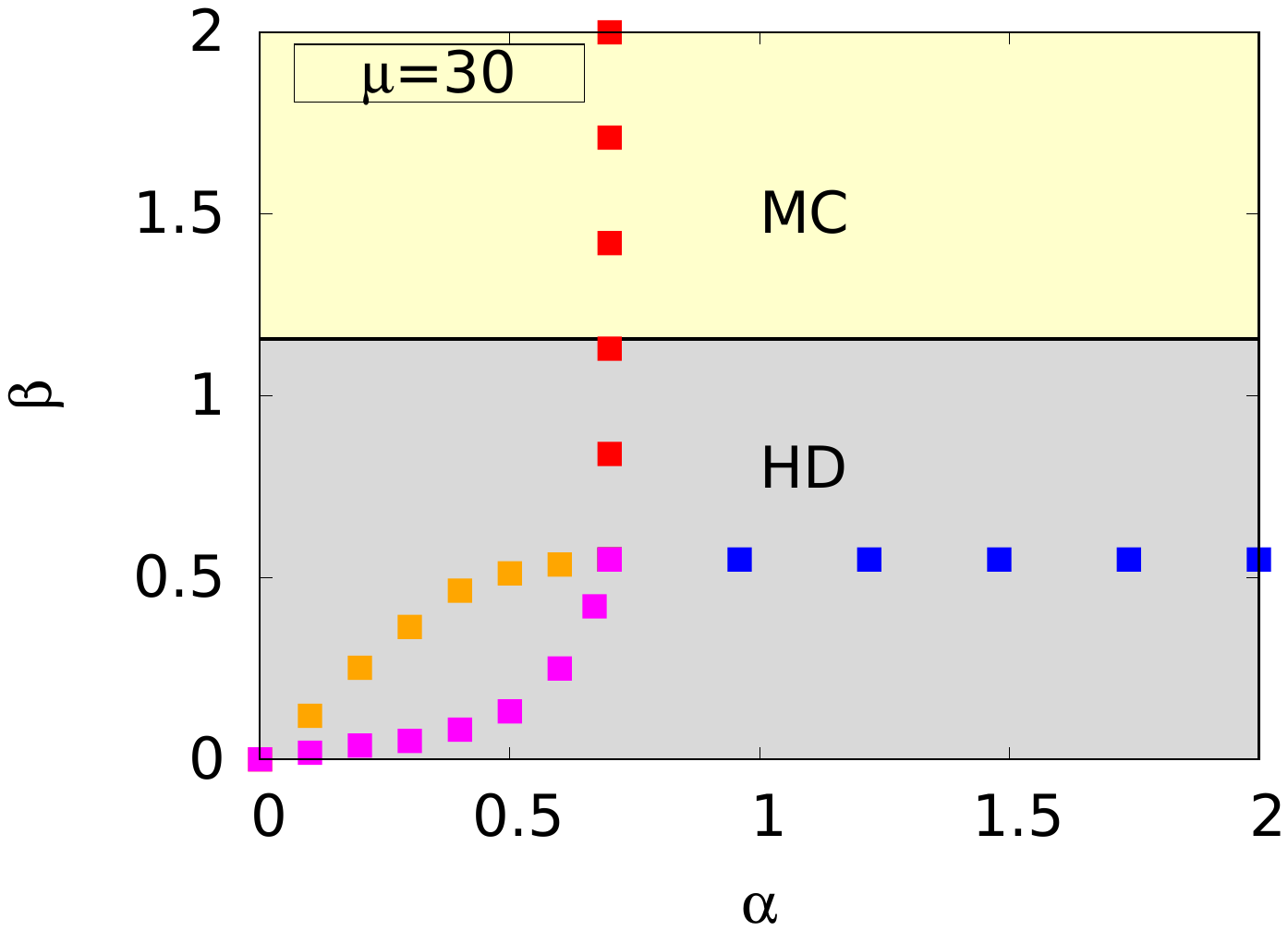}
 \\
\caption{Phase diagrams in the weak coupling limit of the model without particle hole symmetry are shown for various filling factors $\mu$ when exchange rates are $k_{10}=0.01$ and $k_{20}=0.95$. Again in this case, two or four distinct phases are observed subject to the particular value of $\mu$ with $0<\mu<\mu_\text{max}=(2+k_{20}/k_{10})=97$. The phases LD, HD, MC, and DW are represented by green, gray, yellow, and blue regions respectively with solid black line as the phase boundaries, according to MFT and are compared to the MCS phase boundaries indicated by discrete colored points: red (LD-MC), blue (HD-MC), orange (LD-DW), and magenta (HD-DW). (top left) $\mu=0.5$, (top right) $\mu=1$, (bottom left) $\mu=2$ and (bottom right) $\mu=30$. Similar to the phase diagrams in Fig.~\ref{pd weak 3}, significant deviations between MFT and MCS results on the phase boundaries are observed in the phase diagrams. Unlike Fig.~\ref{pd weak 3}, phase diagrams for the present unequal choice of exchange rates are not related by particle-hole symmetry. {The phase diagram for  $\mu=30$ consists of LD and DW phases also but for extremely small $\alpha$ values ($\alpha \lesssim 0.017$), rendering them practically invisible.}}
\label{pd weak 2}
\end{figure*}

\begin{figure*}[htb]
 \includegraphics[width=\columnwidth]{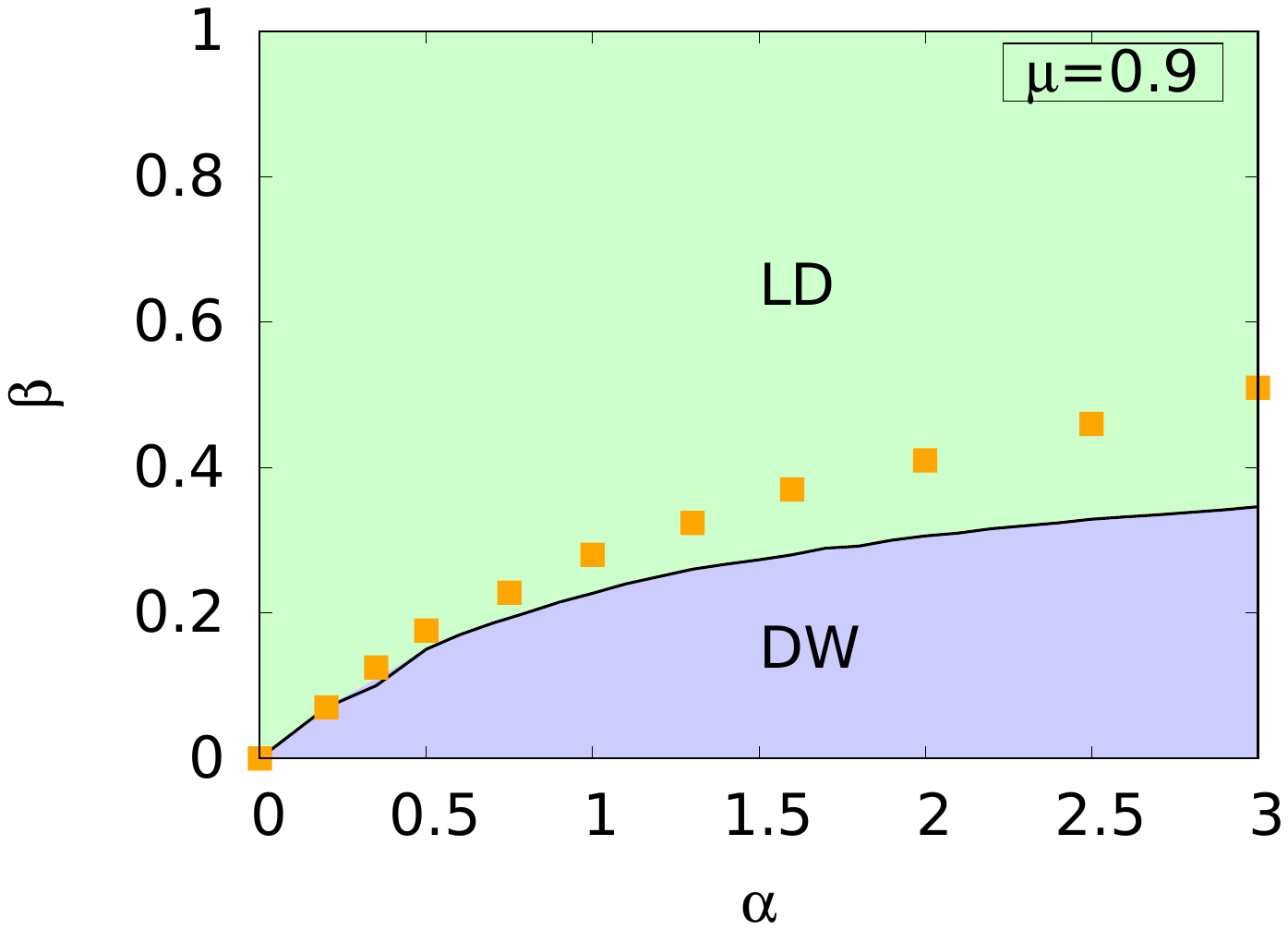}
 \hfill
 \includegraphics[width=\columnwidth]{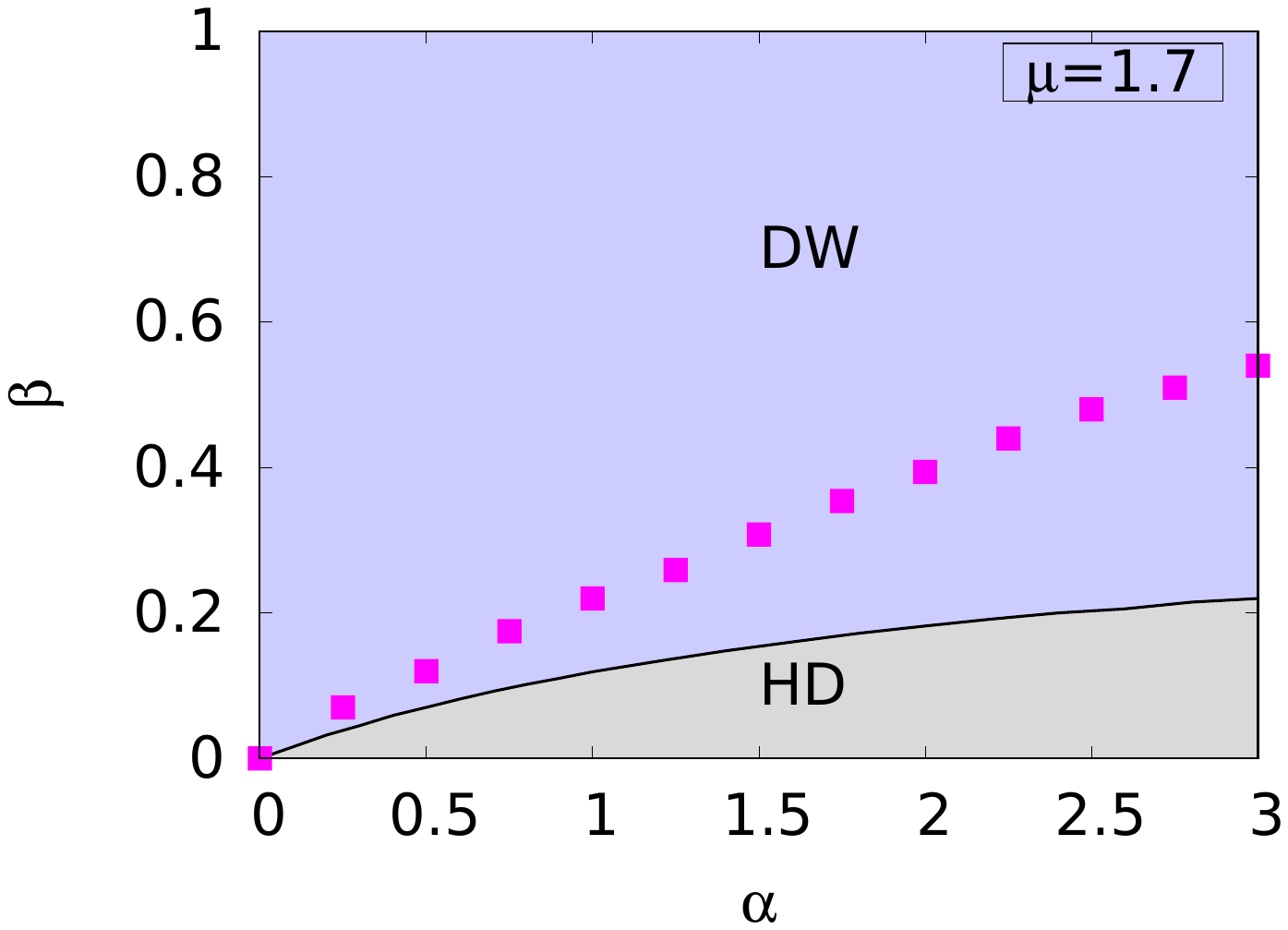}
 \\
\caption{ Phase diagrams in the weak coupling limit of the model without particle hole symmetry are shown when exchange rates are $k_{10}=1$ and $k_{20}=0.1$. Background colors represent different phases and black solid lines are the boundaries separating them, according to MFT; wherein discrete colored points are the MCS phase boundaries. (left) When $\mu=0.9$, only two phases, LD and DW, appear, and (right) when $\mu=1.7$, the phases HD and DW occur. According to the condition in (\ref{cond-for-mc-existence}) for the existence of the MC phase,  $\mu_\text{max}$ should be greater than 3.75 with the chosen exchange rates $k_{10}$ and $k_{20}$, which is not satisfied in the present case where $\mu_\text{max}=(2+k_{20}/k_{10})=2.1$. Consistent with this MFT prediction, the MC phase is absent in our MCS results also for any value of $\mu$ within 0 and 2.1. The MCS phase diagrams show qualitative similarities, but there are quantitative discrepancies compared to the MFT phase diagrams }
\label{pd weak 11}
\end{figure*}

Having defined the model and set up the MFT, we start by determining different phases and phase boundaries in the space of the control parameters $\alpha$ and $\beta$ in the weak coupling case. In this case, with $k_1=k_{10}/L,\,k_2=k_{20}/L$, the general solutions for the reservoir populations $N_1$ and $N_2$ in the MFT are given by
\begin{subequations}
 \begin{equation}
 \label{N1w}
 N_{1}=\frac{k_{20}(N_{0}-N_{T})-LJ_{T}}{k_{10}+k_{20}},
 \end{equation}
 \begin{equation}
 \label{N2w}
 N_{2}=\frac{k_{10}(N_{0}-N_{T})+LJ_{T}}{k_{10}+k_{20}}.
 \end{equation}
 \end{subequations}
Equations~(\ref{N1w}) and (\ref{N2w}) explicitly bring out their $J_T$-dependence, which do not vanish in the limit $L\rightarrow \infty$, since both $N_0,\,N_T$ also scale linearly with $L$. Our results are summarised in Fig.~\ref{pd weak 3}, Fig.~\ref{pd weak 2}, and Fig.~\ref{pd weak 11}, which give the mean-field phase diagrams as a function of the two control parameters $\alpha$ and $\beta$ for a set of representative values of the filling factor $\mu$ and fixed exchange rates $k_{10},\,k_{20}$. For quantitative analysis of the phases and the phase diagram, we proceed as follows, similar as in Refs.\cite{reser1,hinsch,melbinger}.   Before analyzing the structure of phase diagrams of the model under consideration, we revisit the open TASEP. In the open TASEP, one obtains three distinct phases, namely LD, HD, and MC phases, in the parameter space spanned by its (constant) entry and exit rates $\alpha_\text{T}$ and $\beta_\text{T}$ respectively. These phases meet at a common point $(1/2,1/2)$. The conditions under which these phases occur are as follows: $\alpha_\text{T}<\beta_\text{T}$ with $\alpha_\text{T}<1/2$ for LD phase, $\beta_\text{T}<\alpha_\text{T}$ with $\beta_\text{T}<1/2$ for HD phase, and $\alpha_\text{T},\beta_\text{T} \ge 1/2$ for MC phase. The steady state bulk densities in the LD, HD, and MC phases are spatially uniform with $\rho_\text{LD}=\alpha_\text{T}$, $\rho_\text{HD}=1-\beta_\text{T}$, and $\rho_\text{MC}=1/2$. At points on the coexistence line existing between from $(0,0)$ to $(1/2,1/2)$ ($\alpha_\text{T}=\beta_\text{T}<1/2$), the bulk density forms a delocalised domain wall (DDW) which spans the entire length of the lattice with an average position $x=1/2$.  The delocalisation of the domain wall is attributed to the particle non-conservation. Furthermore, the conditions between the transitions between different phases in an open TASEP can be obtained by equating the respective currents in the associated phases~\cite{blythe}. To be precise, the transition between the LD and HD phases occurs when $\alpha_\text{T}=\beta_\text{T}<1/2$. Similarly, the transitions between LD and MC phases, and between HD and MC phases occur when $\alpha_\text{T}=1/2<\beta_\text{T}$ and $\beta_\text{T}=1/2<\alpha_\text{T}$ respectively. The transition from the LD phase to the HD phase is marked by a sharp and abrupt change in the bulk density, signifying a first-order phase transition. In contrast, the transition from either the LD or HD phase to the MC phase involves a gradual and continuous variation in the bulk density, indicating a second-order phase transition.

The obtained phase diagrams of our model are fairly complex, and being parametrised by $\mu$, can have structures very different from the phase diagram of an open TASEP. To obtain the conditions of the ensuing phases and the transitions between them in the present study, we note that in the LD and HD phases, the TASEP densities read $\rho_\text{LD}=\alpha_\text{eff}$, $\rho_\text{HD} = 1 -\beta_\text{eff}$. Then in analogy with open TASEP, we can infer that the transition between the LD and HD phases occur when $\alpha_\text{eff}=\beta_\text{eff}<1/2$. Similarly, the transitions between LD and MC phases, and HD and MC phases occur when $\alpha_\text{eff}=1/2<\beta_\text{eff}$ and $\beta_\text{eff}=1/2<\alpha_\text{eff}$. These considerations are used to obtain the phase diagrams below.

 We study three specific cases, {\em viz.} $k_{10}=k_{20}=0.95$ (Fig.~\ref{pd weak 3}), $k_{10}=0.01, k_{20}=0.95$ (Fig.~\ref{pd weak 2}), and $k_{10}=1, k_{20}=0.1$ (Fig.~\ref{pd weak 11}), and find several notable features of the phase diagrams, summarised as follows:

  (i) $\mathbf{k_{10}=k_{20}=0.95}$: This has particle-hole symmetry. In this case, $\mu_\text{max}=(2+k_{20}/k_{10})=3$  with $\mu=3/2$ as the half-filled limit. Depending on the specific value of $\mu$, either two or four phases can be observed simultaneously in the phase diagrams drawn in the $\alpha-\beta$ plane. Detailed analysis reveals that below a certain threshold value of $\mu$, only the LD and DW phases are present in the phase diagram according to both mean field theory (MFT) and Monte Carlo simulations (MCS); see the phase diagrams given in Fig.~\ref{pd weak 3}. The precise value of this threshold is calculated later in the paper. In this regime, the number of particles is not sufficient to sustain the HD and MC phases, which require a larger particle count. As we increase $\mu$ beyond this threshold, the HD and MC phases gradually appear with increasing regions. Thus, we observe the coexistence of all four phases for these values of $\mu$. The aforementioned characteristics are evident in the phase diagrams for $\mu<3/2$. When $\mu$ exceeds 3/2, we can extrapolate the phase diagram structure based on our observations for $\mu<3/2$ owing to the particle-hole symmetry. In such instances, elevating the value of $\mu$ results in an amplification of the HD phase region while diminishing the LD phase region. Ultimately, surpassing a critical threshold renders the LD phase entirely absent from the phase diagram. However, it may be noted that the phase diagrams obtained from MCS do not quantitatively agree with the MFT-predicted phase diagrams; see Fig.~\ref{pd weak 3}. {For the corresponding density plots showing the quantitative disagreements between the MFT and MCS results, see the plots in Fig.~\ref{role of fluctuation}(top left) and Fig.~\ref{role of fluctuation} (top right).}

  \vspace{1mm}

 (ii) $\mathbf{k_{10}=0.01, k_{20}=0.95}$: For this unequal choice of exchange rates, the particle-hole symmetry is absent. The maximum value of $\mu$ in this case is $\mu_\text{max}=(2+k_{20}/k_{10})=97$ with $\mu=48.5$ as the half-filled limit. For smaller $\mu$ values, the structure of phase diagrams with this unequal particle exchange rates is qualitatively similar to those obtained in case (i) with equal exchange rates; see the phase diagrams in Fig.~\ref{pd weak 2}. Phase diagrams below and above the half-filled limit in this case are not related by any symmetry operation. As in the previous case, quantitative disagreements between the MFT predictions and MCS studies are detected. {For the corresponding density plots showing the quantitative disagreements between the MFT and MCS results, see the plots in Fig.~\ref{role of fluctuation}(bottom left) and Fig.~\ref{role of fluctuation}(bottom right).}

 \vspace{1mm}
 (iii) $\mathbf{k_{10}=1, k_{20}=0.1}$:  This corresponds to $\mu_\text{max}=2.1$. The particle-hole symmetry is not present in this case as well. Upon varying $\mu$ from 0 to 2.1, it is observed that none of the values of $\mu$ result in the simultaneous appearance of all the four phases in the phase diagram; see the phase diagrams in Fig.~\ref{pd weak 11}. This unique characteristic sets it apart from the previous observations in the phase diagram. We present phase diagrams for two specific values of $\mu$. For $\mu=0.9$, only the LD and DW phases are observed, while for $\mu=1.7$, only the HD and DW phases are present. Importantly, in this particular case, the MC phase is non-existent for any admissible value of $\mu<\mu_\text{max}(=2.1)$.

 Detailed quantitative analysis of the MFT phase diagrams are given below.

 \section{STEADY STATE DENSITIES AND PHASE BOUNDARIES}
 \label{ssd and pb}

 \subsection{Low-density phase}
 \label{LD Phase model L weak}

 In the LD phase, the steady state density $\rho_\text{LD}$ is given by
 \begin{equation}
 \label{LD model 1}
 \rho_\text{LD} = \alpha_\text{eff} = \alpha \frac{N_{1}}{L} <\frac{1}{2}.
 \end{equation}
  Substituting $N_{1}$ from Eq.~(\ref{N1}) in Eq.~(\ref{LD model 1}) and identifying the steady current in LD phase as $J_\text{LD}=\rho_\text{LD}(1-\rho_\text{LD})$, one finds the following quadratic equation in $\rho_\text{LD}$:
  \begin{equation}
 \label{ld model 1}
  \rho_\text{LD}=\alpha\bigg[\frac{k_{2}}{k_{1}+k_{2}}(\mu-\rho_\text{LD})-\frac{\rho_\text{LD}(1-\rho_\text{LD})}{L(k_{1}+k_{2})}\bigg],
 \end{equation}
  which, in the weak coupling limit (where $k_{1(2)}=k_{10(20)}/L$), translates into
 \begin{equation}
 \label{ld model 1 weak}
  \rho_\text{LD}=\alpha\bigg[\frac{k_{20}}{k_{10}+k_{20}}(\mu-\rho_\text{LD})-\frac{\rho_\text{LD}(1-\rho_\text{LD})}{k_{10}+k_{20}}\bigg].
 \end{equation}
 \noindent
Eq.~(\ref{ld model 1 weak}) has two solutions:
\begin{align}
  \label{LD density weak model 1 both}
  \begin{split}
  \rho_\text{LD}^{\pm}&=\bigg({\frac{1+k_{20}}{2}+\frac{k_{10}+k_{20}}{2\alpha}}\bigg)   \\
  &\quad \pm\bigg[\bigg({\frac{1+k_{20}}{2}+\frac{k_{10}+k_{20}}{2\alpha}}\bigg)^{2}-\mu k_{20}\bigg]^{\frac{1}{2}}.
  \end{split}
  \end{align}

  \noindent
  When $\mu$ equals zero, the LD phase density must be zero. Therefore, out of the two solutions in (\ref{LD density weak model 1 both}), the physically acceptable solution is the one that vanishes as $\mu\rightarrow 0$, i.e., in the limit of vanishingly small number of particles. We thus get

  \begin{align}
  \label{LD density weak model 1}
  \begin{split}
  \rho_\text{LD}&=\rho_\text{LD}^{-}\\
  &=\bigg({\frac{1+k_{20}}{2}+\frac{k_{10}+k_{20}}{2\alpha}}\bigg)- \\
  &\quad \bigg[\bigg({\frac{1+k_{20}}{2}+\frac{k_{10}+k_{20}}{2\alpha}}\bigg)^{2}-\mu k_{20}\bigg]^{\frac{1}{2}}
  \end{split}
  \end{align}
  as the solution for the LD phase density. Unsurprisingly, $\rho_\text{LD}$ as given in (\ref{LD density weak model 1}) is independent of $\beta$, the exit end rate parameter.

 We can obtain from Eq.~(\ref{LD model 1}) the expression of $N_{1}$; which is related to $N_{2}$ by the PNC equation, which reads $N_{0}=N_{1}+N_{2}+L\rho_\text{LD}$ in the LD phase. Thus, recalling $N_{0}=\mu L$ and the acceptable density in LD phase [Eq. (\ref{LD density weak model 1})], the expressions obtained for $N_{1}$ [using (\ref{LD model 1})] and $N_{2}$ (using PNC after obtaining $N_{1}$) are:
 \begin{eqnarray}
  &&N_1=\frac{L}{\alpha}\rho_\text{LD}, \label{n1-ld-weak}\\
  &&N_2=L\bigg(\mu-\frac{1+\alpha}{\alpha}\rho_\text{LD}\bigg). \label{n2-ld-weak}
 \end{eqnarray}

As $\mu$ increases, more particles are available in the system and in the TASEP lane. Consequently, it is expected that for sufficiently high values of $\mu$, the TASEP lane will transition out of the LD phase. The range of $\mu$ over which the LD phase exists can be obtained from the condition $0< \rho_\text{LD}< 1/2$, which is
\begin{equation}
\label{ld-mu-limit-weak}
0 < \mu < \left(\frac{1}{2} + \frac{1}{4k_{20}} + \frac{k_{10}}{2\alpha k_{20}} + \frac{1}{2\alpha}\right).
\end{equation}
This range indicates that the upper threshold of $\mu$ for the existence of the LD phase is a function of $\alpha$, $k_{10}$, and $k_{20}$. The upper threshold of $\mu$ in (\ref{ld-mu-limit-weak}) is  positive definite for positive parameters $\alpha$, $k_{10}$, and $k_{20}$, and can vary depending on their specific values. When this upper threshold is less than $\mu_\text{max}=(2+k_{20}/k_{10})$, i.e.,
\begin{equation}
 \label{ld-mu-cond-weak-other}
 \mu_\text{max} > \left(\frac{1}{2}+\frac{1}{4k_{20}}+\frac{k_{10}}{2\alpha k_{20}}+\frac{1}{2\alpha}\right),
\end{equation}
 beyond a certain threshold value of $\mu$, the LD phase ceases to exist due to an excess supply of particles. Conversely, when it equals $\mu_\text{max}$, the LD phase can persist for all values of $\mu$ ranging from 0 to $\mu_\text{max}$.

For the specific case of $k_{10} = k_{20} = 0.95$, the range of $\mu$ over which the LD phase exists is  $0 < \mu < (0.76 + 1/\alpha)$. When considering $\mu=1$, the condition $\mu < (0.76 + 1/\alpha)$ indicates the occurrence of the LD phase for any $\alpha<4.17$ according to MFT. Similarly, for the case of $k_{10} = 0.01$ and $k_{20} = 0.95$, the range becomes $0 < \mu < (0.76 + 0.505/\alpha)$, indicating the presence of the LD phase for any $\alpha<2.1$ when $\mu=1$. The phase diagrams in Fig.~\ref{pd weak 3} and Fig.~\ref{pd weak 2}, which include both the MFT and MCS results on the phase diagrams, show the LD phase within these calculated ranges, while also highlighting deviations of the MCS results from MFT predictions. In Fig.~\ref{ld-hd-mc-st-wk-den}, we present the LD phase density profiles in the weak coupling limit for $\mu=0.5$ and $\alpha=0.2,\,1$. The density profiles obtained from MCS agree well with the theoretical predictions of MFT for the lower $\alpha$ values. However, a quantitative discrepancy arises between the two for higher $\alpha$ values.

 \subsection{High-density phase}
 \label{HD Phase model L weak}

Next, we turn to the HD phase where the steady state bulk density is given by
\begin{equation}
 \label{hd weak L}
 \rho_\text{HD} = 1-\beta_\text{eff} = 1-\beta\bigg(1-\frac{N_{2}}{L}\bigg)>\frac{1}{2}.
 \end{equation}
Substitution of the expression of $N_{2}$ from Eq.~(\ref{N2}) into Eq.~(\ref{hd weak L}) leads to a quadratic equation in $\rho_\text{HD}$ given below:
\begin{equation}
\label{hd quad model 1}
\rho_\text{HD}=1-\beta+\beta\bigg[\frac{k_{1}}{k_{1}+k_{2}}(\mu-\rho_\text{HD})+\frac{\rho_\text{HD}(1-\rho_\text{HD})}{L(k_{1}+k_{2})}\bigg],
\end{equation}
which, in the weak coupling limit, reads
\begin{equation}
\label{hd quad weak model 1}
\rho_\text{HD}=1-\beta+\beta\bigg[\frac{k_{10}}{k_{10}+k_{20}}(\mu-\rho_\text{HD})+\frac{\rho_\text{HD}(1-\rho_\text{HD})}{k_{10}+k_{20}}\bigg].
\end{equation}
Solving Eq.~(\ref{hd quad weak model 1}) for $\rho_\text{HD}$ gives two solutions:
\begin{align}
\label{HD density weak model 1 both}
\begin{split}
\rho_\text{HD}^{\pm}&=\bigg(\frac{1-k_{10}}{2}-\frac{k_{10}+k_{20}}{2\beta}\bigg) \pm
\bigg[\bigg(\frac{1-k_{10}}{2}\\
&\quad -\frac{k_{10}+k_{20}}{2\beta}\bigg)^{2}+
\mu k_{10}-\bigg(1-\frac{1}{\beta}\bigg)(k_{10}+k_{20})\bigg]^{\frac{1}{2}}.
\end{split}
\end{align}

\noindent
Between these two solutions, the physically acceptable one is the solution where the TASEP density reaches 1 at $\mu=\mu_{\text{max}}$. We thus get
\begin{align}
\label{HD density weak model 1}
\begin{split}
\rho_\text{HD}&=\rho_\text{HD}^{+}\\
&=\bigg(\frac{1-k_{10}}{2}-\frac{k_{10}+k_{20}}{2\beta}\bigg)+
\bigg[\bigg(\frac{1-k_{10}}{2}\\
&\quad -\frac{k_{10}+k_{20}}{2\beta}\bigg)^{2}+
\mu k_{10}-\bigg(1-\frac{1}{\beta}\bigg)(k_{10}+k_{20})\bigg]^{\frac{1}{2}}
\end{split}
\end{align}
as the acceptable HD phase density. The HD phase density, to no surprise, is independent of $\alpha$. Upon using (\ref{hd weak L}) we derive the expression of $N_{2}$ in terms of $\rho_\text{HD}$, which is connected to $N_{1}$ by the PNC equation given by $N_{0}=N_{1}+N_{2}+L\rho_\text{HD}$ in the HD phase. Consequently, we determine the populations of the reservoirs, $N_{1}$ and $N_{2}$, as follows:
\begin{eqnarray}
  &&N_1=L\bigg(\mu-1+\frac{1}{\beta}-\frac{1+\beta}{\beta}\rho_\text{HD}\bigg),\\
  &&N_2=L\bigg(1-\frac{1}{\beta}+\frac{\rho_\text{HD}}{\beta}\bigg).
 \end{eqnarray}

For the system to sustain the HD phase, an adequate supply of particles is necessary. In cases where the particle supply is insufficient, the HD phase is absent from the phase diagram. The range of $\mu$ that allows for the existence of the HD phase can be determined by considering $1/2<\rho_\text{HD} < 1$. This provides us with the following range for $\mu$ over which the HD phase exists:
\begin{equation}
\label{range of mu for hd phase model 1 weak coupling}
\left(\frac{3}{2} + \frac{k_{20}}{k_{10}} - \frac{1}{4k_{10}} - \frac{k_{20}}{2\beta k_{10}} - \frac{1}{2\beta}\right) < \mu \leq \mu_\text{max}.
\end{equation}
As indicated in (\ref{range of mu for hd phase model 1 weak coupling}), the lower threshold of $\mu$ required for the existence of the HD phase is a function of $\beta$, $k_{10}$, and $k_{20}$. Depending on the values of $\beta$, $k_{10}$, and $k_{20}$, this lower threshold can either be positive or negative. When positive, the lower threshold implies a restricted $\mu$  value below which the HD phase is unobservable. In that case,
\begin{align}
  &\left(\frac{3}{2} + \frac{k_{20}}{k_{10}} - \frac{1}{4k_{10}} - \frac{k_{20}}{2\beta k_{10}} - \frac{1}{2\beta}\right) > 0 \nonumber  \\
  \implies &\mu_\text{max} > \bigg(\frac{1}{2}+\frac{1}{4k_{10}} + \frac{k_{20}}{2\beta k_{10}} + \frac{1}{2\beta}\bigg). \label{hd-mu-lower-cond-weak-other}
 \end{align}
where $\mu_\text{max}=(2+k_{20}/k_{10})$. When it is zero or negative, the HD phase can be obtained for any value of $\mu$, however small it may be.

To illustrate with examples, let us consider the case where $k_{10}=k_{20}=0.95$. With these values of particle exchange rates, the required range of $\mu$ for the existence of the HD phase is $(2.24-1/\beta) < \mu < 3$. When setting $\mu=1$, the HD phase is possible only for $\beta<0.88$, as depicted in the phase diagram of Fig.~\ref{pd weak 3}. In the second case where $k_{10}=0.01, k_{20}=0.95$, the range over which HD phase occurs is calculated as $(71.5-48/\beta)<\mu < 97$. Clearly, when $\mu=1$, HD phase must appear in the region where $\beta < 0.68$, which is consistent with Fig.~\ref{pd weak 2}.  Fig.~\ref{ld-hd-mc-st-wk-den} shows the HD phase density profiles in the weak coupling limit for $\mu=2.5$ and two different values of $\beta$: $\beta=0.1$ and $\beta=1$. The MCS results agree well with the MFT predictions for the lower $\beta$ value, but deviate quantitatively for the higher $\beta$ value.

 \subsection{ Maximal current phase}
 \label{MC Phase model L weak}

  The MC phase is characterised by a steady state bulk density of $\rho_\text{MC}=1/2$ when homogeneous hopping with a rate of 1 occurs throughout the lattice bulk. The PNC relation then reads $N_{0}=N_{1}+N_{2}+L/2$ in the MC phase.
 Similar to the open TASEP, the MC phase in this model arises under the following conditions:

\begin{eqnarray}
  &&\alpha_\text{eff}=\alpha\frac{N_{1}}{L}>\frac{1}{2},\\
  &&\beta_\text{eff}=\beta\bigg(1-\frac{N_{2}}{L}\bigg)>\frac{1}{2}.
 \end{eqnarray}
 Using the condition that at the boundary between the LD and MC phases, $\rho_\text{LD}=1/2$, and at the boundary between the HD and MC phases, $\rho_\text{HD}=1/2$, the LD-MC and HD-MC phase boundaries may be obtained by setting $\rho_\text{LD}=1/2$ and $\rho_\text{HD}=1/2$ in Eqs.~(\ref{LD density weak model 1}) and (\ref{HD density weak model 1}) respectively. We find
 \begin{eqnarray}
  &&\alpha=\frac{k_{10}+k_{20}}{(2\mu-1)k_{20}-\frac{1}{2}}, \label{ldmc bnd wk}\\
  &&\beta=\frac{k_{10}+k_{20}}{(3-2\mu)k_{10}+2k_{20}-\frac{1}{2}}\label{hdmc bnd wk}
 \end{eqnarray}
 respectively, as the LD-MC and HD-MC phase boundaries, which are just straight lines parallel to $\beta$- and $\alpha$-axes according to MFT, respectively, in the $\alpha-\beta$ plane; see Fig.~\ref{pd weak 3} and Fig.~\ref{pd weak 2}.

 The non-negativity of $\alpha$ and $\beta$, when considered in (\ref{ldmc bnd wk}) and (\ref{hdmc bnd wk}), demands the following two conditions on $\mu$:
 \begin{eqnarray}
  &&\mu>\left(\frac{1}{2}+\frac{1}{4k_{20}}\right),\label{mu-gr-wk}\\
  &&\mu < \left(\frac{3}{2}+\frac{k_{20}}{k_{10}}-\frac{1}{4k_{10}}\right) \label{mu-less-wk}
 \end{eqnarray}
respectively. We thus find the lower and upper thresholds of $\mu$ in (\ref{mu-gr-wk}) and (\ref{mu-less-wk}) for MC phase existence. Taken together, the range of $\mu$ over which MC phase appears is as follows:
\begin{equation}
\label{range of mu for mc phase model 1 weak coupling}
\left(\frac{1}{2}+\frac{1}{4k_{20}}\right) < \mu < \left(\frac{3}{2}+\frac{k_{20}}{k_{10}}-\frac{1}{4k_{10}}\right).
\end{equation}
The fact that the lower and upper thresholds of $\mu$ for the MC phase to exist depend only on the particle exchange rates and not on the entrance and exit rate parameters implies that the occurrence of the MC phase is solely a bulk phenomenon and has no connection with the boundary conditions.

To ensure that the thresholds of $\mu$ obtained in (\ref{range of mu for mc phase model 1 weak coupling}) for MC phase existence is meaningful, the upper threshold must be greater than the lower threshold for any value of $k_{10}$ and $k_{20}$:
\begin{equation}
\label{meaningful-m-limits}
\left(\frac{3}{2}+\frac{k_{20}}{k_{10}}-\frac{1}{4k_{10}}\right) > \left(\frac{1}{2}+\frac{1}{4k_{20}}\right).
\end{equation}
Simplifying this inequality, we obtain the following condition for the MC phase to occur:
\begin{equation}
\label{cond-for-mc-existence}
\mu_\text{max} > \bigg[1+\frac{1}{4}\bigg(\frac{1}{k_{10}}+\frac{1}{k_{20}}\bigg)\bigg],
\end{equation}
where $\mu_\text{max}$ is defined as $\mu_\text{max}=(2+k_{20}/k_{10})$.

In the special case where $k_{10}=k_{20}=k_{0}$, the range of $\mu$ over which MC phase comes into existence is determined by the condition (\ref{range of mu for mc phase model 1 weak coupling}), which translates as $(1/2+1/4k_{0})<\mu<(5/2-1/4k_{0})$, clearly demonstrating the particle-hole symmetry of the model with $\mu=3/2$ as the half-filled limit. We consider $k_{10}=k_{20}=0.95$, for which the range of $\mu$ for the existence of the MC phase becomes $0.76<\mu<2.24$ according to MFT. This is supported by MCS in Fig.~\ref{pd weak 3}, where no MC phase is observed at $\mu=0.5$. In the second case, where the exchange rates are highly asymmetric with $k_{10}=0.01$ and $k_{20}=0.95$, the MC phase exists within the range of $0.76<\mu<71.5$, according to MFT. Finally, in the third case when $k_{10}=1$ and $k_{20}=0.1$, MC phase does not exist as $\mu_\text{max}=(2+k_{20}/k_{10})=2.1$, inconsistent with the condition (\ref{cond-for-mc-existence}) which requires $\mu_\text{max}$ to be greater than 3.75 for this choice of $k_{10},\, k_{20}$; see Fig.~\ref{pd weak 11}. The condition for the existence of the MC phase in the limit of $k_{10}\rightarrow 0$ is determined by the upper limit of $\mu$ in (\ref{range of mu for mc phase model 1 weak coupling}), which can be expressed as $[3/2+1/k_{10}(k_{20}-1/4)]$. If $k_{20}$ is greater than 1/4, the MC phase exists due to the HD-MC boundary remaining on the positive side of the $\alpha$-axis. However, if $k_{20}$ is less than 1/4, the MC phase is absent. Therefore, the condition for the MC phase's existence in the limit $k_{10}\rightarrow 0$ is governed by the requirement of $k_{20}>1/4$. Fig.~\ref{ld-hd-mc-st-wk-den} exhibits the MC phase density profile in the weak coupling limit for $\mu=1.5$ and $\alpha=\beta=2.5$.

 \subsection{Domain wall phase}
 \label{SP Phase model L weak}

 In an open TASEP, the LD and HD phases meet when $\alpha_T=\beta_T<1/2$, or $\rho_\text{LD}+\rho_\text{HD}=1$, which is a straight line in the plane of the control parameters $\alpha_T,\,\beta_T$, starting at (0,0) and terminating at (1/2,1/2). Precisely on this line, there is coexistence of the LD and HD phases in the form of a single delocalised domain wall (DDW)~\cite{blythe}. The delocalization is due to particle non-conserving dynamics. { In contrast, global particle number conservation can pin a domain wall, as shown in a variety of TASEP-like models in ring geometries; see, e.g., Refs.~\cite{lebo,hinsch,reser1,reser2,anjan-parna,astik-1tasep}.}

 In the present model, global particle number conservation ensures the ensuing domain wall to be confined at a specific location, say $x_{w}$, in the bulk of TASEP lane.  We thus get a \textit{localised} domain wall (LDW).  How particle number conservation plays out in our model to pin a domain wall, giving its height and location and their quantitative dependence on the model parameters and the associated phase boundaries is a question that we address in this Section. We begin by noting that the non-uniform spatial dependence of density in the DW phase can be represented as follows:
 \begin{equation}
 \label{rho-dw-wk}
 \rho(x)=\rho_\text{LD}+\Theta(x-x_{w})(\rho_\text{HD}-\rho_\text{LD}),
 \end{equation}
 where $\Theta$ is the Heaviside step function defined as $\Theta(x)=1(0)$ for $x>(<)0$. In analogy with an open TASEP, the condition for the DW phase is as follows:
 \begin{equation}
 \label{Domain wall condition}
 \rho_\text{LD}+\rho_\text{HD}=1,
 \end{equation}
 which yields

 \begin{align}
 &\alpha_\text{eff}=\beta_\text{eff} \\
 \implies & \alpha \frac{N_{1}}{L}=\beta \bigg(1-\frac{N_{2}}{L}\bigg) \label{LD-HD coexistence model 1}
 \end{align}

 We now obtain the exact location $x_{w}$ and height $\Delta$ of the LDW in terms of the control parameters.
  In the DW phase, particle number in $T$ can be expressed as
  \begin{equation}
  \label{nt-dw-wk}
 N_{T} = L \int_{0}^{1} \rho(x) dx,
 \end{equation}
   where a multiplicative factor $L$ is introduced in the right-hand side to rescale the integration limit of the position variable $x$. Using (\ref{rho-dw-wk}) and (\ref{LD-HD coexistence model 1}) in (\ref{nt-dw-wk}), we get:
 \begin{equation}
 \label{NT model 1}
 N_{T} = L\bigg[ \alpha \frac{N_{1}}{L} (2x_{w}-1)+1-x_{w}\bigg].
 \end{equation}

 Identifying the steady state TASEP current in the DW phase as $J_{T}=\rho_\text{LD} (1-\rho_\text{LD})$ or $J_{T}=\rho_\text{HD} (1-\rho_\text{HD})$ and substituting the expressions of $N_{T}$ [see Eq.~(\ref{NT model 1})] and $J_{T}$ in Eq.~(\ref{N1}) together with PNC, we obtain the following two equations coupled in $N_{1}/L$ and $x_{w}$:
 \begin{eqnarray}
  &&\frac{N_{1}}{L} \bigg[ 1-\frac{\alpha}{\beta} + \alpha(2x_{w}-1)\bigg]-x_{w}+1=\mu-1, \label{First coupled equation model 1}\\
  &&\begin{split}
  \frac{N_{1}}{L}&=\frac{k_{2}}{k_{1}+k_{2}}\bigg[\mu-\alpha \frac{N_{1}}{L} (2x_{w}-1)-1+x_{w}\bigg]\\
  &\quad -\frac{1}{L(k_{1}+k_{2})}\alpha\frac{N_{1}}{L}\bigg(1-\alpha\frac{N_{1}}{L}\bigg).
 \end{split} \label{Second coupled equation model 1}
 \end{eqnarray}
 While solving Eqs.~(\ref{First coupled equation model 1}) and (\ref{Second coupled equation model 1}) for $N_{1}/L$ and $x_{w}$, we get a qudratic equation for $N_{1}/L$ with two solutions:
 \begin{align}
  \begin{split}
  \bigg(\frac{N_{1}}{L}\bigg)^{\pm}&=\bigg(\frac{1}{2\alpha}+\frac{k_{10}}{2\alpha^{2}}+\frac{k_{20}}{2\alpha\beta}\bigg) \\
  & \quad \pm \bigg[\bigg(\frac{1}{2\alpha}+\frac{k_{10}}{2\alpha^{2}}+\frac{k_{20}}{2\alpha\beta}\bigg)^{2}-\frac{k_{20}}{\alpha^{2}}\bigg]^{\frac{1}{2}}.
  \end{split}
 \end{align}
 The density in the LD part of the DW is thus
 \begin{equation}
  \label{ld-den-dw-phase-weak}
  \rho_\text{LD}=\alpha\frac{N_{1}}{L}=\bigg(\frac{1}{2}+\frac{k_{10}}{2\alpha}+\frac{k_{20}}{2\beta}\bigg)\pm \\
  \bigg[\bigg(\frac{1}{2}+\frac{k_{10}}{2\alpha}+\frac{k_{20}}{2\beta}\bigg)^{2}-k_{20}\bigg]^{\frac{1}{2}}.
 \end{equation}
 At the boundary between the LD and the DW phases, MFT must predict identical (low) density in the bulk of $T$. We now argue that in (\ref{ld-den-dw-phase-weak}) the solution with a negative discriminant is actually the physically acceptable solution. Equating the density in LD phase [Eq.~(\ref{LD density weak model 1})] and density in the LD domain of DW phase [Eq.~(\ref{ld-den-dw-phase-weak}) with negative discriminant], we obtain the LD-DW boundary as follows:
 \begin{align}
  \label{lddw-boundary-1}
  &\bigg({\frac{1+k_{20}}{2}+\frac{k_{10}+k_{20}}{2\alpha}}\bigg) \nonumber\\
  &-\bigg[\bigg({\frac{1+k_{20}}{2}+\frac{k_{10}+k_{20}}{2\alpha}}\bigg)^{2}-\mu k_{20}\bigg]^{\frac{1}{2}} \nonumber\\
  &=\bigg(\frac{1}{2}+\frac{k_{10}}{2\alpha}+\frac{k_{20}}{2\beta}\bigg)-
  \bigg[\bigg(\frac{1}{2}+\frac{k_{10}}{2\alpha}+\frac{k_{20}}{2\beta}\bigg)^{2}-k_{20}\bigg]^{\frac{1}{2}}.
 \end{align}

 The LD-DW boundary equation (\ref{lddw-boundary-1}) can be simplified further (detailed calculation is given in Appendix) after which it reduces to
 \begin{equation}
\frac{N_{1}}{L}\bigg(1+\alpha-\frac{\alpha}{\beta}\bigg)=\mu-1 \label{lddw-boundary-earl}
\end{equation}
where $N_{1}/L$ is given by
 \begin{align}
  \label{n1/l weak L}
  \begin{split}
  \frac{N_{1}}{L}&=\bigg(\frac{1}{2\alpha}+\frac{k_{10}}{2\alpha^{2}}+\frac{k_{20}}{2\alpha\beta}\bigg) \\
  &\quad -  \bigg[\bigg(\frac{1}{2\alpha}+\frac{k_{10}}{2\alpha^{2}}+\frac{k_{20}}{2\alpha\beta}\bigg)^{2}-\frac{k_{20}}{\alpha^{2}}\bigg]^{\frac{1}{2}}.
  \end{split}
  \end{align}
  Therefore, the acceptable solution of $N_{1}/L$ is the one with a negative discriminant; see Eq.~(\ref{n1/l weak L}). Next, the expression for $N_{2}$ can be obtained using Eq.~(\ref{LD-HD coexistence model 1}). One finds

  \begin{align}
  \frac{N_{2}}{L}&=1-\frac{\alpha}{\beta}\frac{N_{1}}{L}\nonumber \\
  &=1-\bigg(\frac{1}{2\beta}+\frac{k_{10}}{2\alpha\beta}+\frac{k_{20}}{2\beta^{2}}\bigg) \nonumber \\
  &\quad +\bigg[\bigg(\frac{1}{2\beta}+\frac{k_{10}}{2\alpha\beta}+\frac{k_{20}}{2\beta^{2}}\bigg)^{2}-\frac{k_{20}}{\beta^{2}}\bigg]^{\frac{1}{2}}. \label{n2/l weak L}
  \end{align}
  Once we obtain the expression for $N_{1}/L$ in Eq.~(\ref{n1/l weak L}), the position $x_{w}$ of the domain wall can be obtained using Eq.~(\ref{First coupled equation model 1}). We find:

  \begin{equation}
 \label{xw model L weak coupling}
  x_{w}=\frac{\bigg(1-\alpha-\frac{\alpha}{\beta}\bigg)\frac{N_{1}}{L}-\mu+2}{1-2\alpha\frac{N_{1}}{L}}.
 \end{equation}

  Having the expression of $N_{1}/L$, it is straightforward to obtain the low and high densities in the DW phase:
\begin{align}
  \rho_\text{LD} &= \left(\frac{1}{2}+\frac{k_{10}}{2\alpha}+\frac{k_{20}}{2\beta}\right)-\left[\left(\frac{1}{2}+\frac{k_{10}}{2\alpha}+\frac{k_{20}}{2\beta}\right)^{2}-k_{20}\right]^{\frac{1}{2}},
 \end{align}
together with $ \rho_\text{HD} = 1-\rho_\text{LD}$.
Now defined as the density difference between high and low density parts, the height ($\Delta$) of the DW comes out to be
  \begin{align}
  \Delta &= \rho_\text{HD}-\rho_\text{LD} \nonumber\\
  &= \bigg[\bigg(1+\frac{k_{10}}{\alpha}+\frac{k_{20}}{\beta}\bigg)^{2}-4k_{20}\bigg]^\frac{1}{2}-\bigg(\frac{k_{10}}{\alpha}+\frac{k_{20}}{\beta}\bigg) \label{delta model L weak coupling}
 \end{align}

 When transitioning from the LD phase to the DW phase, the domain wall should be located at the extreme right end or the exit end of the TASEP lane. Similarly, at the transition from HD to DW phase, one finds the DW to be at the extreme left end or the entrance end. One thus sets $x_{w}=1$ in Eq.~(\ref{xw model L weak coupling}) to obtain the LD-DW phase boundary:
 \begin{eqnarray}
  &&\frac{N_{1}}{L}\bigg(1+\alpha-\frac{\alpha}{\beta}\bigg)=\mu-1, \label{ldsp bnd wk}
 \end{eqnarray}
 and $x_{w}=0$ to obtain the HD-DW phase boundary:
 \begin{eqnarray}
  &&\frac{N_{1}}{L}\bigg(1-\alpha-\frac{\alpha}{\beta}\bigg)=\mu-2, \label{hdsp bnd wk}
 \end{eqnarray}
 where $N_{1}/L$ is given in Eq. (\ref{n1/l weak L}).

 Within the DW phase region, the condition $\alpha_\text{eff}=\alpha N_{1}/L<1/2$, or $(1-2\alpha N_{1}/L)>0$ must be followed. Consequently, for $x_{w}>0$ in Eq.~(\ref{xw model L weak coupling}), we must have its numerator positive:
 \begin{align}
  &\bigg[\bigg(1-\alpha-\frac{\alpha}{\beta}\bigg)\frac{N_{1}}{L}-\mu+2\bigg]>0 \label{num-pos-wk} \\
  \implies &\mu<\bigg[\bigg(1-\alpha-\frac{\alpha}{\beta}\bigg)\frac{N_{1}}{L}+2\bigg]. \label{mu-less-than-wk-dw}
 \end{align}
This sets the upper threshold of $\mu$ for DW phase existence. To determine the lower threshold of $\mu$, we consider the condition $x_{w}<1$ in Eq.~(\ref{xw model L weak coupling}), leading to the following condition:
\begin{equation}
 \label{lower-mu-dw}
 \mu > \bigg[\bigg(1+\alpha-\frac{\alpha}{\beta}\bigg)\frac{N_{1}}{L}+1\bigg].
\end{equation}
Hence, the range of $\mu$ over which the DW phase appears is
\begin{equation}
 \label{range-of-mu-dw}
 \bigg[\bigg(1+\alpha-\frac{\alpha}{\beta}\bigg)\frac{N_{1}}{L}+1\bigg] < \mu < \bigg[\bigg(1-\alpha-\frac{\alpha}{\beta}\bigg)\frac{N_{1}}{L}+2\bigg].
\end{equation}

\begin{figure}[!h]
 \centering
 \includegraphics[width=\columnwidth]{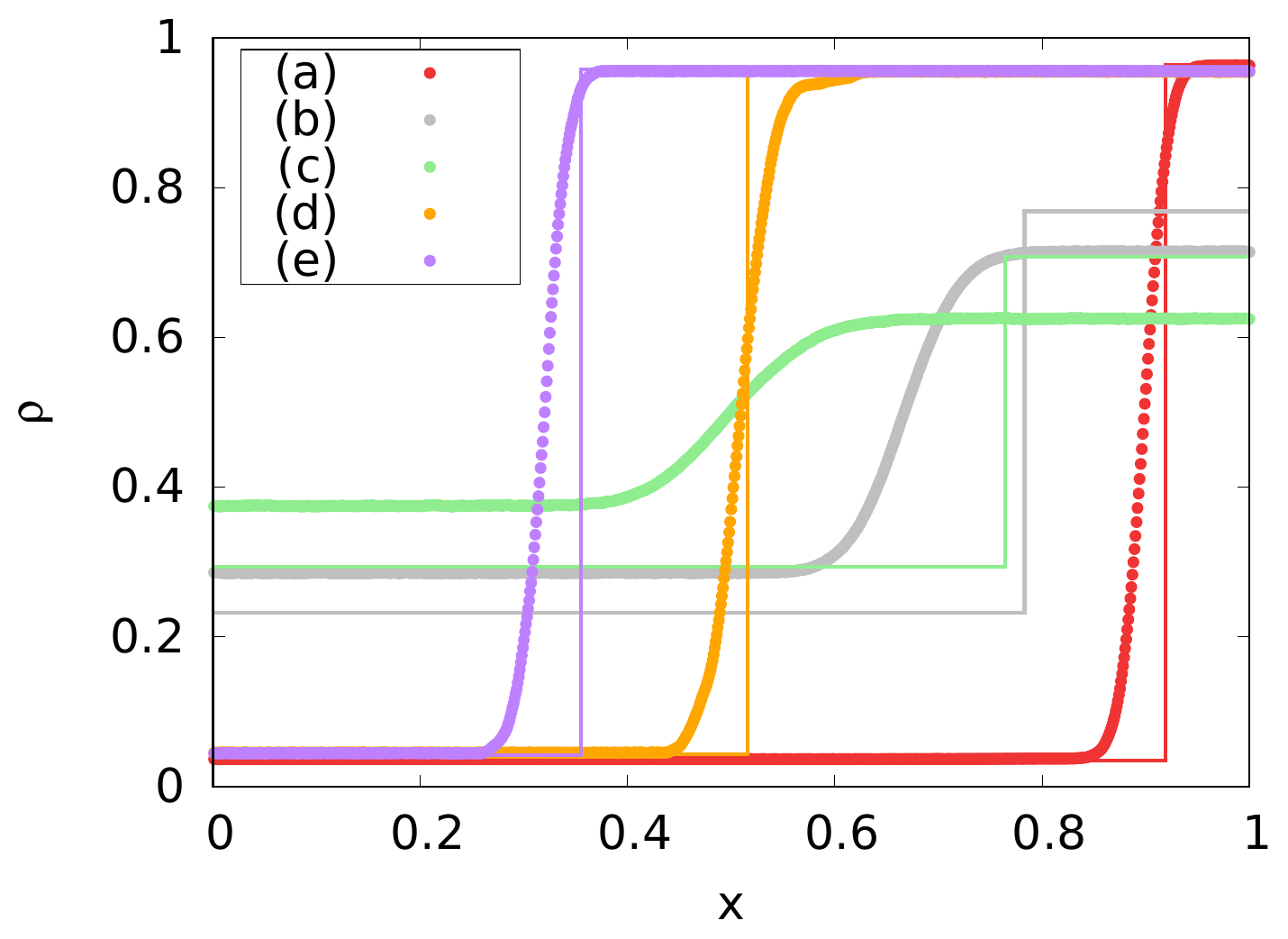}
 \caption{Plots of the density $\rho$ as a function of position $x$ in the DW phase under the weak coupling limit. The solid lines correspond to the  predictions from MFT, while the data points obtained from Monte Carlo simulations (MCS) share the same color as their respective mean-field counterparts. {System size used $L=1000$ and average over $2 \times 10^{9}$ Monte Carlo steps are done.} The other parameter values: \textbf{(a)} $\alpha=0.2$, $\beta=0.045$, $\mu=0.5$, $k_{10}=k_{20}=0.95$; \textbf{(b)} $\alpha=1$, $\beta=0.4$, $\mu=1$, $k_{10}=k_{20}=0.95$; \textbf{(c)} $\alpha=1.5$, $\beta=0.5$, $\mu=1$, $k_{10}=k_{20}=0.95$; \textbf{(d)} $\alpha=0.1$, $\beta=0.07$, $\mu=1.3$, $k_{10}=0.7$, $k_{20}=0.9$; and \textbf{(e)} $\alpha=0.2$, $\beta=0.05$, $\mu=1$, $k_{10}=0.4$, $k_{20}=0.8$. {Mismatch between MFT and MCS results are more pronounced for larger values of $\alpha$ and $\beta$, see sets \textbf{(b)} and \textbf{(c)}.}}
 \label{dw-weak-mu0.5-1}
 \end{figure}

  Fig.~\ref{dw-weak-mu0.5-1} displays density profiles in the DW phase under the weak coupling limit. Increasing values of $\mu$, $\alpha$, and $\beta$ amplify the discrepancies between MFT and MCS, emphasizing the limitations of the mean-field approximation in capturing system behavior.

 \begin{figure*}[htb]
 \includegraphics[width=\columnwidth]{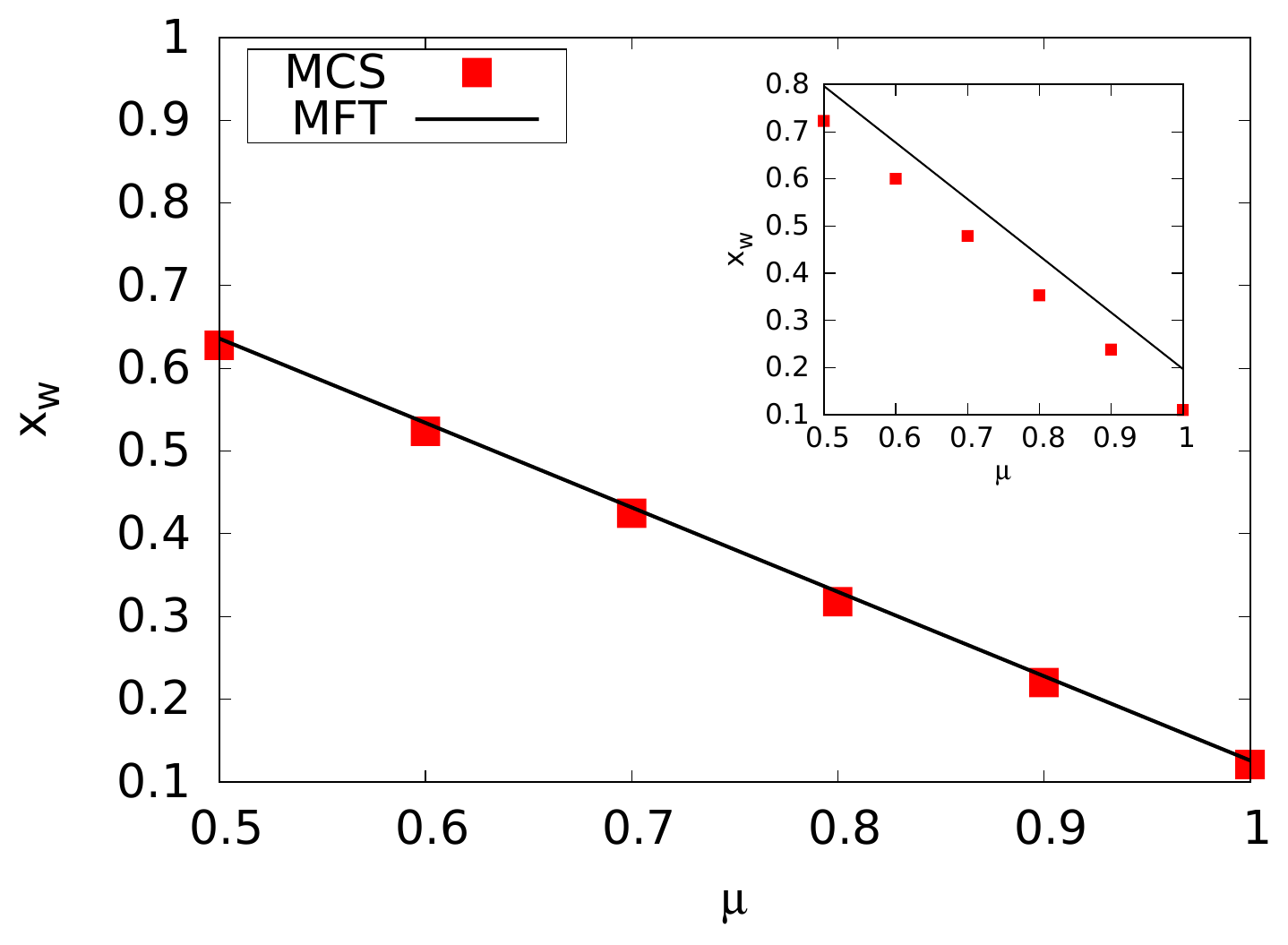}
 \hfill
 \includegraphics[width=\columnwidth]{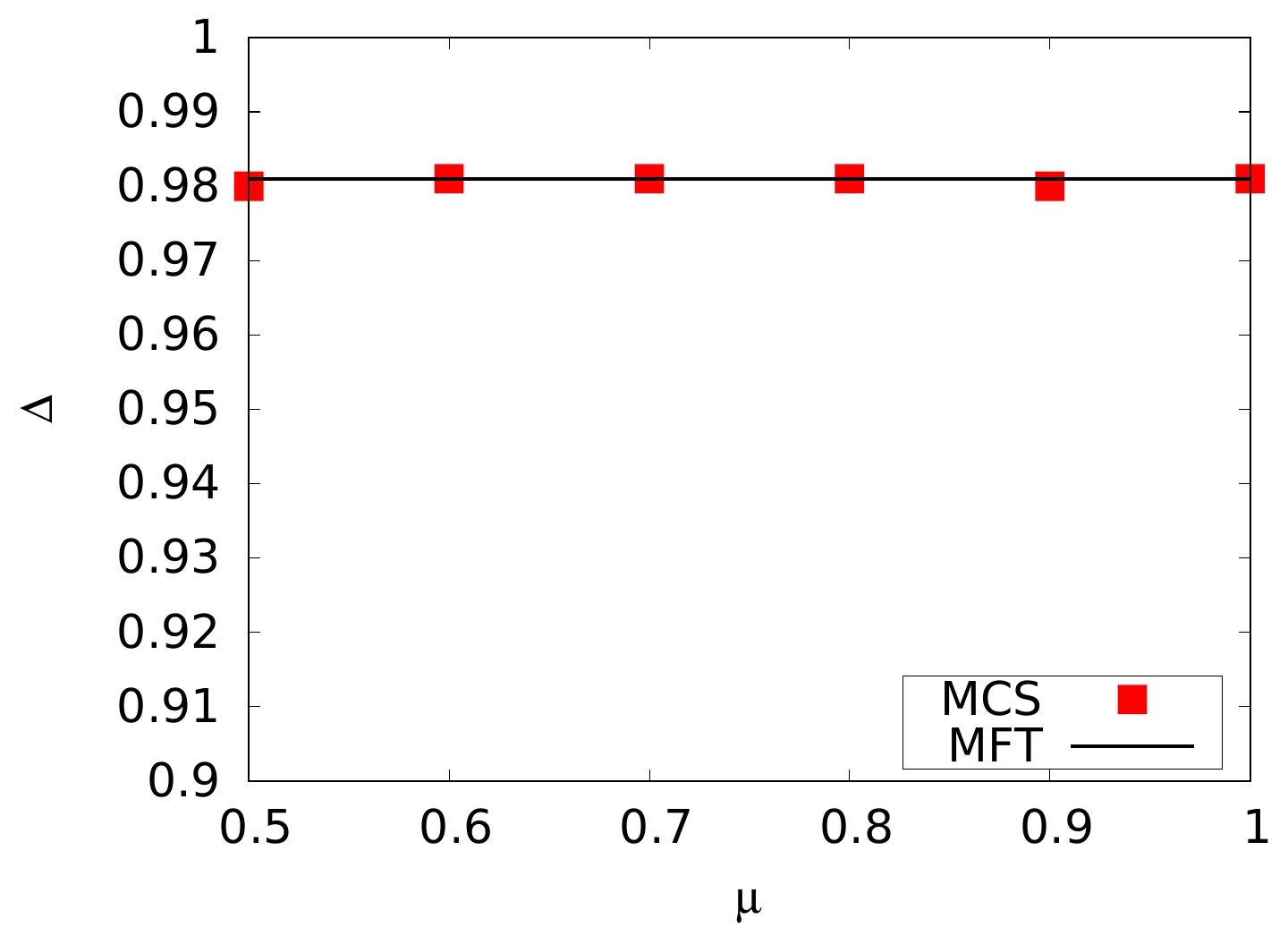}
 \\
\caption{\textbf{(Left)} Plot of the domain wall position $x_{w}$ versus filling factor $\mu$ for a fixed value of exchange rates ($k_{10}=k_{20}=0.95$) in the weak coupling limit of the model. In the main plot, entry and exit rates are $\alpha=0.15$ and $\beta=0.01$, wherein corresponding values of those parameters in the inset plot is set to $\alpha=1$ and $\beta=0.1$. The location of the domain walls decreases linearly with increasing values of $\mu$. Reducing the values of $\alpha$ and $\beta$ improves the agreement between MFT and MCS results.
\textbf{(Right)} Plot of the domain wall height $\Delta$ versus $\mu$ keeping $k_{10}=k_{20}=0.95$ in the weak coupling limit of the model with control parameters $\alpha=0.15$, $\beta=0.01$. The height of the domain wall $\Delta$ remains unchanged with varying $\mu$. Both the MFT and MCS results are in good agreement. Overall, the observed behaviour suggests a movement of the domain wall towards the entrance end of the TASEP lane as the particle supply, represented by $\mu$, increases, without changing the domain wall height.
}
\label{xw-del-vs-mu-weak}
\end{figure*}

 \begin{figure*}[htb]
 \includegraphics[width=\columnwidth]{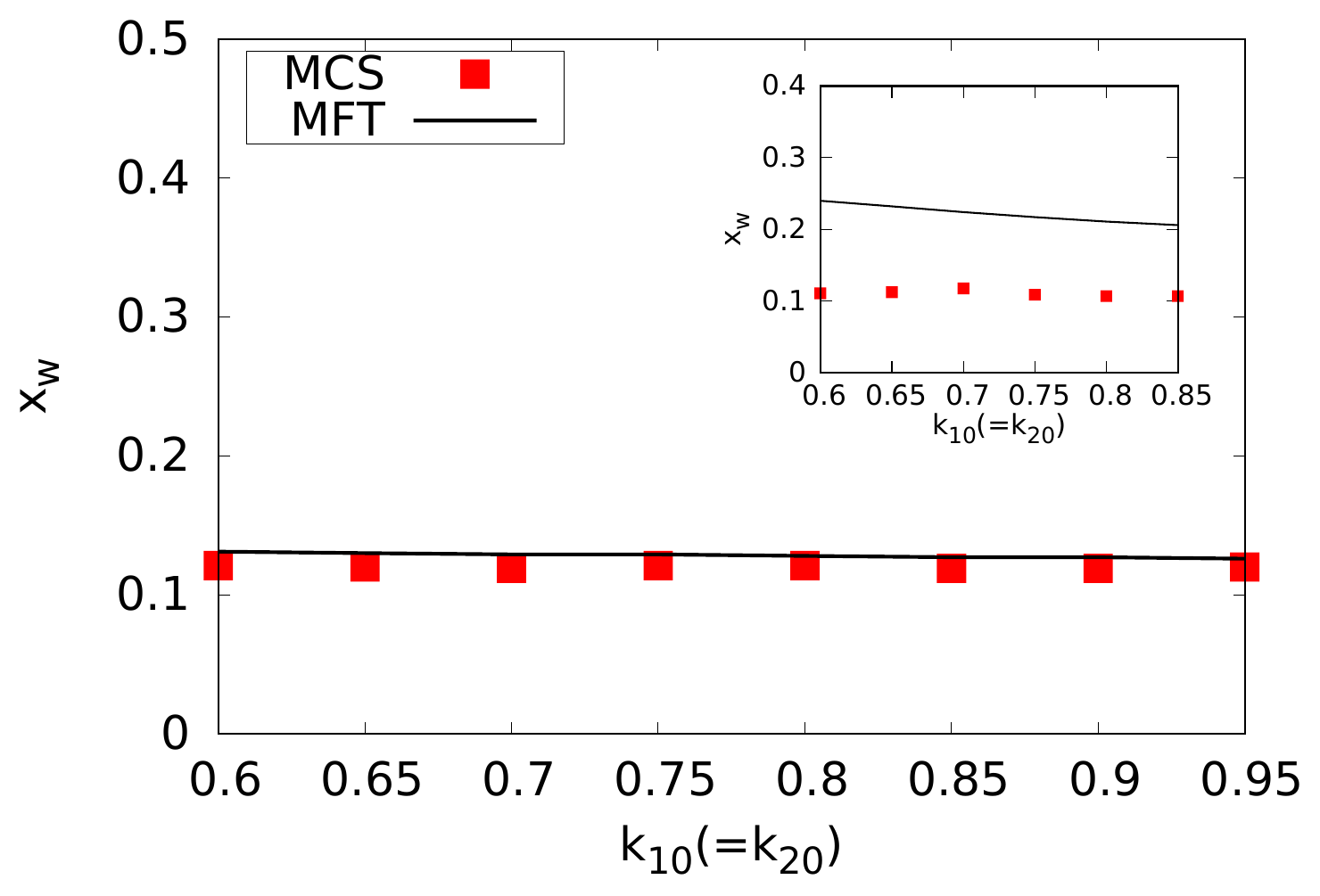}
 \hfill
 \includegraphics[width=\columnwidth]{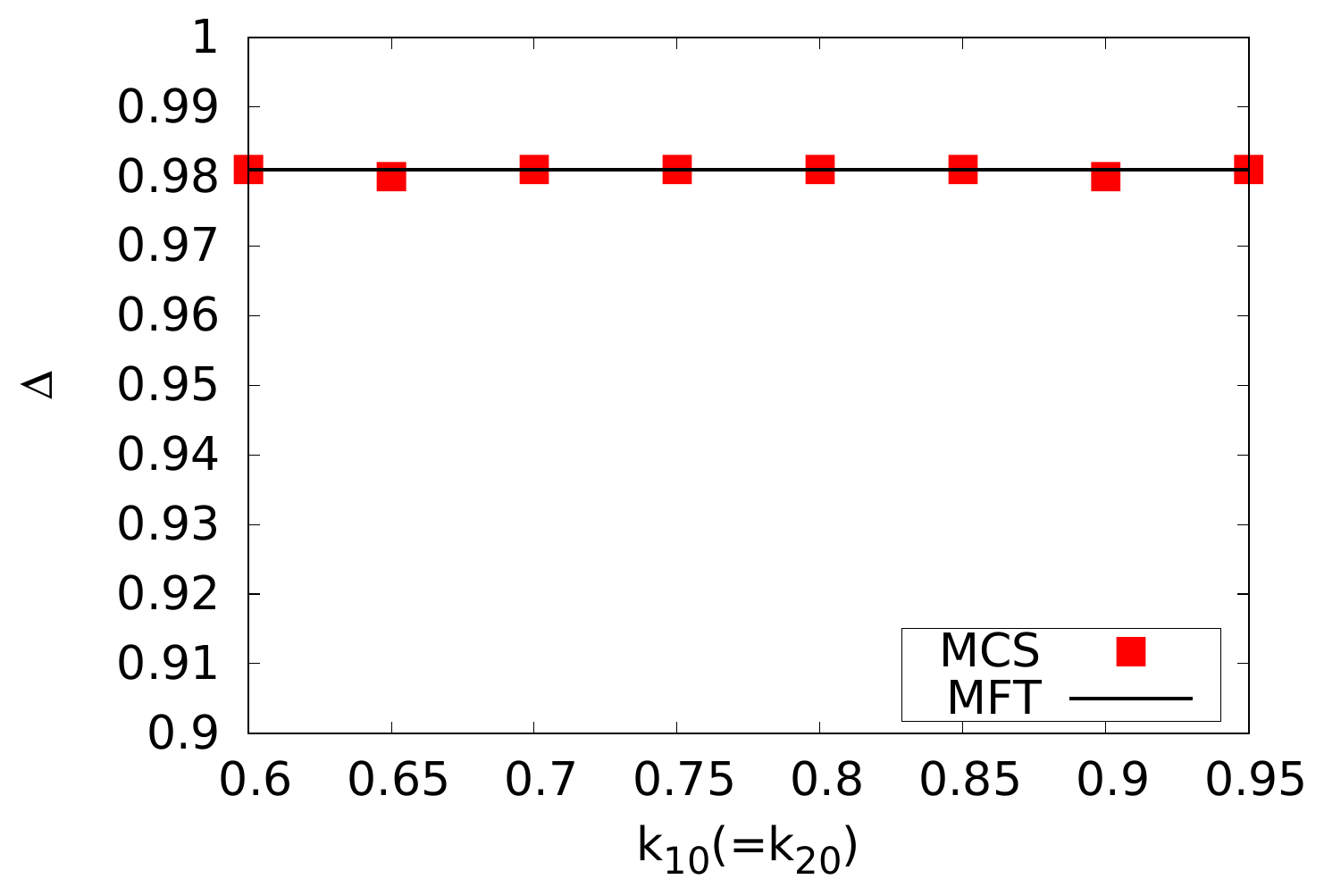}
 \\
\caption{\textbf{(Left)}  Plot of DW position $x_w$ vs. the exchange rates $k_{10} = k_{20}$ for a fixed value of filling factor ($\mu=1$) in the weak coupling limit of the model. The parameters used are $\alpha =0.15$, $\beta = 0.01$ (main plot), and $\alpha =1$, $\beta = 0.1$ (inset plot). Disagreement between the MFT and MCS results are reduced significantly for smaller values of $\alpha$, $\beta$.
\textbf{(Right)} Plot of DW height $\Delta$ with respect to the exchange rates $k_{10} = k_{20}$ in the weak coupling limit of the model for a fixed value of $\mu$. The control parameters employed are $\alpha =0.15$, $\beta = 0.01$, and $\mu = 1$. For the specific values of the parameters employed, the DW position and height may remain unchanged with the variation of exchange rates within the noted range. }
\label{xw-del-vs-k0-weak}
\end{figure*}

 We now consider how the DW location $x_w$ and its height $\Delta$ depend upon $\mu$. Our MF value of $x_w$, as given in Eq.~(\ref{xw model L weak coupling}), gives a linear decrease in $x_{w}$ with increasing $\mu$, while keeping other control parameters, such as $\alpha$, $\beta$, $k_{10}$, and $k_{20}$, unchanged. Further in the MFT, Eq.~(\ref{delta model L weak coupling}) shows that $\Delta$ remains independent of $\mu$ {as the reservoir populations $N_{1}$ and $N_{2}$ are insensitive to any change in $\mu$, see Eqs.~(\ref{n1/l weak L}) and (\ref{n2/l weak L})}. Thus, if there is a higher supply of particles into the TASEP lane, as would happen with a higher $\mu$, the DW position will move towards the entry end of $T$, leaving its height unchanged, to accommodate the additional particles.

Additionally, the relationship between the position or height of the domain wall and the exchange rate parameters $k_{10}$ and $k_{20}$, as given by Eqs.~(\ref{xw model L weak coupling}) and (\ref{delta model L weak coupling}), are  illustrated in Fig.~\ref{xw-del-vs-k0-weak}, where the behavior of LDW position and height assuming equal values for $k_{10}$ and $k_{20}$ are shown. Notably, under these specific conditions, the LDW position and height appear to be quite insensitive to the variations in the exchange rates.

 \subsection{{Phase boundaries meet at a common point}}

 Phase diagrams in Fig.~\ref{pd weak 3} and Fig.~\ref{pd weak 2} reveal the four phases --- LD, HD, MC, and DW --- to meet at a single point named as the multicritical point. This unique point is represented by the coordinates $(\alpha_\text{c},\beta_\text{c})$ in the $\alpha-\beta$ plane. In our MFT, the phase boundaries (\ref{ldmc bnd wk}), (\ref{hdmc bnd wk}), (\ref{ldsp bnd wk}) and (\ref{hdsp bnd wk}) meet at
\begin{equation}
(\alpha_\text{c},\beta_\text{c})=\bigg(\frac{k_{10}+k_{20}}{(2\mu-1)k_{20}-\frac{1}{2}},
\frac{k_{10}+k_{20}}{(3-2\mu)k_{10}+2k_{20}-\frac{1}{2}}\bigg).
\end{equation}
By definition, $\alpha_\text{c}$ and $\beta_\text{c}$ must be positive which puts the multicritical point within the first quadrant of the $\alpha-\beta$ phase diagram. This implies the following range of $\mu$ within which multicritical point exists:
\begin{equation}
\label{mcp-range-wk}
\bigg(\frac{1}{2}+\frac{1}{4k_{20}}\bigg) < \mu < \bigg(\frac{3}{2}+\frac{k_{20}}{k_{10}}-\frac{1}{4k_{10}}\bigg).
\end{equation}

Apparent from (\ref{mcp-range-wk}), the lower and upper thresholds of $\mu$ for the existence of a multicritical point depend only on the exchange rates $k_{10}$ and $k_{20}$. That the upper threshold in (\ref{mcp-range-wk}) is greater than the lower threshold gives the condition
\begin{equation}
\label{cond-mcp}
\mu_\text{max} > \left[1+\frac{1}{4}\left(\frac{1}{k_{10}}+\frac{1}{k_{20}}\right)\right].
\end{equation}
This condition (\ref{cond-mcp}) is consistent with the condition (\ref{cond-for-mc-existence}) for the existence of MC phase, derived earlier.

\begin{figure}[!h]
 \centering
 \includegraphics[width=\columnwidth]{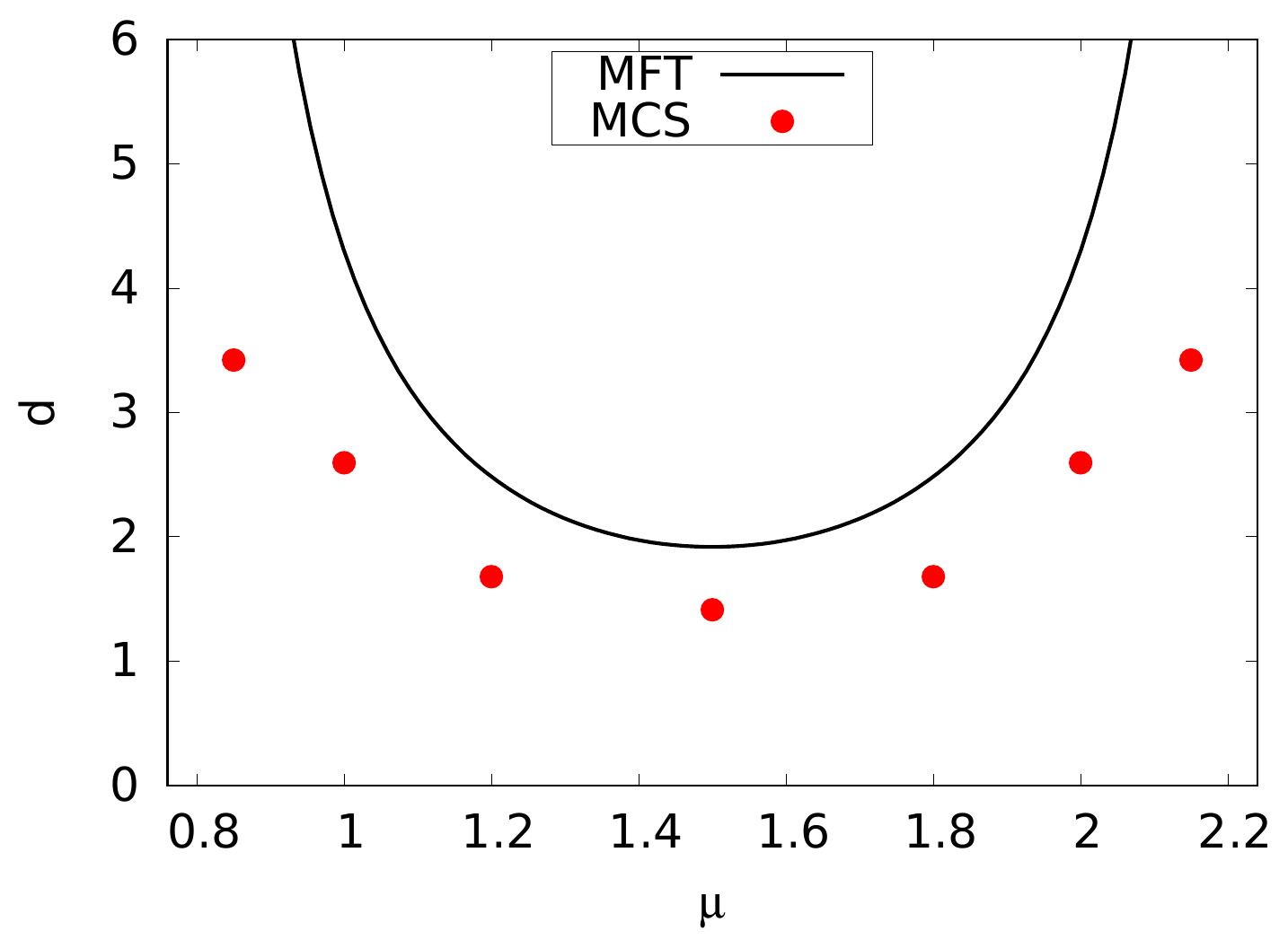}
 \caption{Plot of the distance $d$ between the origin and the multicritical point versus filling factor $\mu$ in the weak coupling limit of the model. Particle exchage rates between reservoirs are taken to be same, i.e., $k_{10}=k_{20}=0.95$. Black solid line and colored discrete points represent the MFT and MCS results respectively. Despite matching qualitatively, there is a quantitative discrepancy between the results from MFT and MCS.}
 \label{d-vs-mu-weak}
 \end{figure}

The distance $d$ between the origin $(0,0)$ and the multicritical point $(\alpha_\text{c},\beta_\text{c})$ is
\begin{equation}
d=\sqrt{\bigg(\frac{k_{10}+k_{20}}{(2\mu-1)k_{20}-\frac{1}{2}}\bigg)^{2}+\bigg(\frac{k_{10}+k_{20}}{(3-2\mu)k_{10}+2k_{20}-\frac{1}{2}}\bigg)^{2}}.
\end{equation}
Fig.~\ref{d-vs-mu-weak} shows the variation of $d$ with $\mu$ according to MFT and MCS. It is important to note that the value of $d$ diverges as $\mu$ approaches $(1/2+1/4k_{20})$ from above and $(3/2+k_{20}/k_{10}-1/4k_{10})$ from below. Thus, when $k_{10}=k_{20}=0.95$, $d$ diverges when $\mu$ goes to 0.76 from above or 2.24 from below according to MFT prediction. The MCS plot, however, shows these divergences in $d$ at a shifted value of $\mu$ from the theoretical predictions.

 \section{PARTICLE-HOLE SYMMETRY AND THE PHASE DIAGRAMS WITH EQUAL EXCHANGE RATES ($\mathbf{k_{10}=k_{20}}$)}
 \label{ph sym and pd}

\begin{figure}[!h]
 \centering
 \includegraphics[width=\columnwidth]{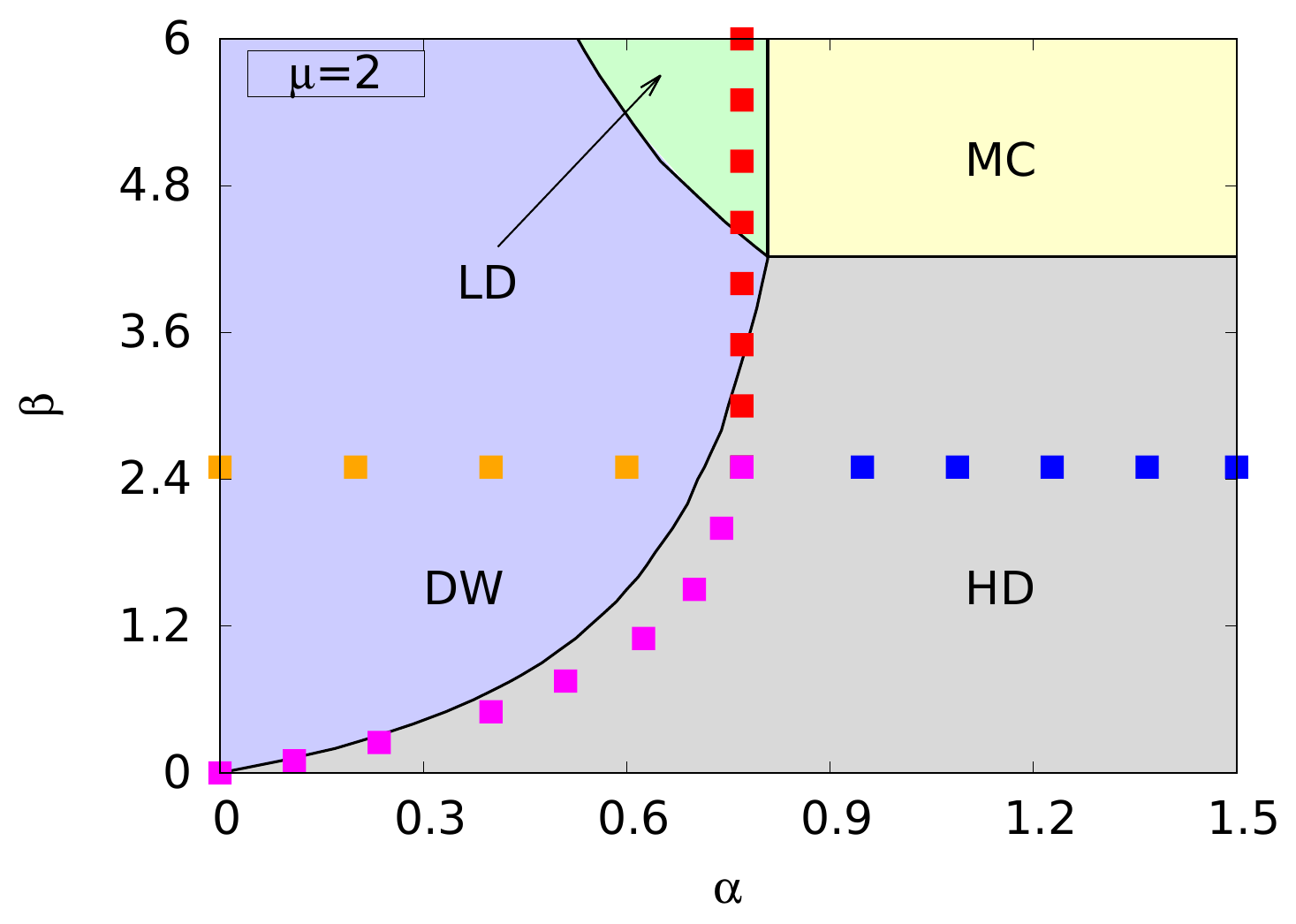}
 \caption{Phase diagram in the weak coupling limit of the model having particle-hole symmetry for $\mu=2$, with equal exchange rates $k_{10}=k_{20}=0.95$. Notably, this phase diagram exhibits a particle-hole symmetry with the phase diagram corresponding to $\mu=1$ in Fig.~\ref{pd weak 3}. All four phases are obtained for this choice of $\mu$. The background colors in the diagram represent different phases, while the black solid lines indicate the boundaries between those phases according to mean-field theory (MFT). To validate these boundaries, MCS  results are included in the form of colored discrete points representing the phase boundaries. }
 \label{ph-wkl}
 \end{figure}

  We consider the case where the particle exchange rates are same in magnitude, i.e., $k_{10}=k_{20}$, for which there is a particle-hole symmetry. The LD-MC phase boundary is interchanged with the HD-MC phase boundary, and the LD-DW phase boundary is interchanged with the HD-DW phase boundary, under the transformations $\mu \leftrightarrow 3-\mu$ and $\alpha \leftrightarrow \beta$; see Eqs.~(\ref{ldmc bnd wk}), (\ref{hdmc bnd wk}), (\ref{ldsp bnd wk}), and (\ref{hdsp bnd wk}). This symmetry around the half-filled limit ($\mu=3/2$) demonstrates the presence of particle-hole symmetry in the phase diagrams. The particle-hole symmetry can also be expressed in terms of the steady state densities in the LD and HD phases for equal exchange rates. One finds the correlation $\rho_\text{LD} \leftrightarrow 1-\rho_\text{HD}$; see Eqs.~(\ref{LD density weak model 1}) and (\ref{HD density weak model 1}) underlying. In Fig.~\ref{ph-wkl}, we present the phase diagram for $\mu=2$ which is connected to the phase diagram for $\mu=1$ shown in Fig.~\ref{pd weak 3} by this symmetry. However, the symmetry breaks when unequal exchange rates are considered (cf. Fig.~\ref{pd weak 2}).

 \section{NATURE OF THE PHASE TRANSITIONS}
 \label{dw nature}

The phases in the phase diagrams in Fig.~\ref{pd weak 3}, Fig.~\ref{pd weak 2}, and Fig.~\ref{pd weak 11} are demarcated by different phase boundaries. We will now
discuss the nature of the transitions across these phase
boundaries. In a TASEP with open boundaries, with the bulk density $\rho$ as the order parameter, the transition between
the LD and HD phases is accompanied by a sudden
jump in $\rho$, implying thus a
discontinuous or first-order transition. Similarly, the transitions between either LD or
HD and MC phases are continuous or second-order transitions, with the
density difference vanishing smoothly across the phase boundaries. The transitions in the present model can be described
in terms of the densities in the TASEP lane $T$ and can be
classified in analogy with open TASEP by noting the
density changes (sudden jump or smooth).
In the present model, all the phase transitions -- LD-DW, HD-DW, LD-MC, and HD-MC -- are continuous in nature, as the density changes continuously and smoothly across all of them. In Fig.~\ref{pd weak 3}, the phase diagram corresponding to $\mu=0.5$ exhibits only one second-order phase transition (LD-DW), whereas phase diagrams with other values of $\mu$ include all four second-order phase transitions (LD-DW, HD-DW, LD-MC, and HD-MC), according to MFT and MCS both. Next in Fig.~\ref{pd weak 2}, the phase diagram for $\mu=0.5$ shows one continuous phase transition (LD-DW), while for $\mu=1$ and $\mu=2$, all four continuous phase transitions are present, as predicted by both MFT and MCS. Interestingly, the phase diagram corresponding to $\mu=30$ exhibits one continuous phase transition (HD-MC) according to MFT but includes all four continuous phase transitions according to MCS. Lastly, in Fig.~\ref{pd weak 11}, phase diagram associated with $\mu=0.9$ and $\mu=1.7$ contains one continuous phase transition, LD-DW and HD-DW respectively, according to both MFT and MCS. For some ranges of $\mu$, there can be all the four phases which meet at a common point; see Fig.~\ref{pd weak 3} and Fig.~\ref{pd weak 2}. This common meeting point is then a {\em multicritical point}.

 \begin{figure}[!h]
 \centering
 \includegraphics[width=0.5\textwidth]{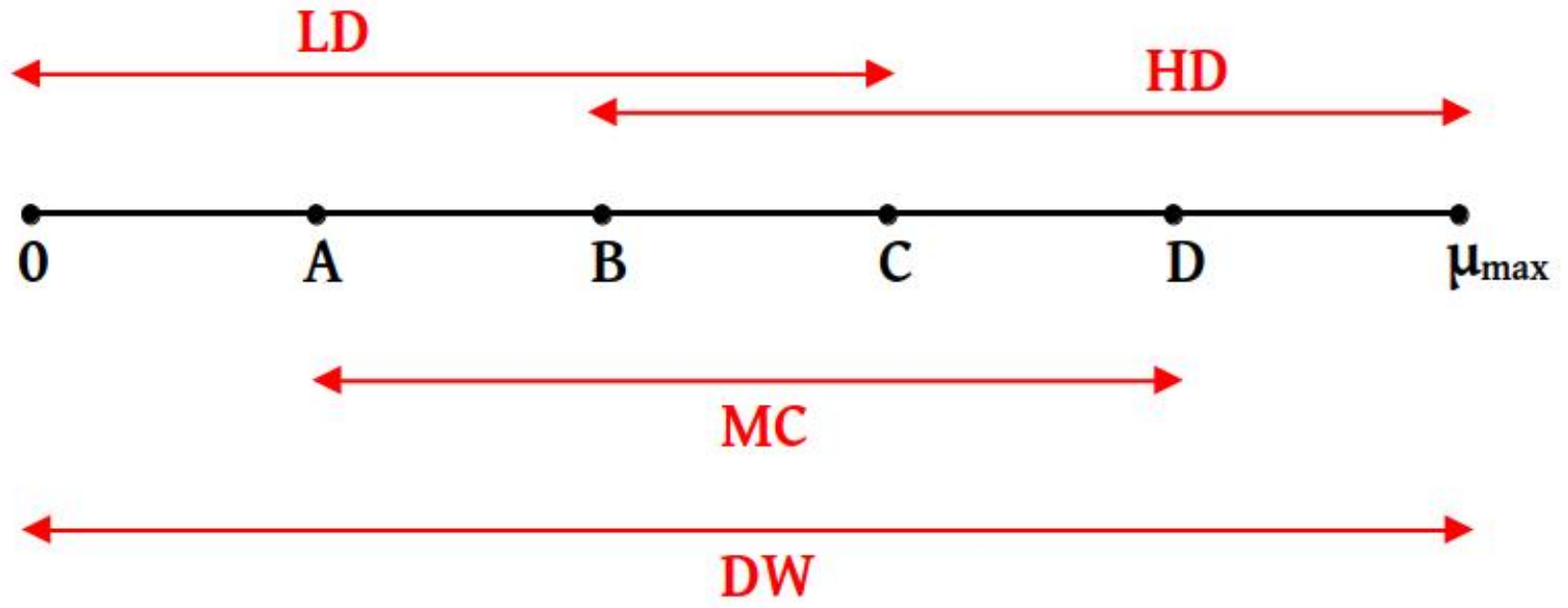}
 \caption{A schematic representation illustrating the ranges of $\mu$ where various phases emerge in the weak coupling limit of the model.}
 \label{mu-wkl-range}
 \end{figure}

  In Fig.~\ref{mu-wkl-range}, occurrence of different phases within a specific range of $\mu$ is tangible. Following are the expressions corresponding to the points on the $\mu$ axis.

 \begin{itemize}
  \item \textbf{A}: $\bigg(\frac{1}{2}+\frac{1}{4k_{20}}\bigg)$

  \item \textbf{B}: $\bigg(\frac{3}{2}+\frac{k_{20}}{k_{10}}-\frac{1}{4k_{10}}-\frac{k_{20}}{2\beta k_{10}}-\frac{1}{2\beta}\bigg)$

  \item \textbf{C}: $\bigg(\frac{1}{2}+\frac{1}{4k_{20}}+\frac{k_{10}}{2\alpha k_{20}}+\frac{1}{2\alpha}\bigg)$

  \item \textbf{D}: $\bigg(\frac{3}{2}+\frac{k_{20}}{k_{10}}-\frac{1}{4k_{10}}\bigg)$

  \item $\mathbf{\mu_\text{max}} =\bigg(2+\frac{k_{20}}{k_{10}}\bigg)$
 \end{itemize}
  These points are the thresholds of $\mu$ for the existence of different phases as obtained in (\ref{ld-mu-limit-weak}), (\ref{range of mu for hd phase model 1 weak coupling}), and (\ref{range of mu for mc phase model 1 weak coupling}). We restate below the conditions (\ref{ld-mu-cond-weak-other}), (\ref{hd-mu-lower-cond-weak-other}), and (\ref{cond-for-mc-existence}), which $\mu_\text{max}$ satisfies.
 \begin{eqnarray}
  &&\mu_\text{max} > \bigg(\frac{1}{2}+\frac{1}{4k_{20}}+\frac{k_{10}}{2\alpha k_{20}}+\frac{1}{2\alpha}\bigg), \label{cond-1} \\
  &&\mu_\text{max} > \bigg(\frac{1}{2}+\frac{1}{4k_{10}} + \frac{k_{20}}{2\beta k_{10}} + \frac{1}{2\beta}\bigg), \label{cond-2}\\
  &&\mu_\text{max} > \bigg[1+\frac{1}{4}\bigg(\frac{1}{k_{10}}+\frac{1}{k_{20}}\bigg)\bigg]. \label{cond-3}
 \end{eqnarray}
 It can be seen that C $\ge$ A and B $\le$ D for any (positive) values of $\alpha$, $\beta$, $k_{10}$, and $k_{20}$. The points B and C, in addition to being dependent on the parameters $k_{10}$ and $k_{20}$, also vary with the values of $\alpha$ and $\beta$. This means that by adjusting $\alpha$ and $\beta$ while $k_{10}$ and $k_{20}$ kept fixed, we can move these points along the $\mu$-axis.

 The point C is set to the right side of the point B. When C $\ge$ B, the following condition emerges:
 \begin{equation}
  \label{cond-4}
  \mu_\text{max}\le\bigg[1+\frac{1}{4}\bigg(\frac{1}{k_{10}}+\frac{1}{k_{20}}\bigg)+\frac{1}{2\alpha}\bigg(1+\frac{k_{10}}{k_{20}}\bigg)+\frac{1}{2\beta}\bigg(1+\frac{k_{20}}{k_{10}}\bigg)\bigg].
 \end{equation}
 For certain positive values of the control parameters $\alpha$, $\beta$, $k_{10}$, and $k_{20}$, the condition (\ref{cond-4}) can go with the conditions (\ref{cond-1}), (\ref{cond-2}), and (\ref{cond-3}). The possibility, C $<$ B, is defied as it leads to the condition:

\begin{equation}
  \label{cond-5}
  \mu_\text{max}>\bigg[1+\frac{1}{4}\bigg(\frac{1}{k_{10}}+\frac{1}{k_{20}}\bigg)+\frac{1}{2\alpha}\bigg(1+\frac{k_{10}}{k_{20}}\bigg)+\frac{1}{2\beta}\bigg(1+\frac{k_{20}}{k_{10}}\bigg)\bigg],
\end{equation}
which may or may not align with the conditions (\ref{cond-1}), (\ref{cond-2}), and (\ref{cond-3}). Thus, the point C can position itself only on the right side of point B or coincide with it, but cannot be at the left side of B.

 Now, let us explore how far the point C (or B) can be moved on the right (or left) side along the $\mu$-axis. Specifically, can point C be placed to the right of point D, or can point B be located to the left of point A? To help answer it, we, upon examining the expressions for points C and D, deduce that when C is positioned to the left of D, following condition for $\alpha$ has to be fulfilled:
\begin{equation}
\label{c-less-than-d}
\alpha > \frac{\bigg(1+\frac{k_{10}}{k_{20}}\bigg)}{2\bigg[1+\frac{k_{20}}{k_{10}}-\frac{1}{4}\bigg(\frac{1}{k_{10}}+\frac{1}{k_{20}}\bigg)\bigg]}.
\end{equation}
Moreover, to ensure the occurrence of the HD and MC phases, the corresponding boundary equation (\ref{hdmc bnd wk}) requires $\beta$ to be positive, which is fulfilled when $\mu <$ D. Now, at the boundary between LD and MC phases, $\alpha$ is given by (\ref{ldmc bnd wk}), substituting which in (\ref{c-less-than-d}) we get $\mu <$ D. If we consider positioning the point C on the right side of point D, it will lead to the condition:

\begin{equation}
\label{c-greater-than-d}
\alpha < \frac{\bigg(1+\frac{k_{10}}{k_{20}}\bigg)}{2\bigg[1+\frac{k_{20}}{k_{10}}-\frac{1}{4}\bigg(\frac{1}{k_{10}}+\frac{1}{k_{20}}\bigg)\bigg]},
\end{equation}
which corresponds to $\mu >$ D, or to put in other way, $\beta<0$. This is unphysical. These considerations restrict the location of point C in between the points A and D. In a way similar to what is done considering the points C and D, we can also conclude the location of point B to be constrained between the points A and D.

In summary, keeping the exchange rates $k_{10}$ and $k_{20}$ fixed, the positions of points A, D, and $\mu_\text{max}$ remain unchanged along the $\mu$-axis. On the other hand, points B and C have the flexibility to slide between A and D by tuning the values of $\alpha$ and $\beta$, with the constraint that C cannot be positioned to the left of B.

 A comprehensive summary of the different phases, domain walls, phase boundaries, and multicritical points can be found in Table~\ref{tab1}.

\begin{widetext}

\begin{table}[h!]
 \begin{center}
 \begin{tabular} { |p{0.8cm}|p{7.2cm}|p{3cm}|p{2cm}|p{2cm}|p{2cm}| }
  \hline
 \textbf{Row no.} & \textbf{Range of $\mu$} & \textbf{Phases} & \textbf{Domain walls} & \textbf{Phase boundaries} & \textbf{Multicritical points (MCPs)}  \\
 \hline
 1 & $0<\mu<\bigg(\frac{1}{2}+\frac{1}{4k_{20}}\bigg)$ & LD and DW  & One LDW & One second-order & None \\
 \hline
 2 & $\bigg(\frac{1}{2}+\frac{1}{4k_{20}}\bigg)<\mu<\bigg(\frac{3}{2}+\frac{k_{20}}{k_{10}}-\frac{1}{4k_{10}}-\frac{k_{20}}{2\beta k_{10}}-\frac{1}{2\beta}\bigg)$ & LD, HD, MC, and DW & One LDW & Four second-order & One four-phase MCP\\
 \hline
 3 & $\bigg(\frac{3}{2}+\frac{k_{20}}{k_{10}}-\frac{1}{4k_{10}}-\frac{k_{20}}{2\beta k_{10}}-\frac{1}{2\beta}\bigg)<\mu<\bigg(\frac{1}{2}+\frac{1}{4k_{20}}+\frac{k_{10}}{2\alpha k_{20}}+\frac{1}{2\alpha}\bigg)$ & LD, HD, MC, and DW & One LDW & Four second-order & One four-phase MCP\\
 \hline
 4 & $\bigg(\frac{1}{2}+\frac{1}{4k_{20}}+\frac{k_{10}}{2\alpha k_{20}}+\frac{1}{2\alpha}\bigg)<\mu<\bigg(\frac{3}{2}+\frac{k_{20}}{k_{10}}-\frac{1}{4k_{10}}\bigg)$ & LD, HD, MC, and DW & One LDW & Four second-order & One four-phase MCP\\
 \hline
 5 & $\bigg(\frac{3}{2}+\frac{k_{20}}{k_{10}}-\frac{1}{4k_{10}}\bigg)<\mu<(2+\frac{k_{20}}{k_{10}})$ & HD and DW & One LDW & One second-order & None\\
 \hline
 \end{tabular}
 \caption{ Table summarizing the occurence of certain phases and phase boundaries over a range of filling factor $\mu$ according to MFT in the weak coupling limit case. Particle-hole symmetry holds when a symmetric choice of exchange rates is considered, meaning $k_{10}=k_{20}$.}
 \label{tab1}
 \end{center}

\end{table}

\end{widetext}

\begin{figure*}[htb]
 \includegraphics[width=\columnwidth]{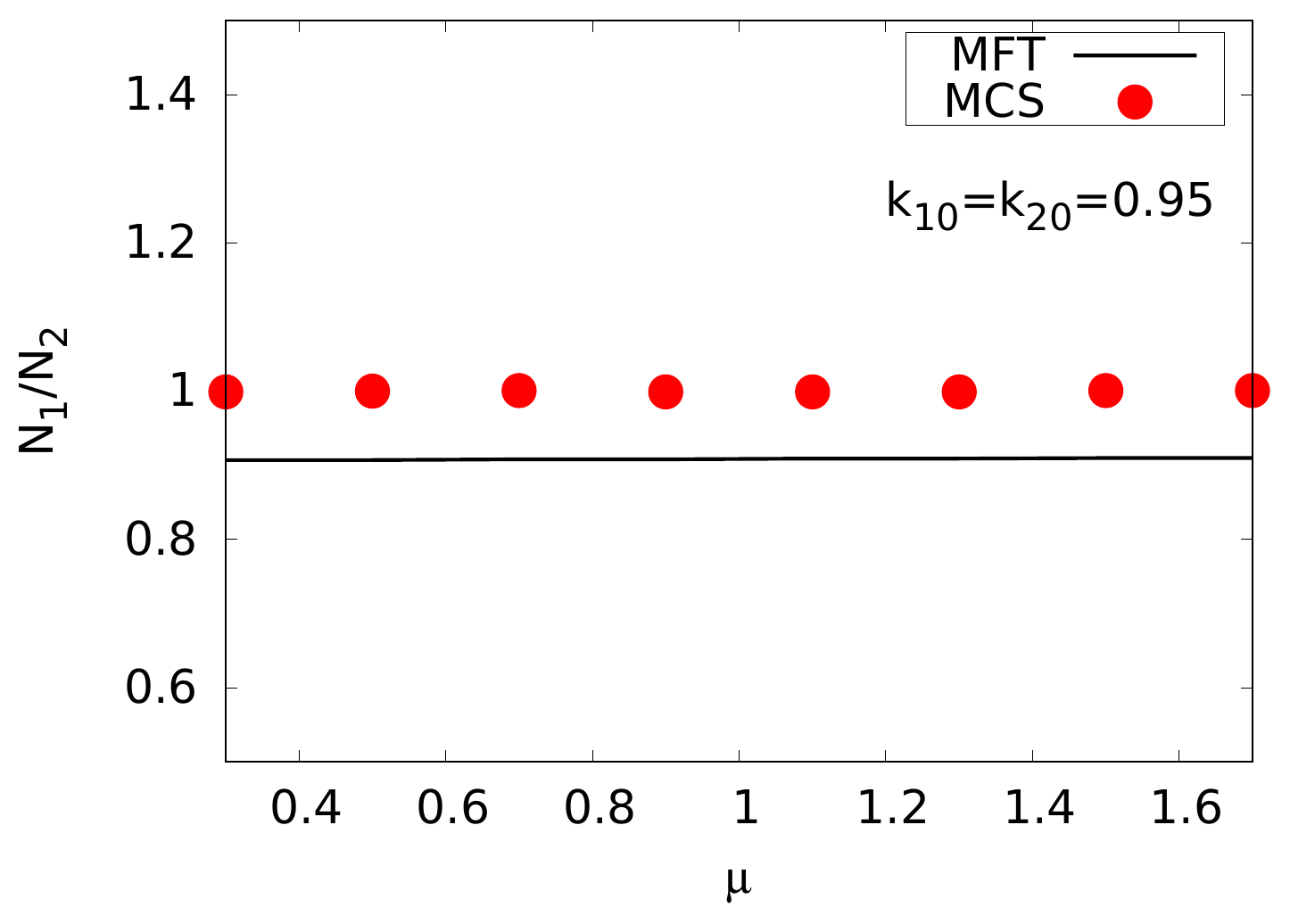}
 \hfill
 \includegraphics[width=\columnwidth]{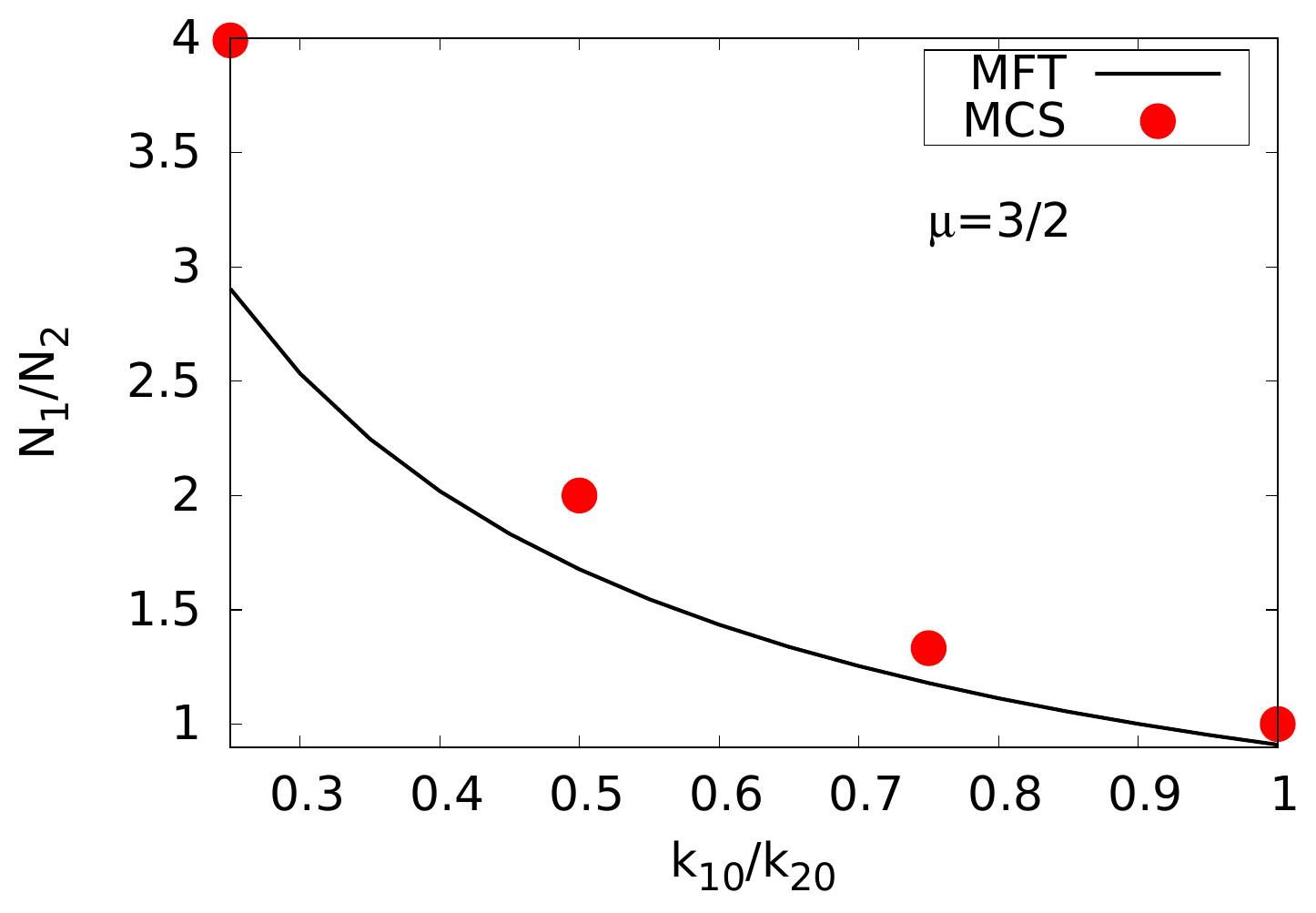}
 \\
\caption{\textbf{(Left)} Variation of the reservoir population ratio $N_{1}/N_{2}$ with the filling factor $\mu$ for fixed values of the exchange rates $k_{10}=k_{20}=0.95$ in the weak coupling limit of the model. \textbf{(Right)} Variation of the reservoir population ratio $N_{1}/N_{2}$ with the exchange rate ratio $k_{10}/k_{20}$ for a fixed $\mu=3/2$ in the weak coupling limit. For both plots, the entry and exit rates are chosen such that the TASEP lane always remains in the LD phase. The values of $\alpha=0.1$ and $\beta=1$ are used. There is a small mismatch between the results obtained from MFT and MCS in the left plot. However, in the right plot, the quantitative disagreement between MFT and MCS increases with decreasing exchange rate ratio. }
\label{n1/n2-vs-mu-and-k10/k20}
\end{figure*}

\section{Role of fluctuations}
\label{Role of fluctuations}

In the above Sections, we have found that there are significant discrepancies between the MFT predictions on the phase boundaries and the corresponding MCS results in the weak coupling case. In contrast, as shown in Appendix, the agreement between   MFT and MCS predictions in the strong coupling case is excellent. We attribute these discrepancies to larger fluctuations in the weak coupling case relative to the strong coupling case. We further find that these discrepancies persist even under significantly longer time averages. In this context, we note that while MFTs are adhoc approximations, they often work well, so long as fluctuations, neglected in MFT, are not important. This holds true in equilibrium systems away from critical points~\cite{chaikin}. Close to the critical points, mean-field theories do not give quantitatively (or in some cases even qualitatively) correct results in equilibrium systems at low dimensions~\cite{chaikin}. Similarly, there are examples of TASEP-like systems, where MFT predictions do not quantitatively match with the corresponding MCS results; see, e.g., Refs.~\cite{tirtha-anjan,waclaw}. In particular in Ref.~\cite{tirtha-anjan}, such mismatches between MFT and MCS results are clearly argued to be due to the fluctuation effects.

\begin{figure*}[htb]
 \includegraphics[width=\columnwidth]{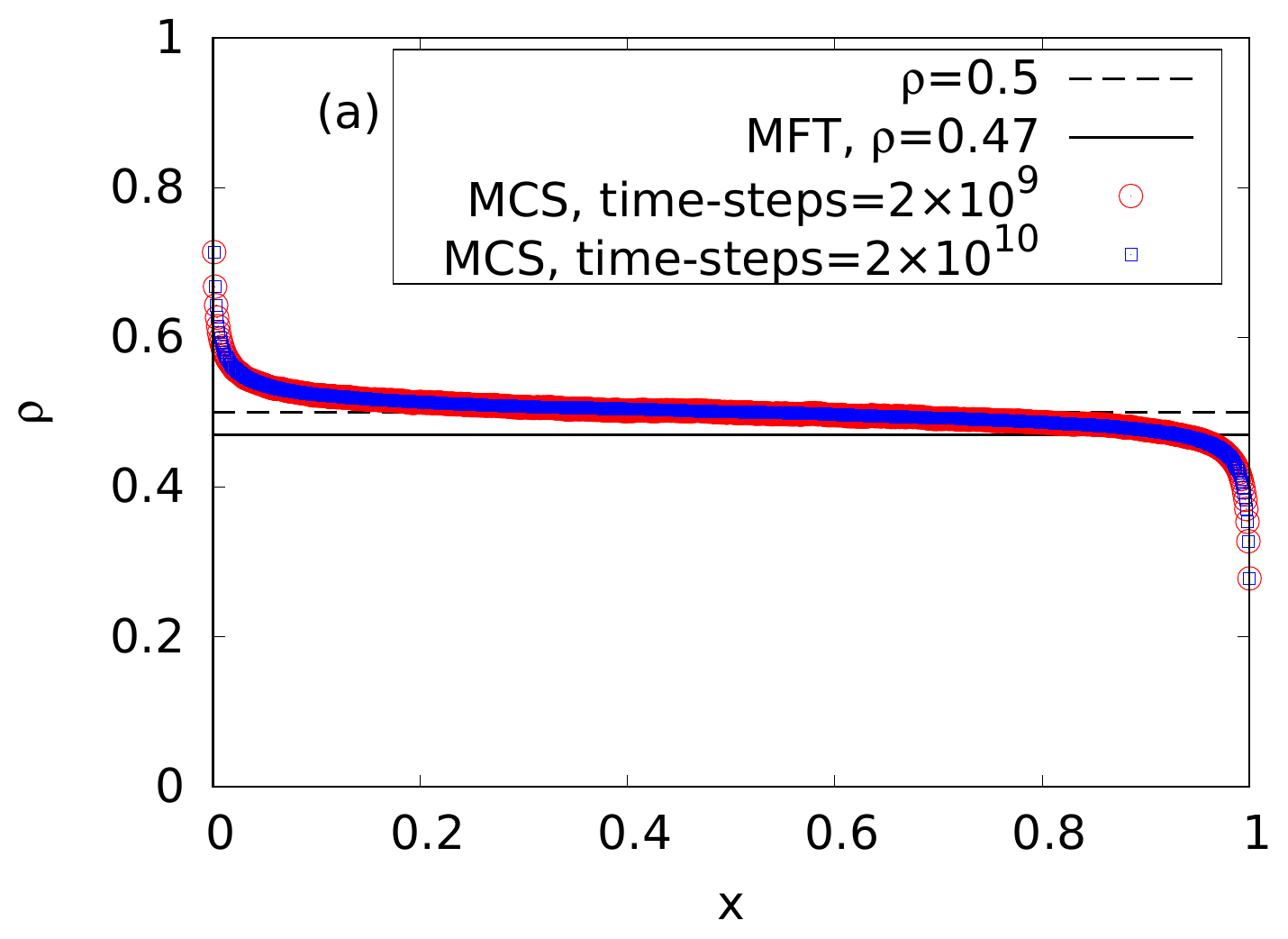}
 \hfill
 \includegraphics[width=\columnwidth]{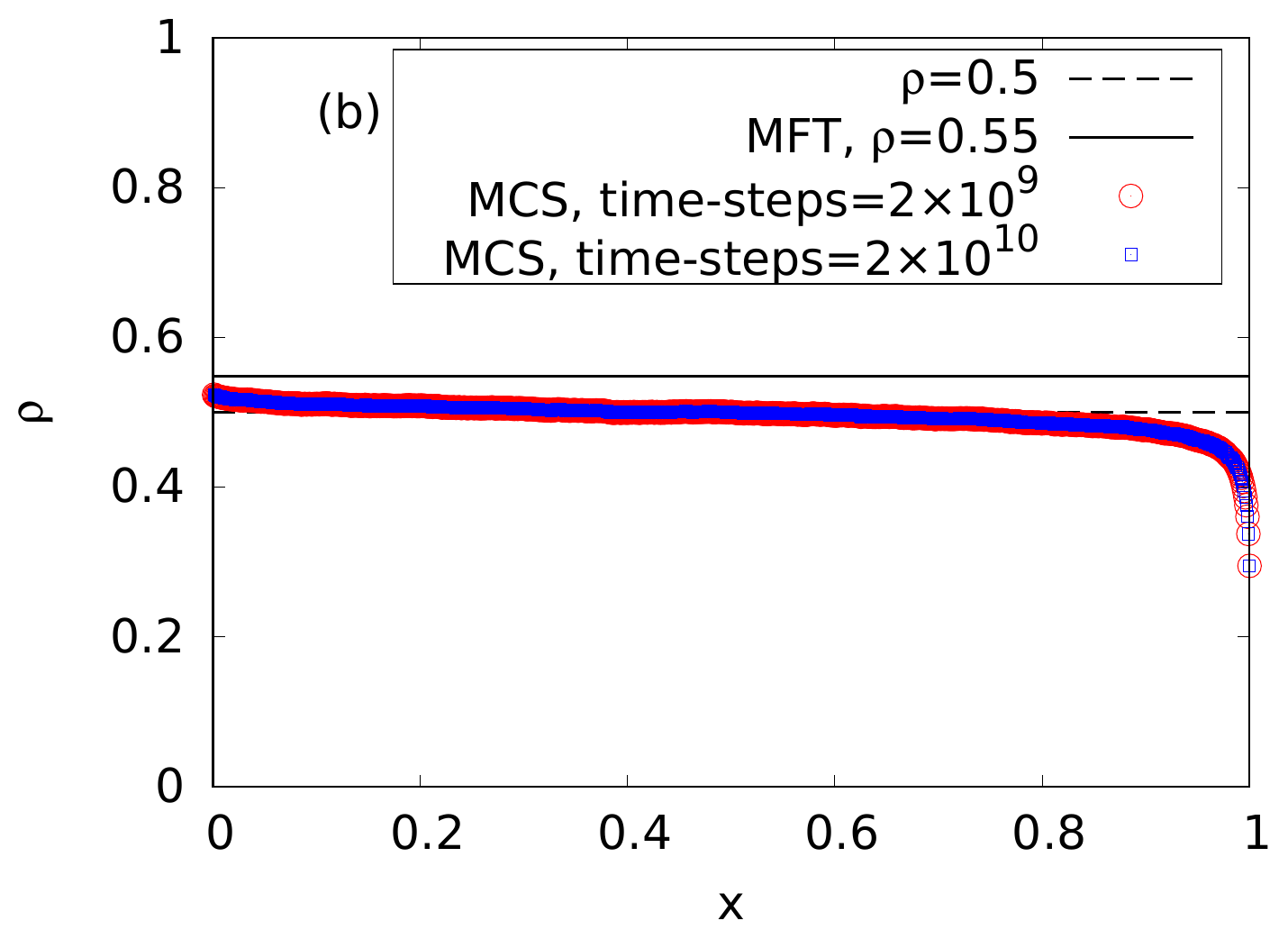}
 \\
 \includegraphics[width=\columnwidth]{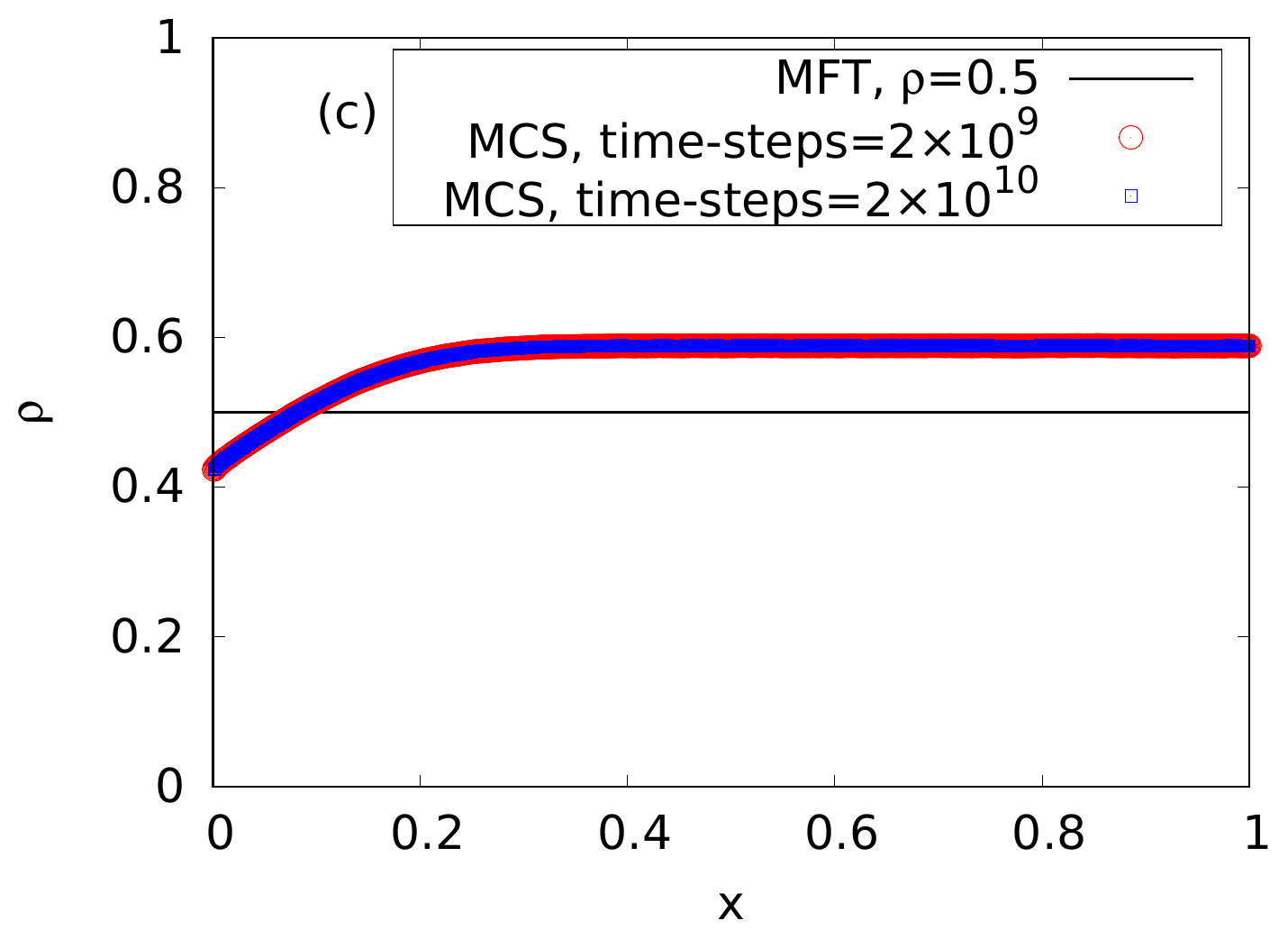}
 \hfill
 \includegraphics[width=\columnwidth]{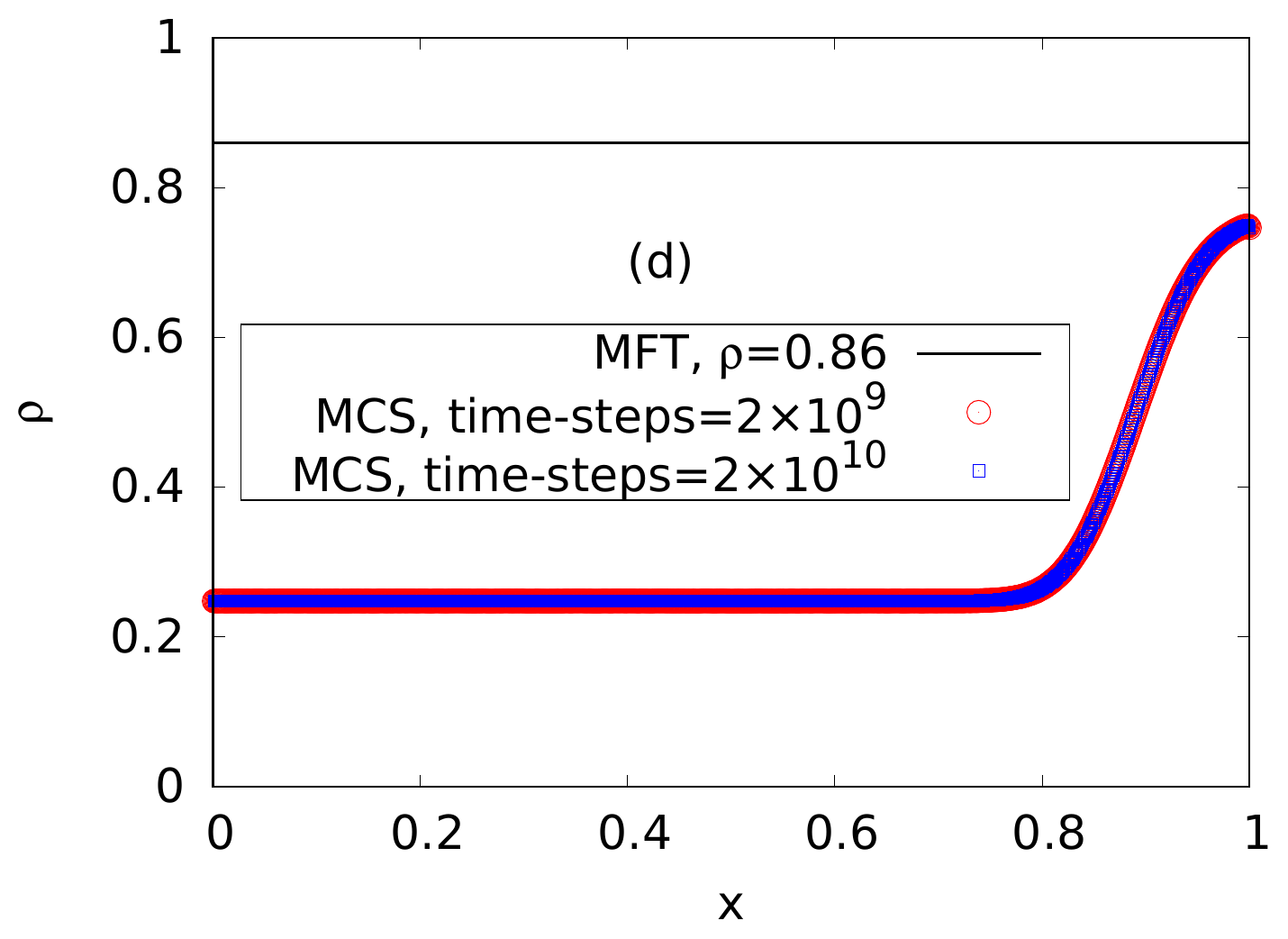}
 \\
 \caption{ Discrepancies between MFT and MCS results in steady state TASEP density plots: {\textbf{(top left)}  $\alpha=3.5$, $\beta=1.2$, $\mu=1$, $k_{10}=k_{20}=0.95$. With these, MFT predicts the system to be in LD phase with density $\rho=0.47$, while according to MCS simulation results, MC phase with density $\rho=0.5$ is obtained. \textbf{(top right)}  $\alpha=2$, $\beta=1.15$, $\mu=1.5$, $k_{10}=k_{20}=0.95$. According to MFT, the TASEP is in its HD phase with density $\rho=0.55$, whereas simulation results show MC phase is achieved. \textbf{(bottom left)} Parameter values are $\alpha=0.42$, $\beta=0.72$, $\mu=2$, $k_{10}=0.01$, $k_{20}=0.95$. MFT predicts MC phase while MCS studies give a domain wall for these parameters. \textbf{(bottom right)} $\alpha=0.35$, $\beta=0.25$, $\mu=30$, $k_{10}=0.01$, $k_{20}=0.95$. MFT predicts  HD phase, whereas MCS results indicate a domain wall under the specified parameters. System size $L=1000$. Overall, increasing the number of time steps in simulation has no significant effect on the results obtained, thus confirming the intrinsic discrepancies between our MFT and MCS results in the weak coupling limit (see text).}
}
\label{role of fluctuation}
\end{figure*}

In Fig.~\ref{role of fluctuation}, we show a few representative steady state density plots obtained from our MFT analysis and MCS runs with $2\times 10^9$ Monte-Carlo steps (red circles) and $2\times 10^{10}$ Monte-Carlo steps (blue squares) respectively. {System size is $L=1000$ and parameter values are mentioned in the caption. In Fig.~\ref{role of fluctuation}(top left), MFT and MCS predict the TASEP to be in LD phase and MC phase respectively. For the parameters in Fig.~\ref{role of fluctuation}(top right), MFT suggests TASEP to be in its  HD phase, while MCS suggests MC phase. Next, in Fig.~\ref{role of fluctuation}(bottom left) MFT and MCS studies indicate the appearence of MC and DW phases respectively.  Finally in Fig.~\ref{role of fluctuation}(bottom right), MFT predicts HD phase while MCS studies suggest DW phase in the TASEP for the parameters specified. 
We note that the MCS results for $2\times 10^9$ Monte-Carlo steps and $2\times 10^{10}$ Monte-Carlo steps are nearly overlapping in both the plots. We thus conclude from Fig.~\ref{role of fluctuation} by increasing time-steps better agreement between MFT and MCS results are not obtained. Further, in the density plots of Fig.~\ref{dw-weak-mu0.5-1} and Fig.~\ref{ld-hd-mc-st-wk-den}, agreements between MFT and the corresponding MCS results are quantitatively better for relatively smaller values of $\alpha$ and $\beta$.

Next, we heuristically argue that these fluctuation-induced shifts in the phase boundaries relative to their MFT predictions in the weak coupling is not a finite size effect, i.e., they should persist even in the thermodynamic limit $L\rightarrow \infty$. To this end, we construct a set of scaling level arguments as given below.

Due to the fluctuations, the effective value of the reservoir populations will differ from their values in MFT. We thus generalise the relations $\alpha_\text{eff}=\alpha N_1/L$ and $\beta_\text{eff}=\beta(1-N_{2}/L)$ to
\begin{equation}
 \tilde\alpha_\text{eff}=\frac{\alpha}{L}N_1^\text{eff}, \hspace{5mm} \tilde\beta_\text{eff}=\beta\bigg(1-\frac{N_{2}^\text{eff}}{L}\bigg)
\end{equation}
with
\begin{equation}
 N_1^\text{eff}=N_1 +\delta N_1, \hspace{5mm} N_2^\text{eff}=N_2 +\delta N_2,
\end{equation}
where $N_1$, $N_{2}$ are to be understood as the solution for the populations of reservoirs $R_1$, $R_{2}$ in MFT [i.e., solutions in Eqs.~(\ref{N1w}) and (\ref{N2w})], and $\delta N_1$, $\delta N_2$ are the corresponding deviations due to the fluctuations from their respective mean-field values, and can have any sign.  Hence, $\tilde\alpha_\text{eff}-\alpha_\text{eff}$ and $\beta_\text{eff}-\tilde\beta_\text{eff}$ can be positive or negative. In our MCS studies, $\tilde\alpha_\text{eff}$ and $\tilde\beta_\text{eff}$ are the effective entry and exit rates. 

In the weak coupling case, both $N_1,\,N_2$ have $J_T$-dependent contributions that scale with $J_T$ as $LJ_{T}$ [cf. Eq.~(\ref{N1w}) and Eq.~(\ref{N2w})]. Here, $J_{T}$, the TASEP current, is a ${\cal O}(1)$ quantity (since its maximum value is 1/4). So the fluctuations in $J_{T}$ should also be ${\cal O}(1)$. This means $\delta N_1,\,\delta N_2$, the fluctuations in $N_1,\,N_2$ must be ${\cal O}(L)$. Indeed, the ratio of the reservoir populations is not a fixed number but depends upon $J_T$, meaning the two populations can fluctuate relative to each other. This immediately gives $\tilde\alpha_\text{eff}-\alpha_\text{eff}\sim {\cal O}(1)$ and $\beta_\text{eff}-\tilde\beta_\text{eff}\sim {\cal O}(1)$ even in the thermodynamic limit. Furthermore, the sum of the two reservoir populations $N_R$ is independent of $J_T$ and hence should have typical fluctuations of the size ${\cal O}\sqrt {N_R}\sim {\cal O}(\sqrt L)\ll$ fluctuations in $N_1$ or $N_2$ in the limit of large $L$. Since $N_1+N_2=N_R$, we must have $\delta N_1\sim -\delta N_2$ to the leading order in $L$ for large $L$. This in turn means when $N_1^\text{eff}$ is larger (smaller) than $N_1$, $N_2^\text{eff}$ is smaller (larger) than $N_2$, giving   $\tilde\alpha_\text{eff}>(<) \alpha_\text{eff}$ and $\tilde\beta_\text{eff} >(<) \beta_\text{eff}$. This then generally implies that in the weak coupling case the effects of fluctuations survive in the thermodynamic limit, and if the transition from LD  to another phase takes place at a value of $\alpha$ lower (higher) than that predicted in MFT, then the corresponding transition from HD to another phase should take place at a value of $\beta$  lower (higher) than predicted in MFT, a feature generally observed in our phase diagrams in Fig.~\ref{pd weak 3} and Fig.~\ref{pd weak 2}, and is consistent with the particle-hole symmetry when $k_1=k_2$. In contrast, in the strong coupling regime $N_1,\,N_2$ are independent of $J_T$ in the thermodynamic limit  and the particle exchange dynamics between the reservoirs being effectively an equilibrium dynamics, should have $\delta N_1,\,\delta N_2\sim {\cal O}(\sqrt L)$ only. Since the MF expressions for $N_1,\,N_2$ scale as $L$,  $\tilde\alpha_\text{eff}\rightarrow\alpha_\text{eff}$ and $\tilde\beta_\text{eff}\rightarrow\beta_\text{eff}$  in the thermodynamic limit, giving good quantitative agreement between MFT and MCS results in the strong coupling case. This is consistent with our findings in the Appendix. Lastly, our analysis above can be used to qualitatively explain the observation made in the density plots of Fig.~\ref{dw-weak-mu0.5-1} and Fig.~\ref{ld-hd-mc-st-wk-den}. For sufficiently small $\alpha,\,\beta$, the TASEP channel is likely to be in
its low LD phase (i.e., its bulk density very small) or high HD (i.e., its bulk density very high, approaching unity). In either case, the steady TASEP current $J_T$ should be quite small. Since it is $J_T$ in   Eq.~(\ref{N1w}) and Eq.~(\ref{N2w}) that is responsible for the fluctuation effects in the weak coupling case that persists even in the thermodynamic limit, smaller values of $J_T$ for smaller $\alpha,\,\beta$ should indicate relatively smaller extent of fluctuations in $J_T$, giving smaller quantitative deviations from the MFT results. This is clearly consistent with trends observed in the density plots of Fig.~\ref{dw-weak-mu0.5-1} and Fig.~\ref{ld-hd-mc-st-wk-den}.


}

 \section{CONCLUSION AND OUTLOOK}
 \label{conclusion and outlook}

  In this article, we have studied how the interplay between the finite availability and carrying capacity of particles at
different parts of a spatially extended system, and diffusion between these parts can control the steady state currents and steady state density
profiles in a quasi-1D current-carrying channel connecting the different parts of the
system. To address this issue, we propose and study a conceptual model that is composed of two reservoirs $R_1$ and $R_2$ of finite capacities and connected by a single TASEP channel at its entry and exit ends respectively. The reservoirs are allowed to exchange particles among them instantaneously, which models particle diffusion between them. The latter process ensures that there is a finite steady state current in the system. The model has five free parameters -- the two parameters $\alpha,\beta$ defining the entry and exit rates, the two rates $k_1,k_2$ for the particle exchanges between the reservoirs, and the filling fraction $\mu$. In addition, there are two functions, $f$ and $g$, that control the effective entry and exit rates, respectively. To simplify the subsequent analysis, we have chosen very simple forms for the functions $f$ and $g$, such that reservoir $R_2$ has a maximum capacity of $L$ particles, wheres reservoir $R_1$ has a capacity $k_2L/k_1$, where $L$ is the size of the TASEP channel. When $k_1=k_2$, the model admits a particle-hole symmetry, which is absent for $k_1\neq k_2$. We have used simple MFT and MCS to obtain the phase diagrams in the $\alpha-\beta$ plane and also the steady state density profiles, parametrised by all the other parameters. We have chosen $f$ to be monotonically rising, whereas $g$ is monotonically decreasing. We have studied our model in two distinct limits, when the particle exchange process competes with the TASEP current, and when it overwhelms the latter. We call the former {\em weak coupling limit} and the latter {\em strong coupling limit} of the model.
Nonetheless, the model displays unexpected phase behavior not generally observed in the existing models for TASEPs with finite resources. First and foremost, depending upon the values of $\mu$, the model can either be in two or four phases, with continuous transitions between them. We have identified different threshold values of $\mu$ for the existence of the various phases. Secondly, the occupations $N_1$ and $N_2$ of the two reservoirs $R_1$ and $R_2$ are generally unequal. In fact, the populations of the two reservoirs could be preferentially controlled, i.e., getting them relatively populated or depopulated, by appropriate choice of the above model parameters. This in turn can lead to possible  population imbalances in the steady states and consequently highly inhomogeneous particle distributions between different parts of the systems.

We have used analytical mean-field theory together with stochastic Monte Carlo simulations studies for our work. The mean-field theory predictions agree quantitatively with the corresponding MCS results in the strong coupling limit, but we find significant quantitative mismatch between the two in the weak coupling limit. We attribute this mismatch to the stronger fluctuations in the weak coupling limit of the model that are argued to remain relevant even in the thermodynamic limit.  Our results have been obtained with specific choices for $f$ and $g$. While other choices for $f$ and $g$ should quantitatively change the phase diagram, so long as $f$ and $g$ remain monotonically increasing and decreasing functions of their arguments, and allow for maximum finite capacities of the reservoirs, we expect our results should hold qualitatively.

 From general nonequilibrium statistical perspectives, our model serves as a simple conceptual model that reveals how a coupling between a TASEP, a paradigmatic nonequilibrium process, and reservoirs exchanging particles, a simple equilibrium process, can produce nontrivial steady states and phase diagrams for the TASEP. We also highlight how fluctuations significantly affect the MCS simulation results, when the particle exchanges are {\em weak}, but not when it is {\em strong}, giving impetus to further quantitative studies on the effects of fluctuations on the nonequilibrium steady states in the thermodynamic limit.

We have studied the simplest case with only one TASEP channel connecting two reservoirs. One can consider more than one TASEP channels and more than two reservoirs with diffusion between them. Our MFT can be in principle extended to such a system, whose precise mathematical form will of course depend upon the actual connectivity of the TASEP channels and the reservoirs. In such a situation, it is expected that the system may display more than one delocalised domain walls, one each in each channel, instead of a single LDW as here, for some choices of the model parameters. Stochastic simulations should be helpful to validate these qualitative expectations.

\section{Acknowledgement}

AB thanks SERB (DST), India for partial financial support through the CRG scheme [file: CRG/2021/001875].

 \appendix


 \section{Details on the LD-DW phase boundary in the weak coupling case}

    Here we show explicitly how Eq.~(\ref{lddw-boundary-1}) reduces to Eq.~(\ref{lddw-boundary-earl}). Denoting $a=\bigg({\frac{1+k_{20}}{2}+\frac{k_{10}+k_{20}}{2\alpha}}\bigg)$ and $b=\bigg(\frac{1}{2}+\frac{k_{10}}{2\alpha}+\frac{k_{20}}{2\beta}\bigg)$, we rewrite Eq.~(\ref{lddw-boundary-1}):
   \begin{align}
    &a-\sqrt{a^{2}-\mu k_{20}}=b-\sqrt{b^{2}-k_{20}} \nonumber \\
    \implies &\frac{4}{k_{20}}[(\mu+1)ab-a^{2}-\mu b^{2}]=(\mu-1)^{2}. \label{simp1}
   \end{align}
   Having the definition of $a$ and $b$, we compute the term inside the square bracket in Eq.~(\ref{simp1}). With that being done, the resulting equation can be expressed as a quadratic equation in $\mu$ as follows:
   \begin{align}
    &\mu^{2}-\mu\bigg(3+\frac{1+k_{10}}{\alpha}-\frac{1-k_{20}}{\beta}+\frac{k_{10}}{\alpha^{2}}-\frac{k_{20}}{\beta^{2}}-\frac{k_{10}-k_{20}}{\alpha\beta}\bigg) \nonumber \\
    &+2+k_{20}+\frac{1+k_{10}+2k_{20}}{\alpha}-\frac{1+k_{20}}{\beta} \nonumber \\
    &+(k_{10}+k_{20})\bigg(\frac{1}{\alpha^{2}}-\frac{1}{\alpha\beta}\bigg)=0. \label{simp2}
   \end{align}
   Next, defining $c=\bigg(\frac{1}{2}+\frac{k_{10}}{2\alpha}+\frac{k_{20}}{2\beta}\bigg)$, we recast Eq.~(\ref{lddw-boundary-earl}) as
   \begin{align}
    &\bigg(c-\sqrt{c^{2}-k_{20}}\bigg)\bigg(1+\frac{1}{\alpha}-\frac{1}{\beta}\bigg)=\mu-1 \nonumber \\
    \implies &c^{2}-k_{20}=\bigg(c-\frac{\mu-1}{1+\frac{1}{\alpha}-\frac{1}{\beta}}\bigg)^{2} \nonumber \\
    \implies &\bigg[2c\bigg(1+\frac{1}{\alpha}-\frac{1}{\beta}\bigg)-\mu+1\bigg](\mu-1)=k_{20}\bigg(1+\frac{1}{\alpha}-\frac{1}{\beta}\bigg)^{2}. \label{simp3}
   \end{align}
   Again, with defined $c$, we compute the term inside the square bracket in Eq.~(\ref{simp3}). Then, after some rearrangement of terms, the resulting equation can be seen as a quadratic equation in $\mu$ identical with Eq.~(\ref{simp2}).


 \section{STRONG COUPLING LIMIT}
 \label{mft strong coupling}

In this Section, we consider the strong coupling limit of the model, i.e., when $k_1,\,k_2\sim {\cal O}(1)$ and independent of $L$. We shall see below that in this limit the direct particle exchanges between the reservoirs dominate. Therefore in the strong coupling limit, the particle numbers in the two reservoirs $R_1$ and $R_2$ maintain a fixed ratio for a given set of particle exchange rates (see below). This makes the ensuing MFT algebraically simpler and  shows good agreement with the corresponding MCS studies results, in contrast to the weak coupling limit of the model. We set up the MFT following the logic used to develop the MFT in the weak coupling limit.

 \subsection{Mean-field phase diagrams and steady state densities}
 \label{MF pd strong coupling}

\begin{figure*}[htb]
 \includegraphics[width=\columnwidth]{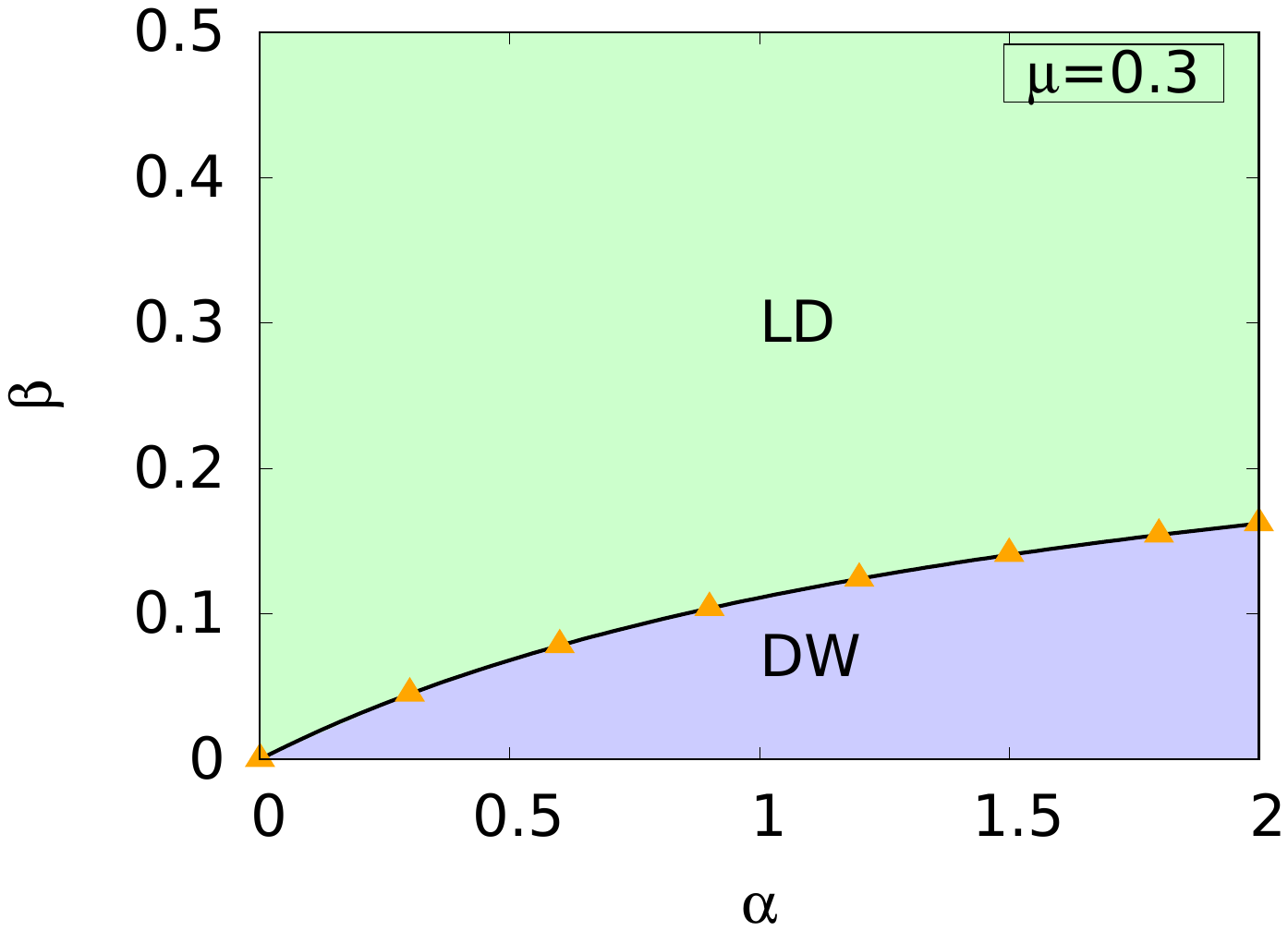}
 \hfill
 \includegraphics[width=\columnwidth]{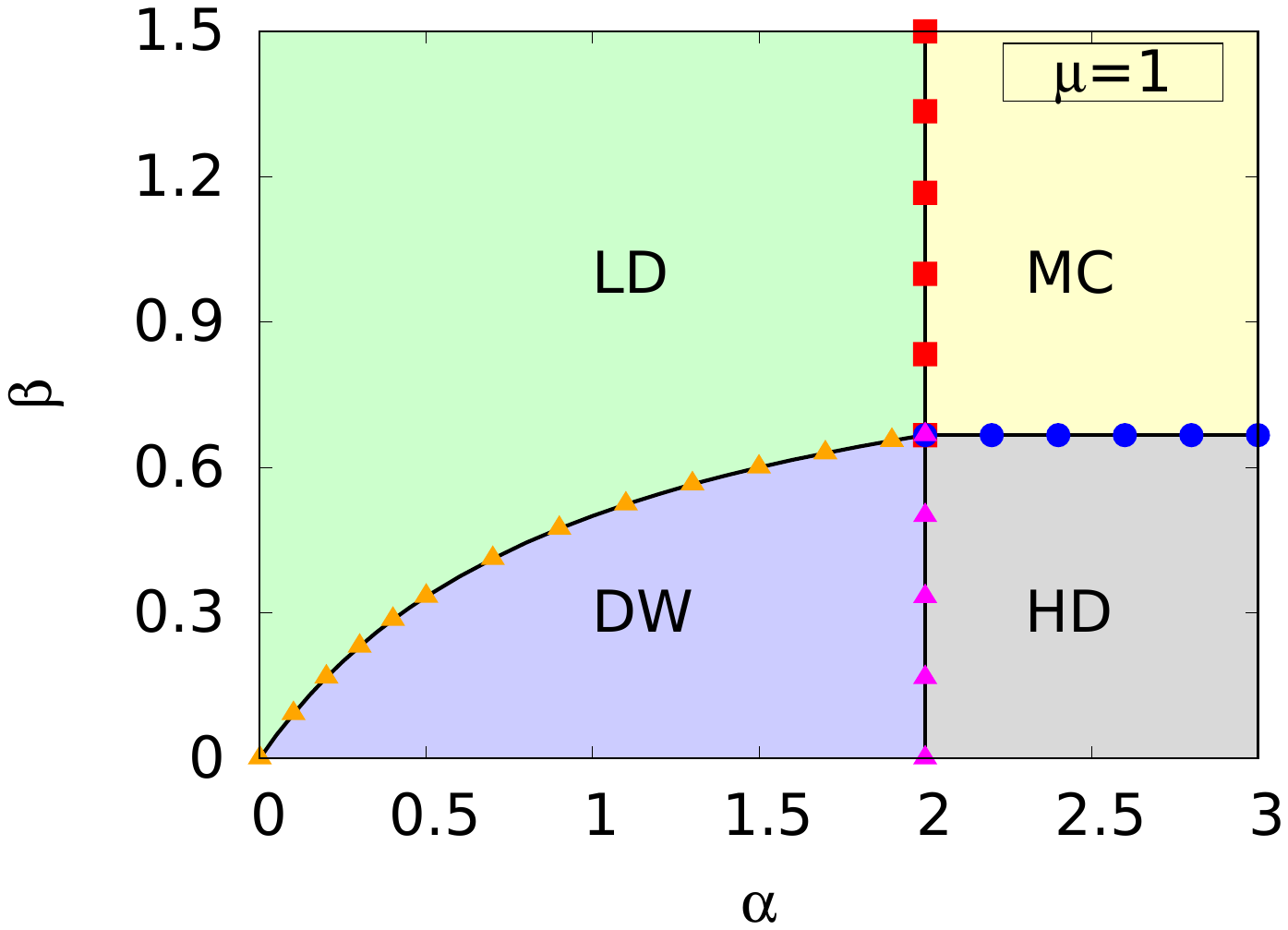}
 \\
 \includegraphics[width=\columnwidth]{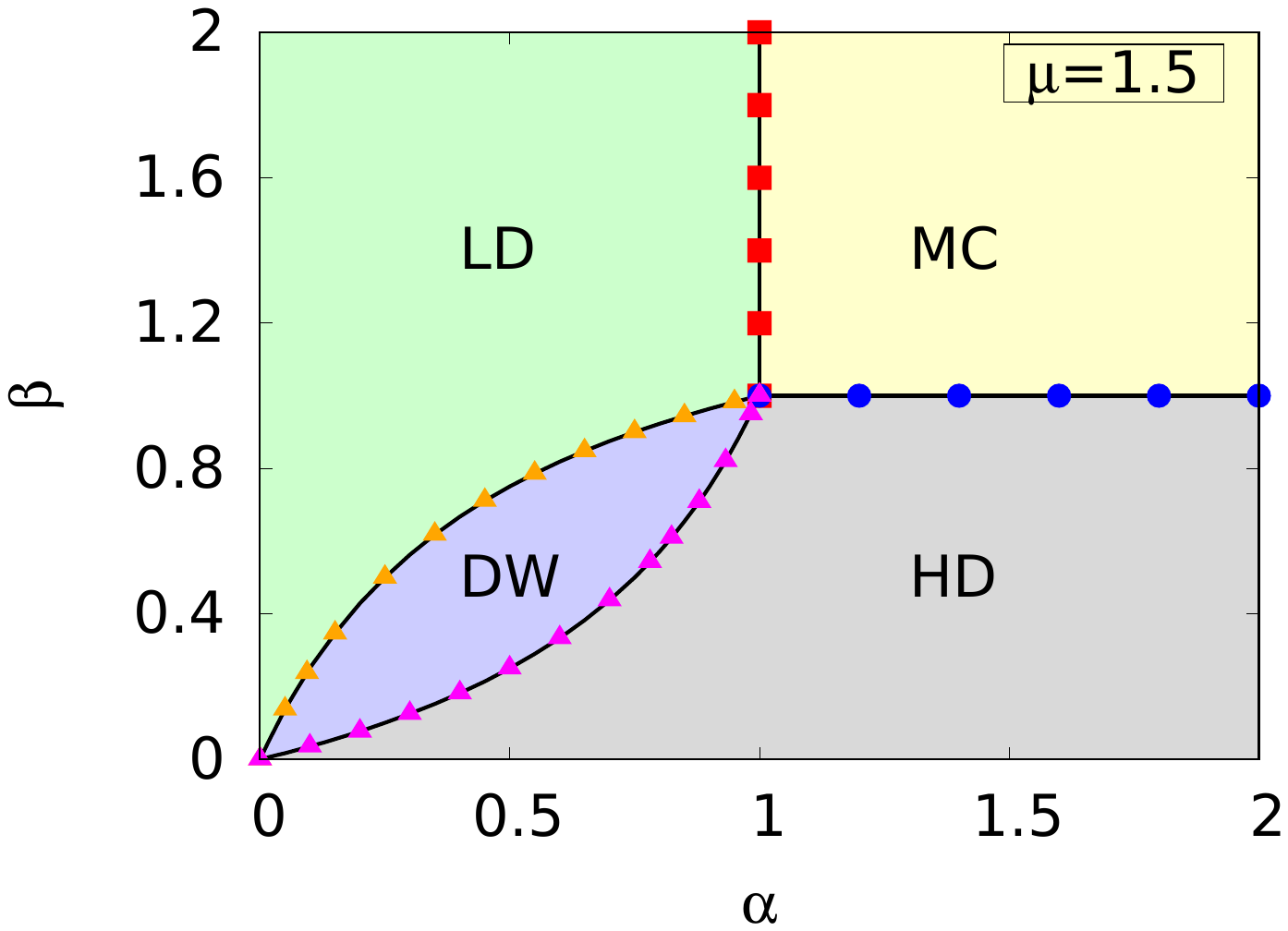}
 \hfill
 \includegraphics[width=\columnwidth]{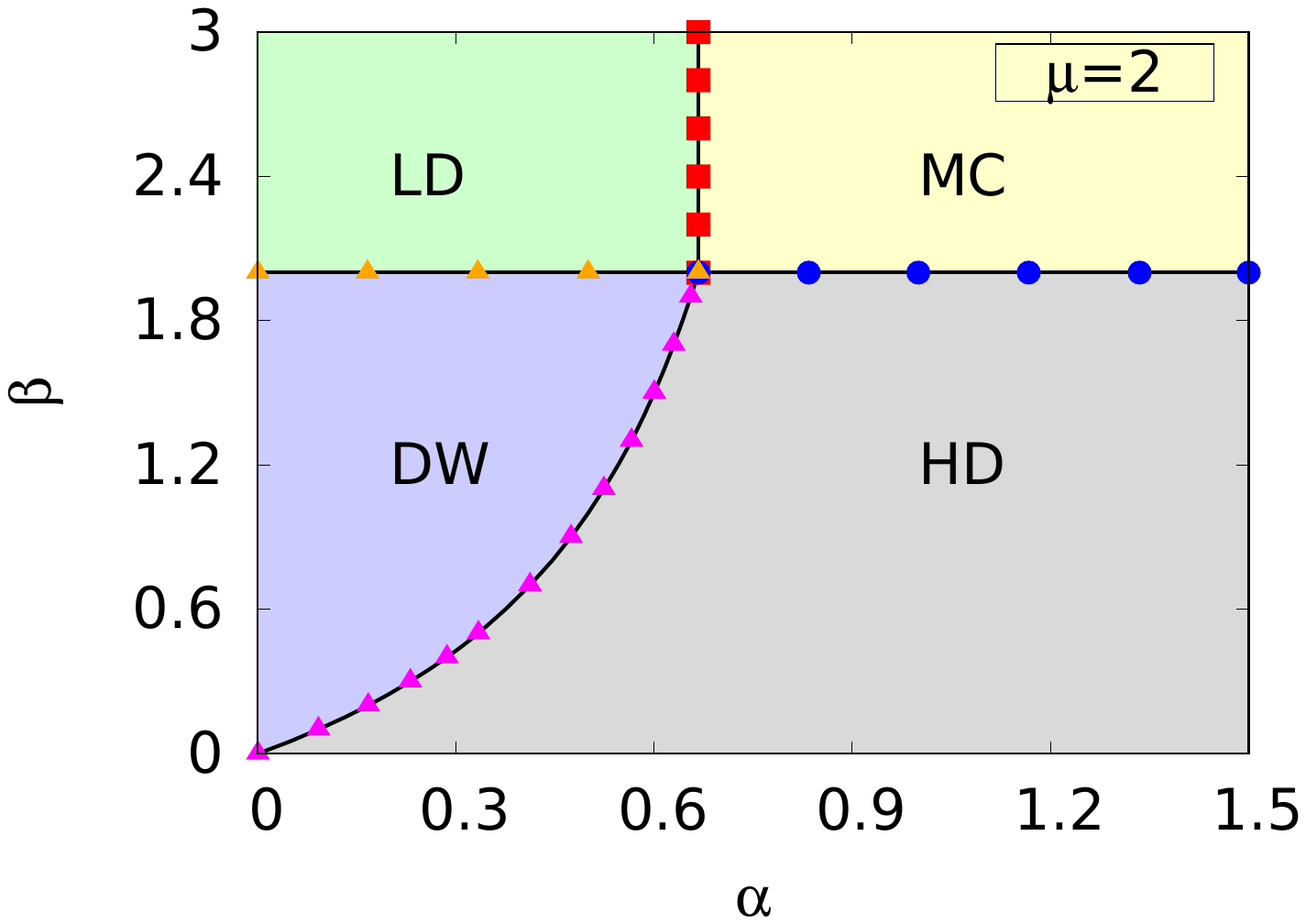}
 \\
\caption{ Phase diagrams of the model in the strong coupling limit with particle-hole symmetry for different $\mu$ with equal exchange rates ($k_{1}=k_{2}=0.95$) are shown. Different colours in the background represents the region covered by distinct phases according to MFT, with black solid lines being the MFT phase boundaries. To corroborate them, MCS phase boundaries in the form of discrete colored points are shown. Both are in excellent agreement. Depending on the specific value of $\mu$ which varies from 0 to $\mu_\text{max}=(2+k_{2}/k_{1})=3$,
the phase diagrams contain either two or four distinct phases.(top left) $\mu=0.3$, (top right) $\mu=1$, (bottom left) $\mu=1.5$ and (bottom right) $\mu=2$. With $k_{1}=k_{2}=0.95$, particle-hole symmetry connects the phase diagrams for $\mu>3/2$ with the ones for $\mu<3/2$, $\mu=3/2$ being the half-filled limit.}
\label{pd strong}
\end{figure*}

\begin{figure*}[htb]
 \includegraphics[width=\columnwidth]{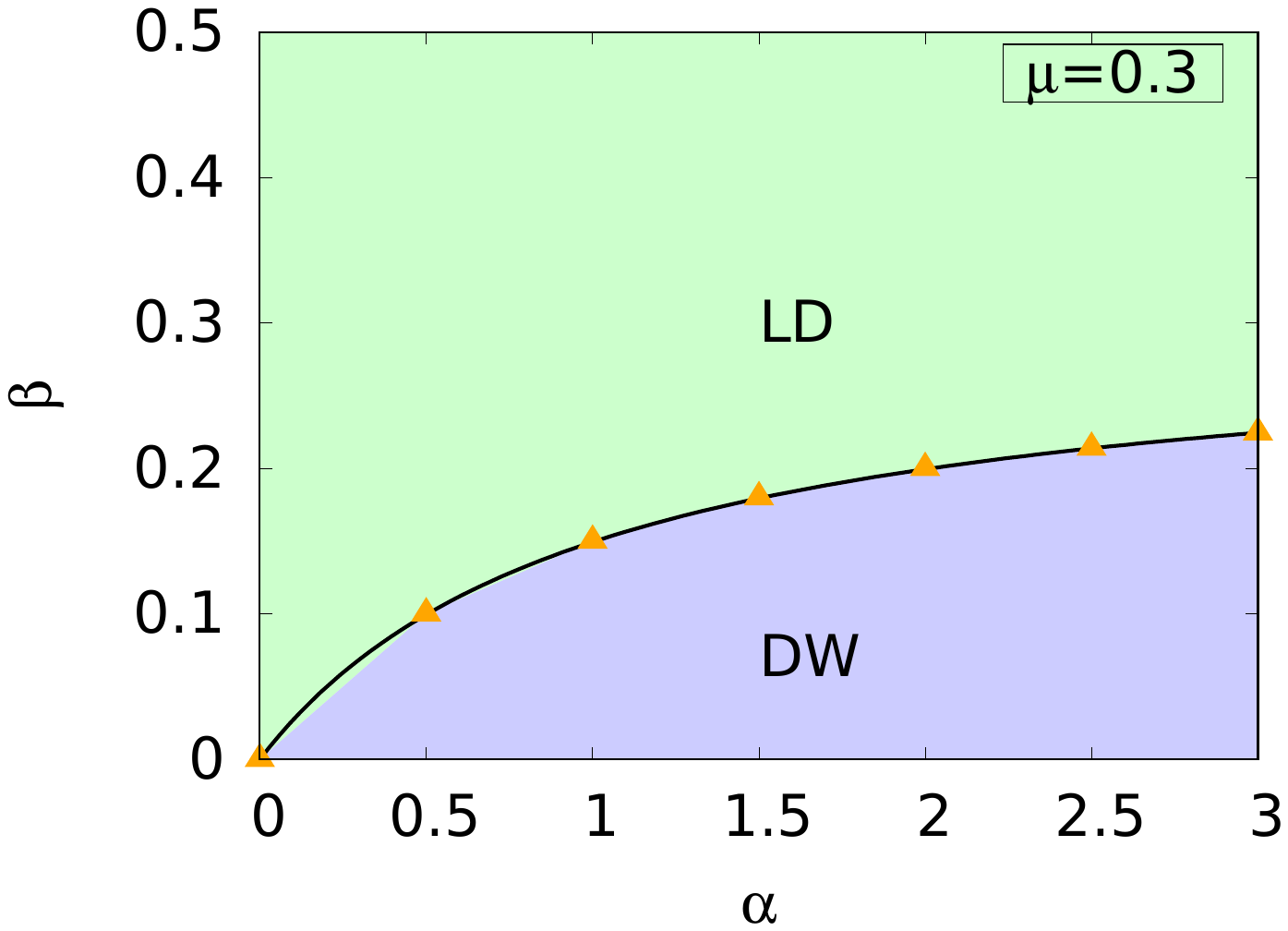}
 \hfill
 \includegraphics[width=\columnwidth]{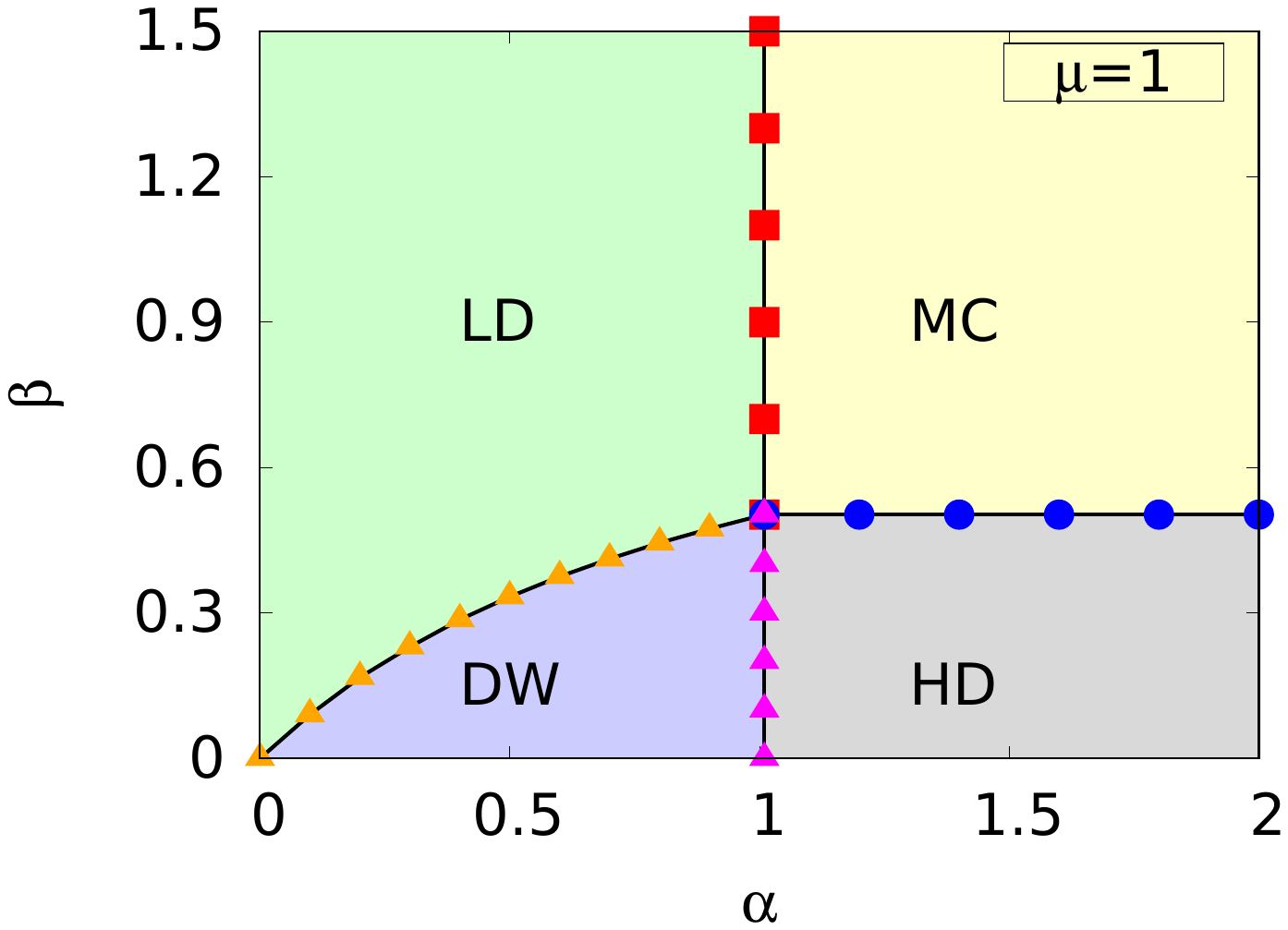}
 \\
 \includegraphics[width=\columnwidth]{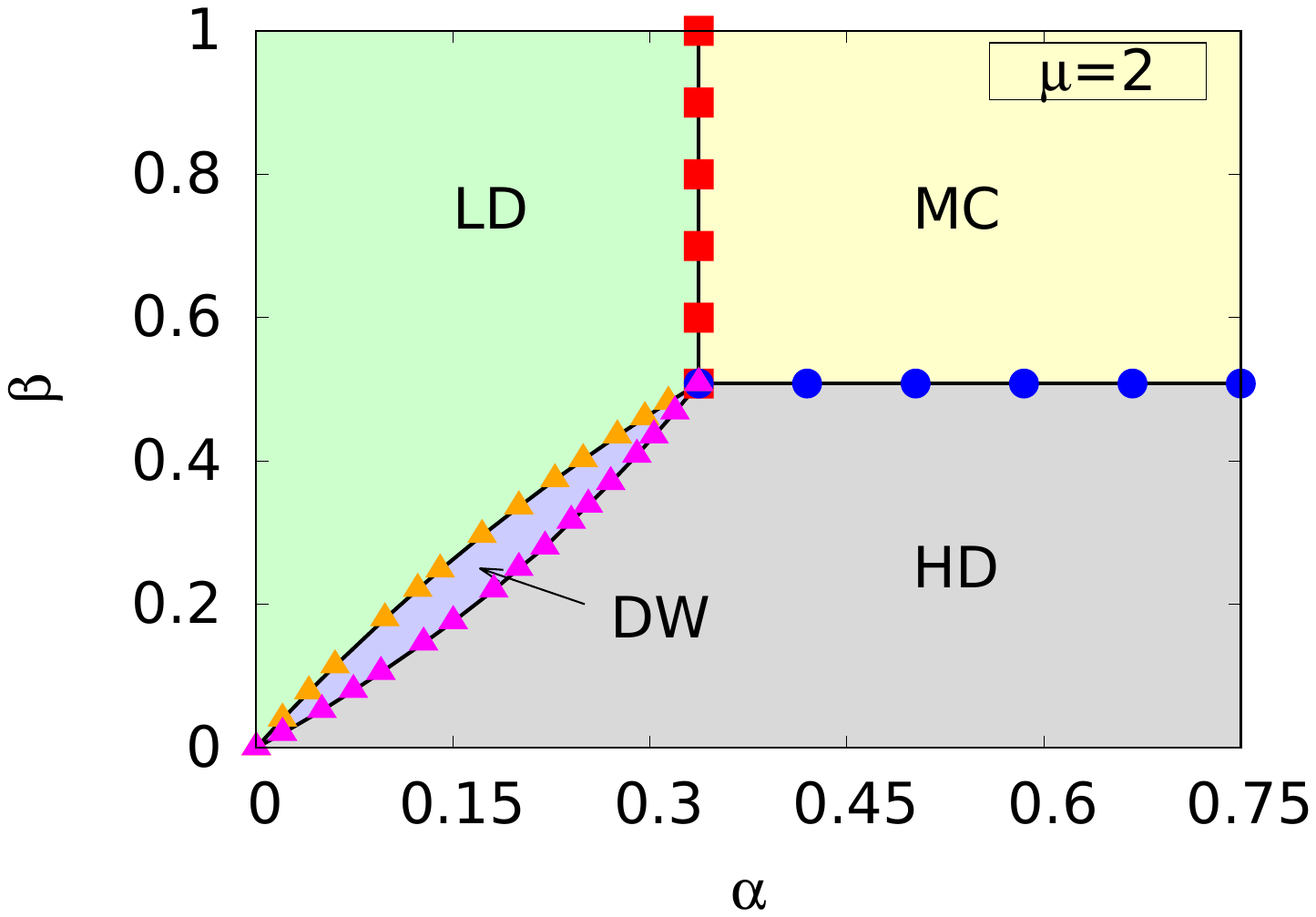}
 \hfill
 \includegraphics[width=\columnwidth]{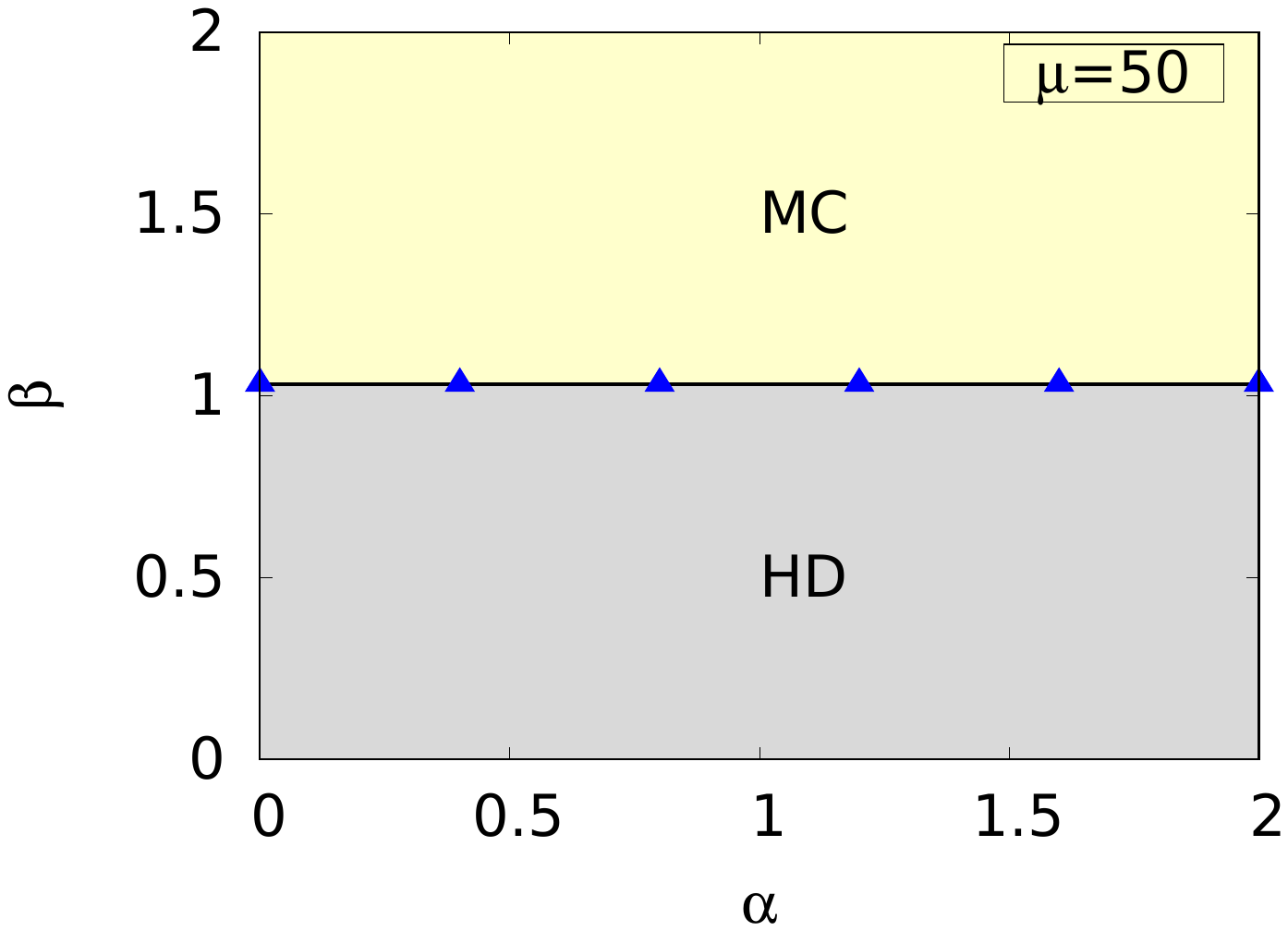}
 \\
\caption{ Phase diagrams of the model in the strong coupling limit without particle-hole symmetry for different $\mu$ with unequal exchange rates ($k_{1}=0.01, k_{2}=0.95$) are shown. The background colours indicate distinct phases according to MFT with black lines being the MFT phase boundaries, whereas the discrete coloured points are the phase boundaries obtained by MCS. They are in good agreement with each other. The maximum value of $\mu$ in this case is calculated as $\mu_\text{max}=(2+k_{2}/k_{1})=97$. (top left) $\mu=0.3$, (top right) $\mu=1$, (bottom left) $\mu=2$, (bottom right) $\mu=50$. 
With a significantly large value of $\mu$, such as $\mu=50$, only two phases, namely HD and MC, are observed.}
\label{pd strong asymmetric}
\end{figure*}

 For specificity, we consider two cases, one with equal exchange rates $k_{1}=k_{2}=0.95$ with $\mu_\text{max}=3$, another with unequal exchange rates $k_{1}=0.01,k_{2}=0.95$ with $\mu_\text{max}=97$. The phase diagrams are shown in Fig.~\ref{pd strong} and Fig.~\ref{pd strong asymmetric}, respectively.

 We start with the MFT Eqs.~(\ref{rho-sols}) and (\ref{flux-bal-sol}) given above. As explained in Section~\ref{steady-state density model L} above, in the strong coupling limit
 \begin{equation}
  \frac{N_2}{N_1}=\frac{k_1}{k_2}
 \end{equation}
with $L\rightarrow \infty$. This considerably simplifies the MFT as discussed below in details.

 \vspace{2mm}
 \subsubsection{{ Low-density phase}}
 \label{ld phase strong}

 We proceed as in the weak coupling case.
 In the limit $L \rightarrow \infty$, the last term in the right-hand side of Eq.~(\ref{ld model 1}) with the order of $\sim \mathcal{O}(1/L)$ vanishes and $\rho_\text{LD}$ can be approximated as a function of the TASEP parameters $\alpha$, $\mu$, $k_{1}$, and $k_{2}$ as following:
  \begin{equation}
 \label{rhold strong model 1}
  \rho_\text{LD} \approx \frac{\alpha k_{2} \mu}{k_{1}+(1+\alpha)k_{2}}.
 \end{equation}
  Eq.~(\ref{rhold strong model 1}) gives a linear dependence of $\rho_\text{LD}$ on $\mu$ with $\rho_\text{LD} = 0$ when $\mu = 0$. The form of $\rho_\text{LD}$ in Eq.~(\ref{rhold strong model 1}) allows for expressions to be obtained for the population of the two reservoirs:
 \begin{eqnarray}
  &&N_{1}=\frac{Lk_{2} \mu}{k_{1}+(1+\alpha)k_{2}}, \label{N1 strong}\\
  &&N_{2}=\frac{Lk_{1} \mu}{k_{1}+(1+\alpha)k_{2}}. \label{N2 strong}
 \end{eqnarray}
As mentioned above, the relative population of the two reservoirs becomes $N_{1}/N_{2}=k_{2}/k_{1}$,  independent of other control parameters $\alpha$, $\beta$, and $\mu$.

 With rising $\mu$, the availability of particles in the system, and hence to the TASEP lane, increases, and eventually, for high enough $\mu$, LD phase disappears. By definition, $0 < \rho_\text{LD} <1/2$, we get the following range of $\mu$ over which LD phase appears:
 \begin{equation}
  \label{range of mu for LD strong}
  0 < \mu<\bigg(\frac{1}{2}+\frac{k_{1}}{2\alpha k_{2}}+\frac{1}{2\alpha}\bigg).
 \end{equation}
  As shown in (\ref{range of mu for LD strong}), the upper threshold of $\mu$ for the LD phase explicitly depends on positive parameters $\alpha$, $k_{1}$, and $k_{2}$. Consequently, the upper threshold is also positive. For certain values of these parameters, it is possible for the upper threshold to be less than $\mu_\text{max}=(2+k_{2}/k_{1})$, thus limiting the existence of the LD phase up to that threshold:

 \begin{equation}
  \label{ld-upper-mu-st-other}
  \mu_\text{max} > \bigg(\frac{1}{2}+\frac{k_{1}}{2\alpha k_{2}}+\frac{1}{2\alpha}\bigg).
 \end{equation}
 When the upper threshold equals $\mu_\text{max}$, LD phase exists for any $\mu$ value.

In Fig.~\ref{ld-hd-mc-st-wk-den}, LD phase density profiles in the strong coupling limit are obtained for $\mu=1$ with two distinct values of $\alpha$: $\alpha=0.3$ and $\alpha=0.5$. Both the MFT and MCS density profiles agree with each other.

 \vspace{2mm}
 \subsubsection{{ High-density phase}}
 \label{hd phase strong}

 We again follow the logic outlined in the weak coupling case.
 One obtains the following expression for $\rho_\text{HD}$ solving Eq.~(\ref{hd quad model 1}) in TL ($L\rightarrow\infty$):
  \begin{equation}
 \label{hd strong model 1}
  \rho_\text{HD} \approx \frac{\beta k_{1} \mu+(k_{1}+k_{2})(1-\beta)}{k_{1}(1+\beta)+k_{2}},
 \end{equation}
 which shows a linear dependence of $\rho_\text{HD}$ in $\mu$ with $\rho_\text{HD} = 1$ as $\mu = \mu_\text{max}=(2+k_{2}/k_{1})$. The corresponding reservoir populations $N_{1}$ and $N_{2}$ are given by:
 \begin{eqnarray}
  &&N_{1}=\frac{Lk_{2}(\mu-1+\beta)}{k_{1}(1+\beta)+k_{2}}, \label{N1 hd strong}\\
  &&N_{2}=\frac{Lk_{1}(\mu-1+\beta)}{k_{1}(1+\beta)+k_{2}} \label{N2 hd strong}
 \end{eqnarray}
 once again yielding $N_{1}/N_{2}=k_{2}/k_{1}$, as expected in the strong coupling limit of the model.

 Below a certain value of $\mu$, there are not enough particles in the system to maintain the HD phase. This lower threshold value of $\mu$ is obtained by the definition of $\rho_\text{HD}$ ($1/2<\rho_\text{HD}< 1$).  Following is the range, over which the HD phase is likely to be found:
 \begin{equation}
  \label{range of mu for HD strong}
  \bigg(\frac{3}{2}+\frac{k_{2}}{k_{1}}-\frac{k_{2}}{2\beta k_{1}}-\frac{1}{2\beta}\bigg)<\mu < \bigg(2+\frac{k_{2}}{k_{1}}\bigg).
 \end{equation}

 \noindent
  Thus, the lower threshold of $\mu$ for HD phase existence depends on $\beta$, $k_{1}$, and $k_{2}$ explicitly and can be positive (for which HD phase sustains up to a limited value of $\mu$), or zero as well as negative (for which HD phase can be obtained for any value of $\mu$ between 0 and $\mu_\text{max}$, even at the lower values). When the lower threshold in (\ref{range of mu for HD strong}) is positive, we have :
 \begin{equation}
  \label{hd-lower-mu-st-other}
  \mu_\text{max}>\bigg(\frac{1}{2}+\frac{k_{2}}{2\beta k_{1}}+\frac{1}{2\beta}\bigg).
 \end{equation}

 In Fig.~\ref{ld-hd-mc-st-wk-den}, the HD phase density profiles in the strong coupling limit for $\mu=1$ with two different values of $\beta$ ($\beta=0.2$ and $\beta=0.4$) exhibit good agreement between the MFT and MCS results.

 \subsubsection{{ Maximal current phase}}
 \label{mc phase strong}

 The steady state bulk density in the MC phase with unit hopping rate is $\rho_\text{MC}=1/2$. The MC phase occurs when $\alpha_\text{eff}>1/2$ and $\beta_\text{eff}>1/2$ simultaneously, yielding the following:

 \begin{eqnarray}
  &&\alpha_\text{eff}=\alpha\frac{N_{1}}{L}>\frac{1}{2},\\
  &&\beta_\text{eff}=\beta\bigg(1-\frac{N_{2}}{L}\bigg)>\frac{1}{2}.
 \end{eqnarray}

 Now, substitution of $\rho_\text{LD}=1/2$ in Eq.~(\ref{rhold strong model 1}) and $\rho_\text{HD}=1/2$ in Eq.~(\ref{hd strong model 1}) allows one to obtain the LD-MC and HD-MC phase boundaries respectively. Thus we get the LD-MC phase boundary to be

 \begin{equation}
  \alpha=\frac{k_{1}+k_{2}}{(2\mu-1)k_{2}}, \label{ldmc strong}
 \end{equation}
 and the HD-MC phase boundary as
\begin{equation}
  \beta=\frac{k_{1}+k_{2}}{(3-2\mu)k_{1}+2k_{2}}. \label{hdmc strong}
 \end{equation}
 Clearly, the LD(HD)-MC phase boundaries in the strong coupling limit are straight lines parallel to $\beta(\alpha)$-axis (cf. Fig.~\ref{pd strong} and Fig.~\ref{pd strong asymmetric}), identical to the weak coupling limit case. Range of $\mu$ over which the MC phase appears is determined by exploiting the non-negativity of $\alpha$ and $\beta$ in (\ref{ldmc strong}) and (\ref{hdmc strong}), which turns out to be the following:
 \begin{equation}
 \label{mc phase range mu strong}
 \frac{1}{2} < \mu < \bigg(\frac{3}{2}+\frac{k_{2}}{k_{1}}\bigg).
 \end{equation}
 Particularly for $k_{1}=k_{2}=0.95$, (\ref{mc phase range mu strong}) evaluates to $1/2<\mu<5/2$, implying the MC phase cannot be present outside this window for the choice of exchange rates mentioned. Fig.~\ref{pd strong} illustrates the same where no MC phase for $\mu=0.3$ exists in the phase diagram. In Fig.~\ref{ld-hd-mc-st-wk-den}, we present the steady state density profile in the MC phase for $\mu=1$ with $\alpha=2.2$ and $\beta=1.5$.

 \subsubsection{ Domain wall phase }
 \label{sp phase strong}

 To begin with, consider the coupled equations (\ref{First coupled equation model 1}) and (\ref{Second coupled equation model 1}). In TL, neglecting the last part containing the coefficient of term of order $\mathcal{O}(1/L)$ in the right-hand side of (\ref{Second coupled equation model 1}), one solves these coupled equations in the strong coupling limit case to obtain the following:

 \begin{eqnarray}
  &&\frac{N_{1}}{L}=\frac{k_{2}}{k_{1}+k_{2}\frac{\alpha}{\beta}}, \label{n1l strong model 1}\\
  &&x_{w}=\frac{\bigg(1-\alpha-\frac{\alpha}{\beta}\bigg)\bigg(\frac{k_{2}}{k_{1}+k_{2}\frac{\alpha}{\beta}}\bigg)-\mu+2}{1-\frac{2\alpha k_{2}}{k_{1}+k_{2}\frac{\alpha}{\beta}}}. \label{xw strong model 1}
 \end{eqnarray}
 The population $N_{2}$ of reservoir $R_{2}$ is related to the population $N_{1}$ of reservoir $R_{1}$ by the condition $\alpha_\text{eff}=\beta_\text{eff}$ in DW phase, and is calculated as:

 \begin{equation}
  \label{n2-dw-strong}
  \frac{N_{2}}{L}=1-\frac{k_{2}}{k_{1}\frac{\beta}{\alpha}+k_{2}}
 \end{equation}
 The steady state densities at the low and high density regions of the domain wall can be determined using the expression of $N_{1}/L$ obtained in Eq.~(\ref{n1l strong model 1}). We find

 \begin{eqnarray}
  &&\rho_\text{LD}=\alpha \frac{N_{1}}{L}= \frac{\alpha k_{2}}{k_{1}+k_{2}\frac{\alpha}{\beta}}, \label{ld-den-dw-st}\\
  &&\rho_\text{HD}=1-\rho_\text{LD}=1-\frac{\alpha k_{2}}{k_{1}+k_{2}\frac{\alpha}{\beta}}. \label{Hd-den-dw-st}
 \end{eqnarray}

 \noindent
 We also find the DW height ($\Delta$):
 \begin{equation}
 \label{dw height strong model 1}
  \Delta=\rho_\text{HD}-\rho_\text{LD}=1-\frac{2\alpha \beta k_{2}}{\beta k_{1}+\alpha k_{2}}.
 \end{equation}

 In the strong coupling limit, the range of $\mu$ for the existence of the domain wall phase can be obtained similarly to the weak coupling case. The range is given by:

\begin{equation}
\bigg[\bigg(1+\alpha-\frac{\alpha}{\beta}\bigg)\frac{N_{1}}{L}+1\bigg] < \mu < \bigg[\bigg(1-\alpha-\frac{\alpha}{\beta}\bigg)\frac{N_{1}}{L}+2\bigg],
\end{equation}
where $N_{1}/L$ is calculated in Eq.~(\ref{n1l strong model 1}).

  \begin{figure}[!h]
 \centering
 \includegraphics[width=\columnwidth]{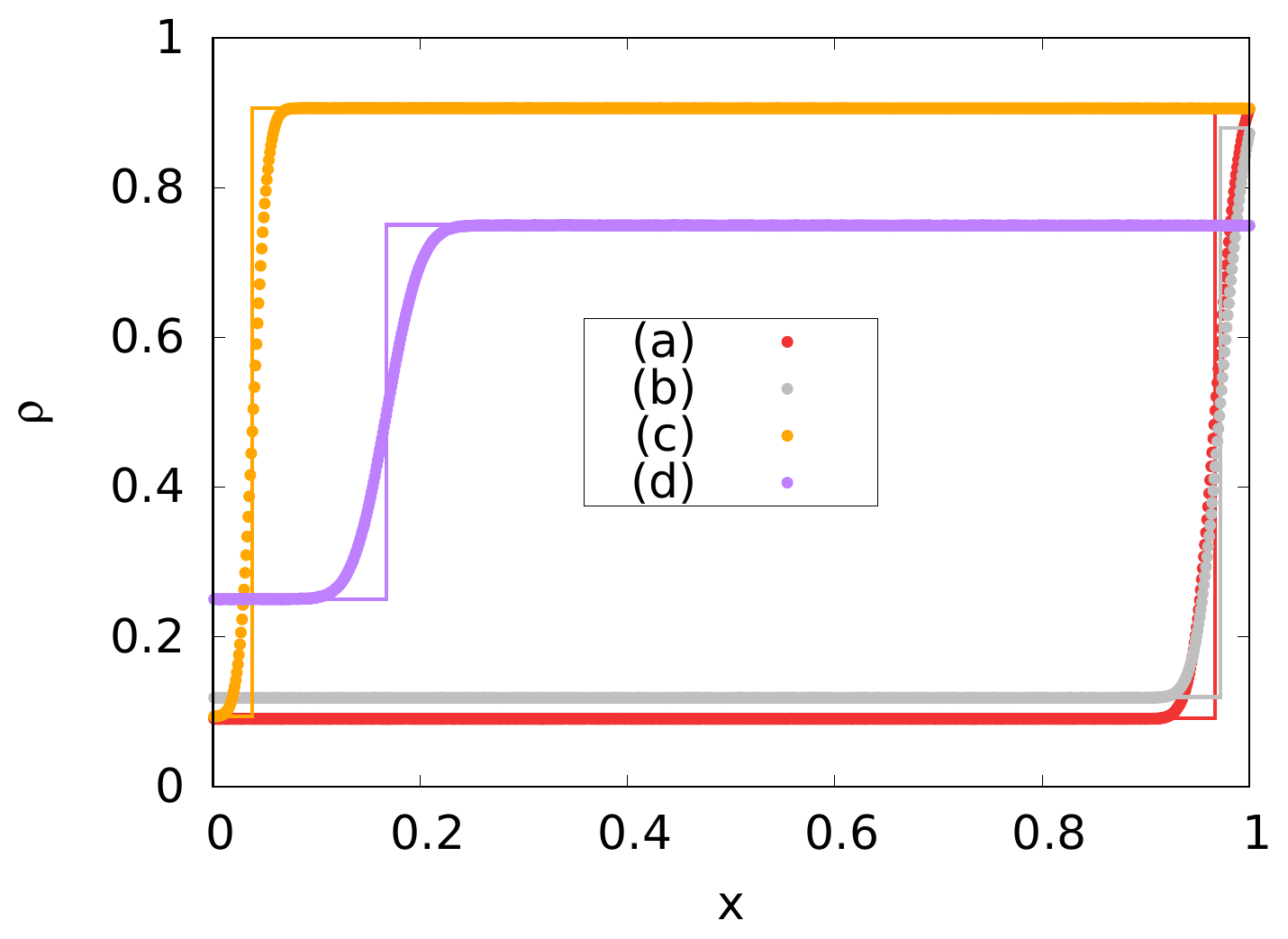}
 \caption{Plots illustrating the density $\rho$ as a function of position $x$ in the DW phase under the strong coupling limit. MFT predictions (colored solid lines) are confirmed with great accuracy by MCS results (colored discrete points), both sharing the same color for a particular set of parameter values. {System size  $L=1000$ and time-average over $2 \times 10^{9}$ Monte Carlo steps are done.} Values of these parameters are as follows: \textbf{(a)} $\alpha=1$, $\beta=0.1$, $\mu=0.3$, $k_{1}=k_{2}=0.95$; \textbf{(b)} $\alpha=1.5$, $\beta=0.13$, $\mu=0.3$, $k_{1}=k_{2}=0.95$; \textbf{(c)} $\alpha=1.5$, $\beta=0.1$, $\mu=1$, $k_{1}=k_{2}=0.95$; and \textbf{(d)} $\alpha=1.5$, $\beta=0.3$, $\mu=1$, $k_{1}=k_{2}=0.95$.}
 \label{dw-strong-mu0.5-1}
 \end{figure}

\begin{figure*}[htb]
 \includegraphics[width=\columnwidth]{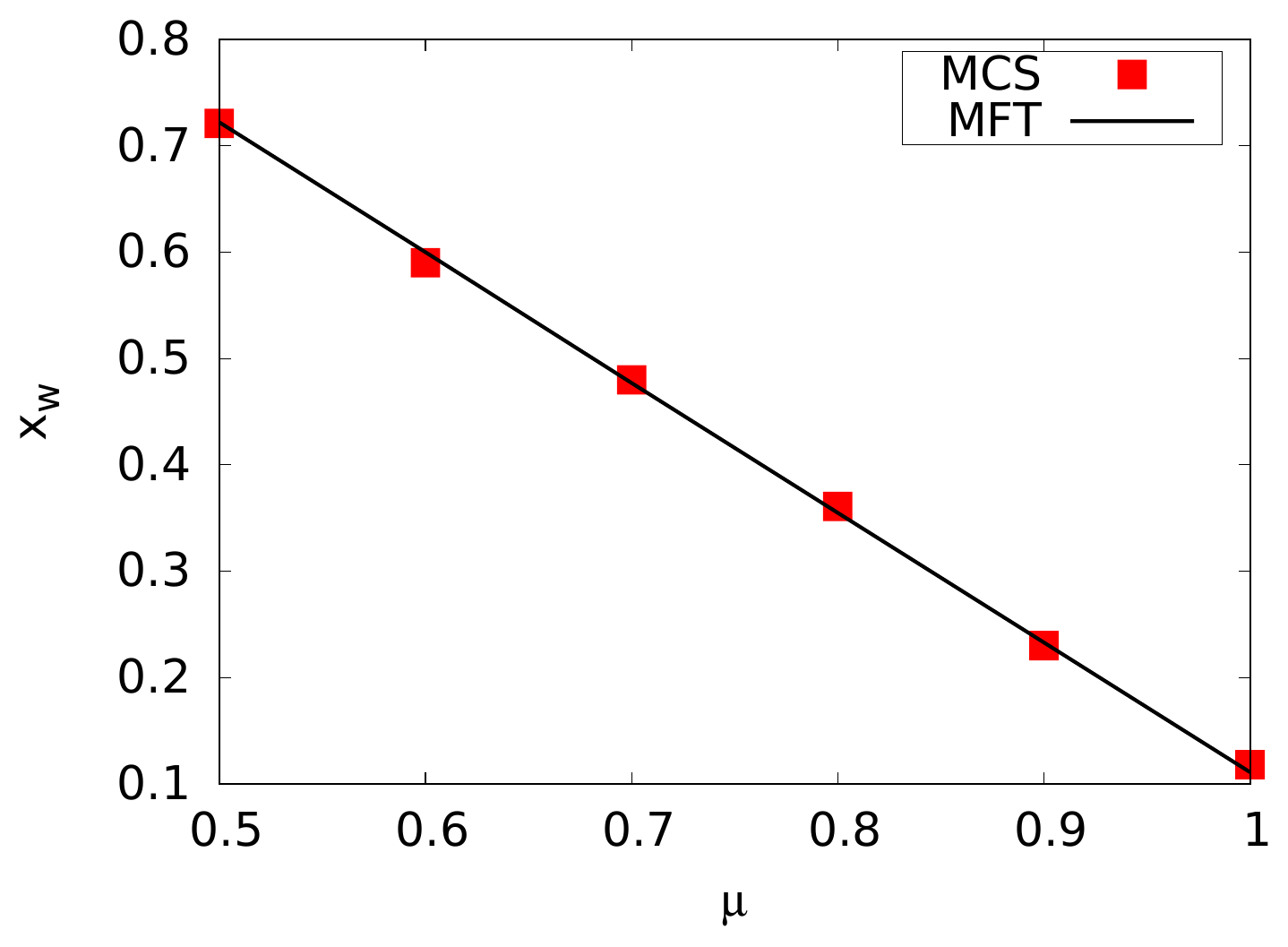}
 \hfill
 \includegraphics[width=\columnwidth]{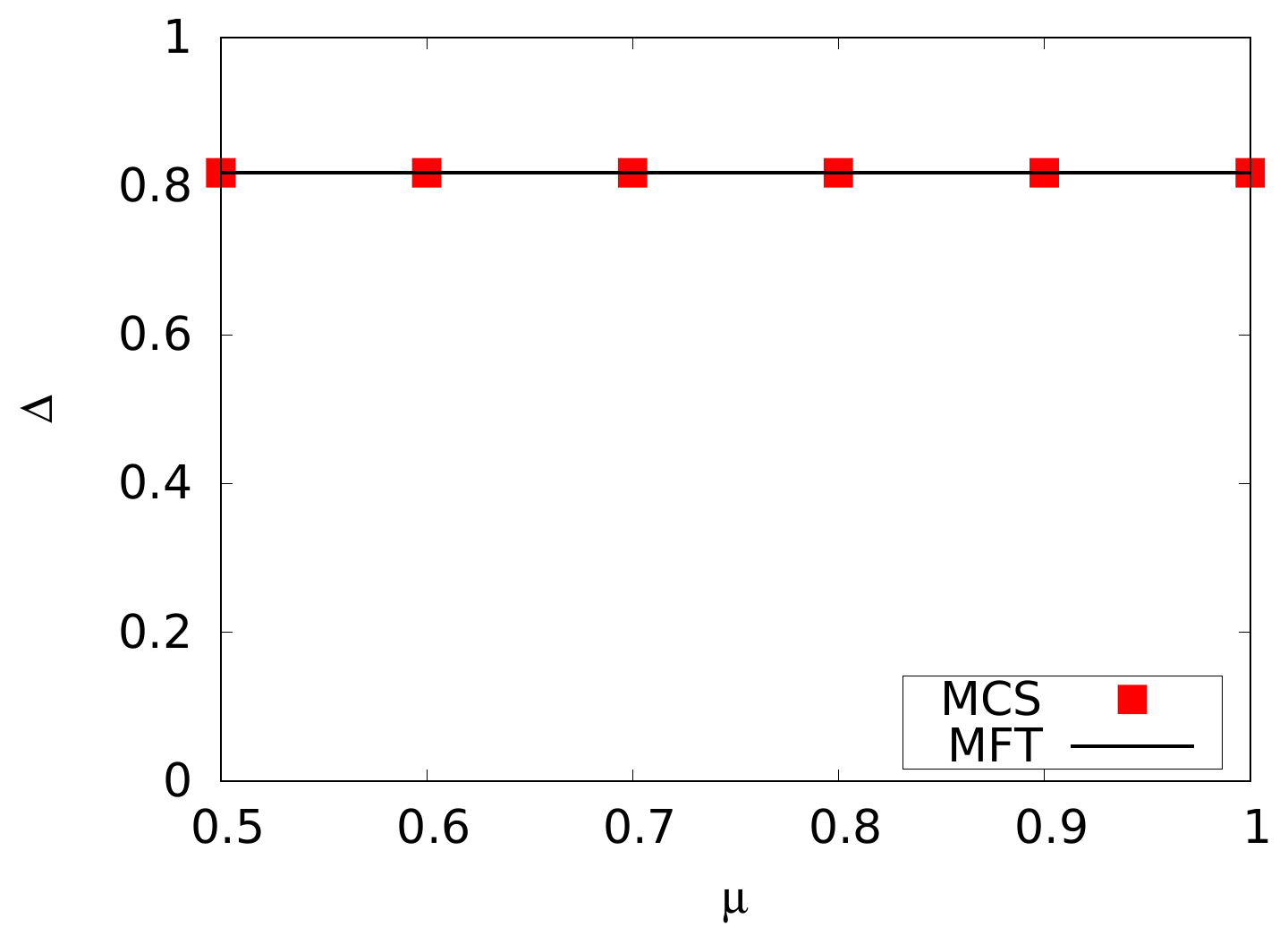}
 \\
\caption{\textbf{(Left)} Plot of the domain wall position $x_{w}$ versus $\mu$ in the strong coupling limit of the model. The entry and exit rates are $\alpha=1$ and $\beta=0.1$, and the exchange rates are $k_{1}=k_{2}=0.95$. The locations of the domain wall shows a linear decrease with increasing values of $\mu$. The results obtained from both MFT and MCS show similar trends and demonstrate good quantitative agreement.
\textbf{(Right)} Plot of the domain wall height $\Delta$ versus $\mu$ in the strong coupling limit of the model with control parameters $\alpha=1$, $\beta=0.1$, and $k_{1}=k_{2}=0.95$. The height of the domain wall remains  constant with varying $\mu$. The MFT and MCS results are in reasonable agreement.}
\label{xw-del-vs-mu-st}
\end{figure*}

\begin{figure*}[htb]
 \includegraphics[width=\columnwidth]{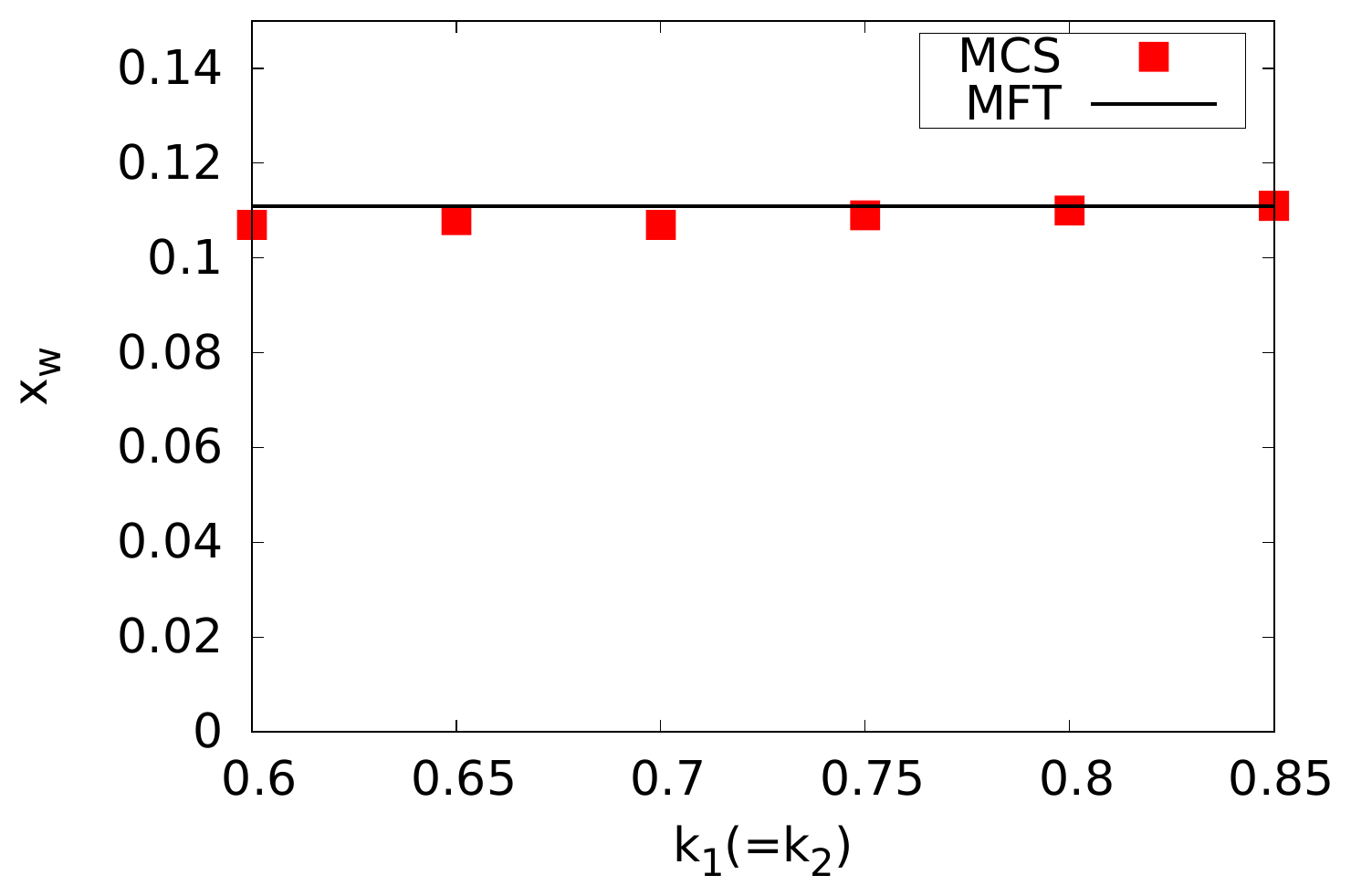}
 \hfill
 \includegraphics[width=\columnwidth]{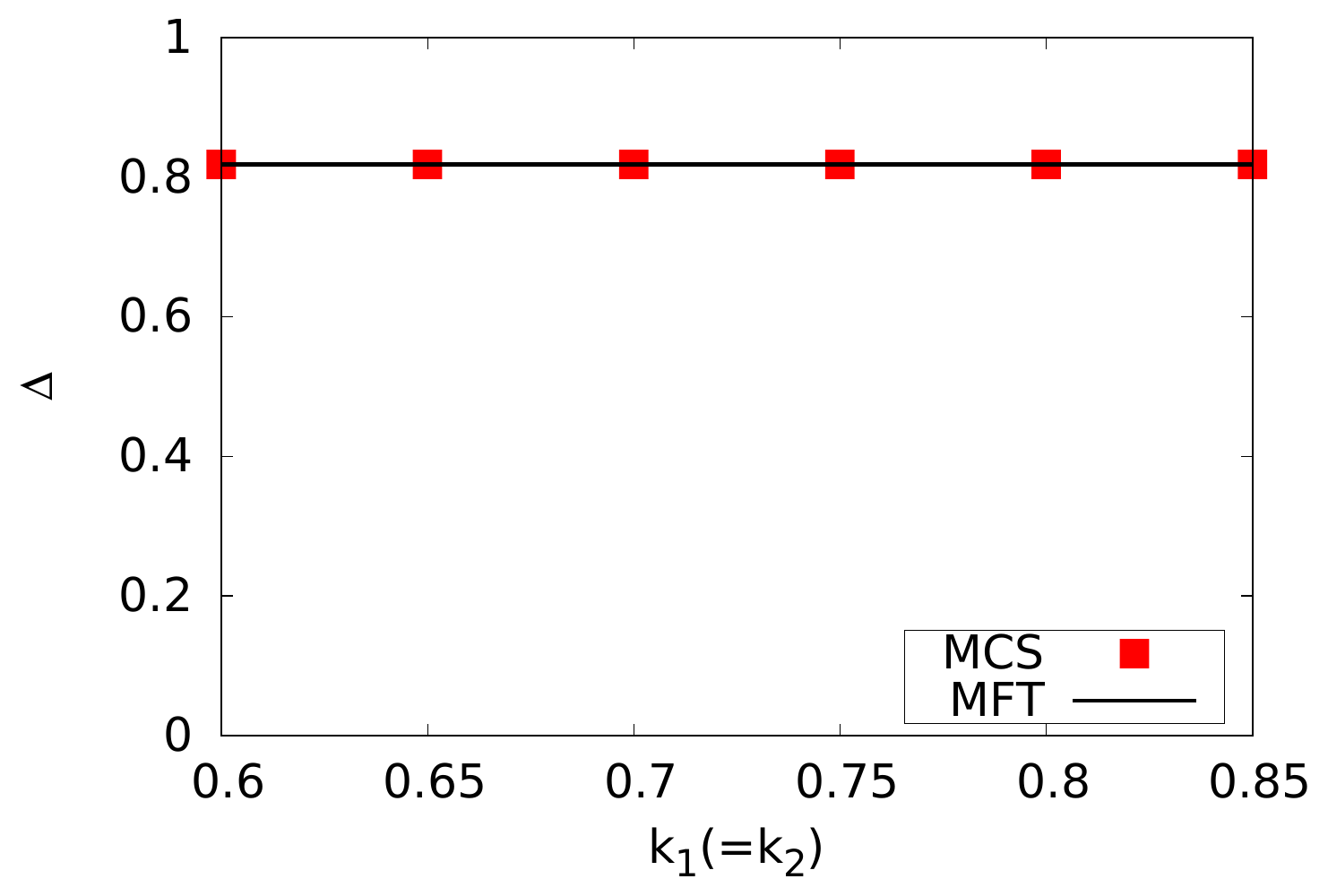}
 \\
\caption{\textbf{(Left)} The left plot illustrates the relationship between the position of a domain wall ($x_w$) and the exchange rates $k_1 = k_2$ in the strong coupling regime of the model. Further $\alpha= 1$, $\beta=0.1$, and $\mu=1$. Specific to these parameter values, the location of the domain wall remains constant as the exchange rates vary within the observed range. The results obtained from MFT and MCS show good agreement. \textbf{(Right)} The right plot displays the variation of the domain wall height $\Delta$ with  the exchange rates $k_1 = k_2$ in the strong coupling regime of the model. Further, $\alpha=1$, $\beta=0.1$ and $\mu=1$. The height of the domain wall remains constant as long as the exchange rates are equal. The results from MFT and MCS exhibit a good agreement. }
\label{xw-del-vs-k1k2-st}
\end{figure*}

 In Fig.~\ref{dw-strong-mu0.5-1}, DW phase density profiles in the strong coupling limit are obtained. Both the MFT and MCS results demonstrate remarkable consistency and conformity. Additionally, how the DW position ($x_{w}$) and height ($\Delta$) varies with $\mu$ and $k_{1},k_{2}$ is illustrated in Fig.~\ref{xw-del-vs-mu-st} and Fig.~\ref{xw-del-vs-k1k2-st}.

\begin{figure*}[htb]
 \includegraphics[width=\columnwidth]{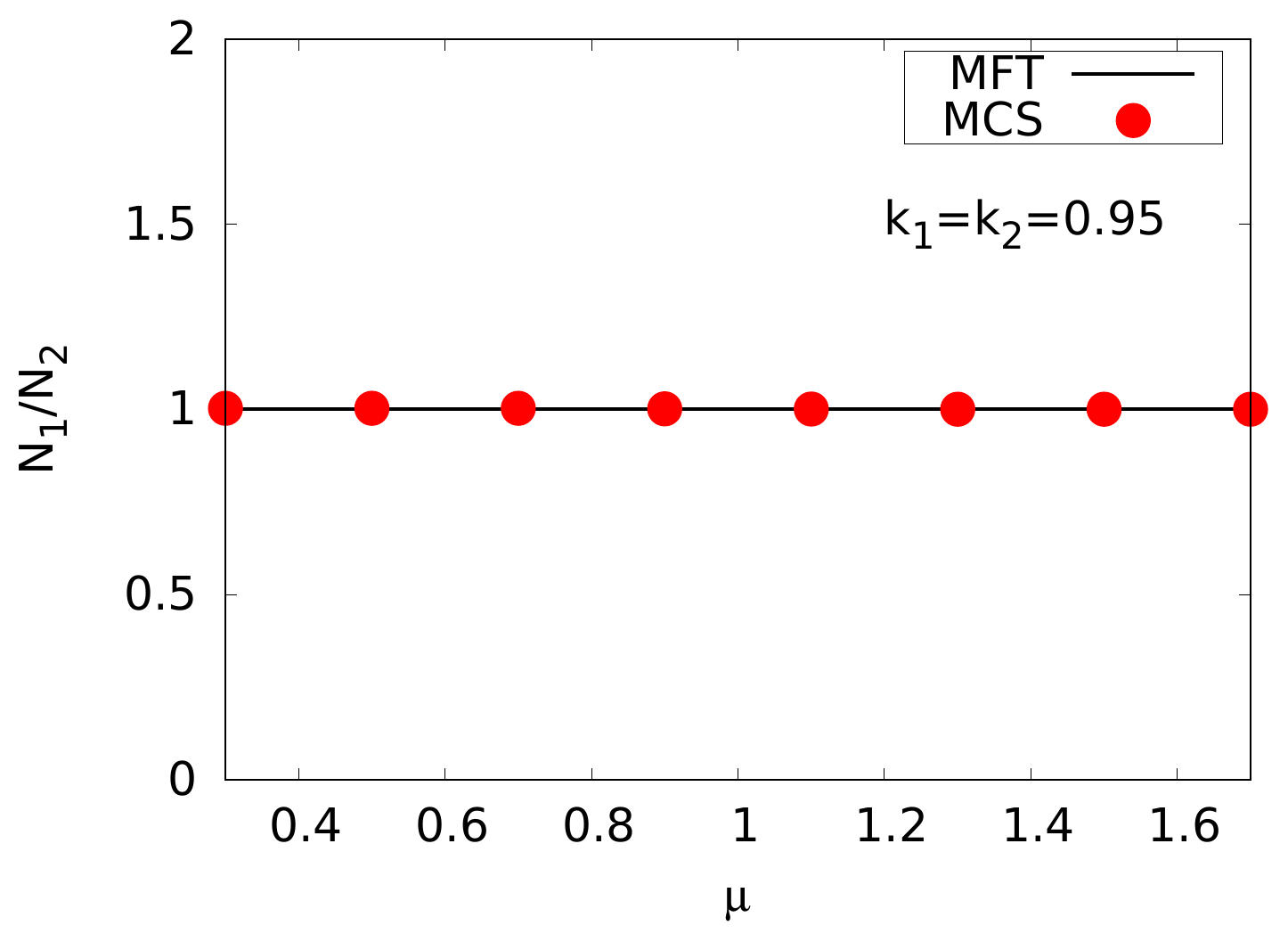}
 \hfill
 \includegraphics[width=\columnwidth]{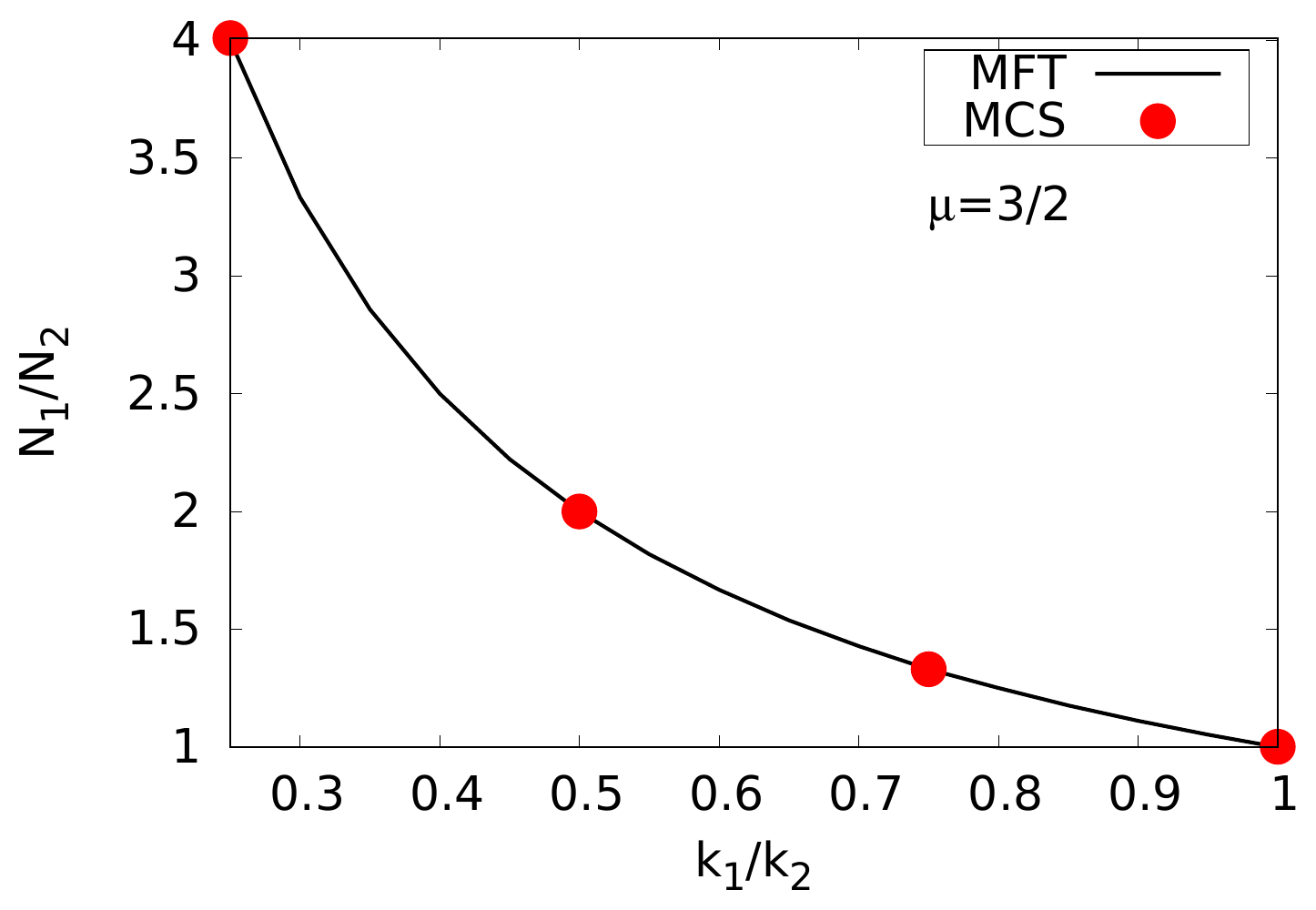}
 \\
\caption{ \textbf{(Left)} The plot shows how the reservoir population ratio $N_{1}/N_{2}$ varies with  $\mu$ in the strong coupling limit of the model. The exchange rates $k_{1}=k_{2}0.95$. The parameters $\alpha,\,\beta$ are chosen to ensure that the TASEP lane remains in the LD phase. The values of $\alpha=0.1$ and $\beta=1$ are used. There is good agreement between MFT and MCS results.
\textbf{(Right)} The plot illustrates the dependence of the reservoir population ratio $N_{1}/N_{2}$ on the exchange rate ratio $k_{1}/k_{2}$ for a fixed $\mu=3/2$ in the strong coupling limit. The parameters $\alpha,\,\beta$ are again chosen to keep the system in the LD phase with $\alpha=0.1$ and $\beta=1$. Both MFT and MCS results agree with each other.}
\label{n1/n2-vs-mu-and-k1/k2}
\end{figure*}

 To get the LD-DW and HD-DW phase boundaries, one sets $x_{w}=1$ and $x_{w}=0$ respectively in Eq.~(\ref{xw strong model 1}). Thus, with $x_{w}=1$, the LD-DW boundary is determined to be
  \begin{equation}
  \bigg(1+\alpha-\frac{\alpha}{\beta}\bigg)\bigg(\frac{k_{2}}{k_{1}+\frac{\alpha}{\beta}k_{2}}\bigg)=\mu-1. \label{ldsp boundary strong}
 \end{equation}
Likewise with $x_w=0$, the HD-DW boundary is obtained as
 \begin{equation}
  \bigg(1-\alpha-\frac{\alpha}{\beta}\bigg)\bigg(\frac{k_{2}}{k_{1}+\frac{\alpha}{\beta}k_{2}}\bigg)=\mu-2. \label{hdsp boundary strong}
 \end{equation}

\begin{figure}[!h]
 \centering
 \includegraphics[width=\columnwidth]{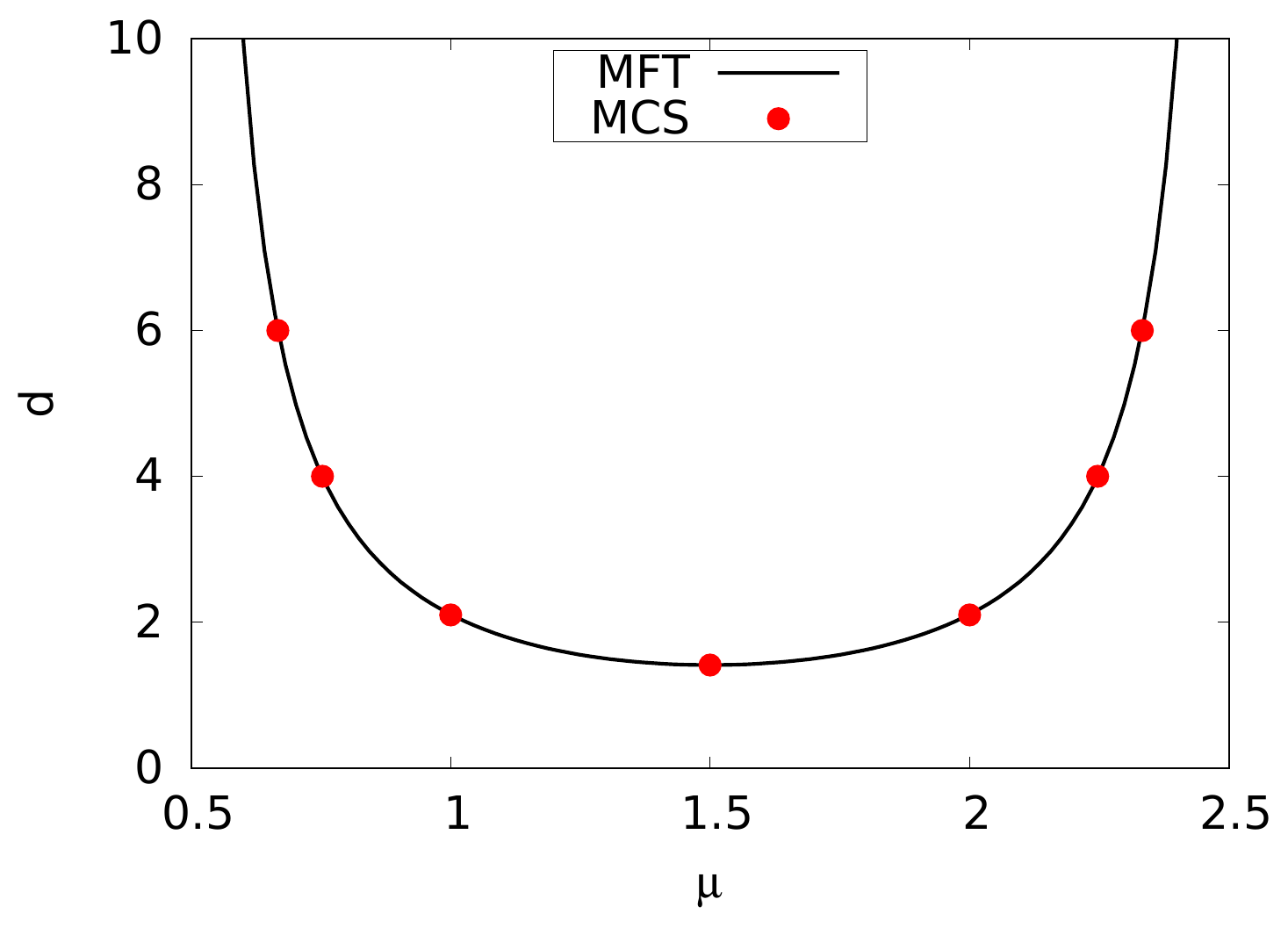}
 \caption{Plot of the distance $d$ between the origin and the multicritical point versus filling factor $\mu$ in the strong coupling limit of the model. Particle exchnage rates between reservoirs $k_{1}=k_{2}=0.95$. Black solid line and the discrete colored points represent the MFT and MCS results respectively, which are in excellent agreement. As $\mu \rightarrow 1/2$ from above or $\mu \rightarrow 5/2$ from below, $d$ diverges.}
 \label{d-vs-mu-str}
 \end{figure}
 {See the phase diagrams in Fig.~\ref{pd strong} and Fig.~\ref{pd strong asymmetric} for these boundaries.}

 Analogous to the weak coupling limit, multiple phases can share a common point in the strong coupling limit too. {This multicritical point can be found as the meeting point of the boundaries (\ref{ldmc strong}), (\ref{hdmc strong}), (\ref{ldsp boundary strong}), and (\ref{hdsp boundary strong}) and is given by the coordinate}

 \begin{equation}
  (\alpha_\text{c},\beta_\text{c}) = \left(\frac{k_{1}+k_{2}}{(2\mu-1)k_{2}},\frac{k_{1}+k_{2}}{(3-2\mu)k_{1}+2k_{2}}\right),
 \end{equation}
 which exists over the range $1/2<\mu<(3/2+k_{2}/k_{1})$. The distance between the origin $(0,0)$ and the multicritical point is
 \begin{equation}
  d=\sqrt{\left(\frac{k_{1}+k_{2}}{(2\mu-1)k_{2}}\right)^{2}+\left(\frac{k_{1}+k_{2}}{(3-2\mu)k_{1}+2k_{2}}\right)^{2}}.
 \end{equation}
  The distance $d$ diverges as $\mu$ approaches $1/2$ from above and $(3/2+k_{2}/k_{1})$ from below. The plot of $d$ versus $\mu$ is presented in Fig.~\ref{d-vs-mu-str} with symmetric choice of exchange rates ($k_{1}=k_{2}=0.95$), where MFT and MCS results match well.

  \vspace{2mm}
 \subsection{Nature of the phase transitions}
 \label{pt-strong}

 \begin{figure}[!h]
 \centering
 \includegraphics[width=0.5\textwidth]{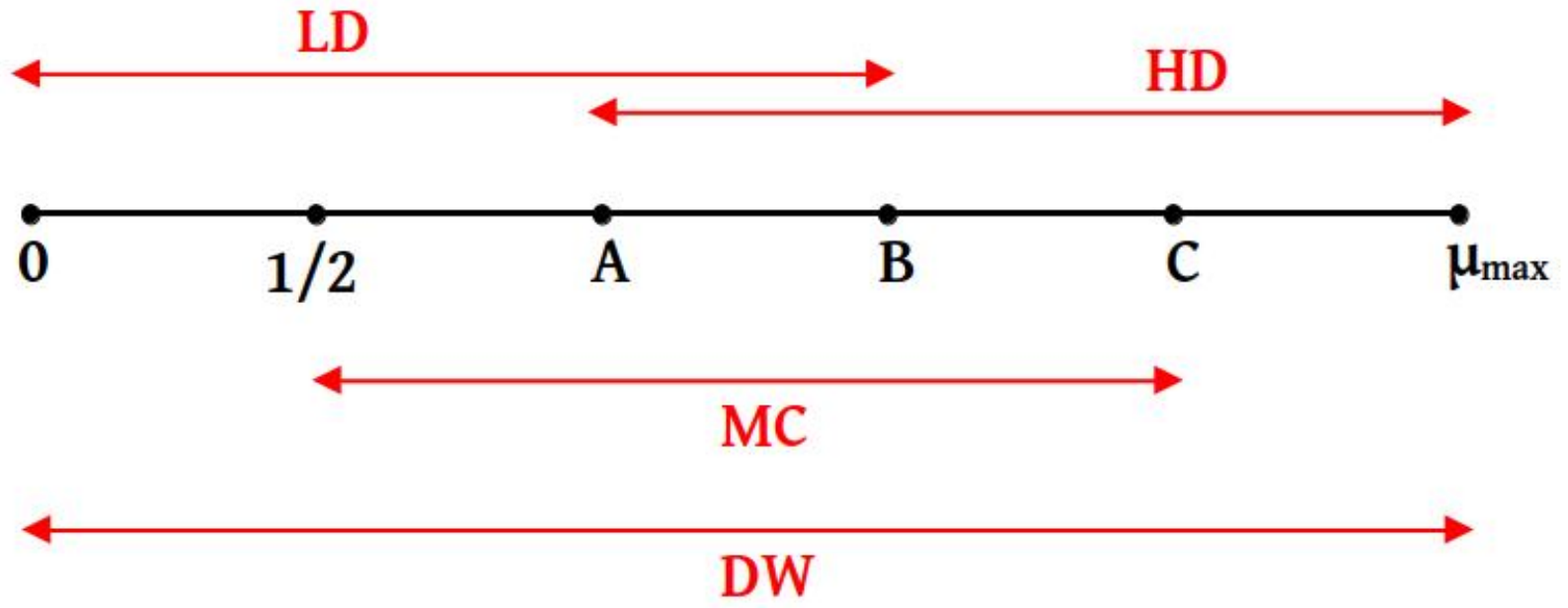}
 \caption{A schematic representation illustrating the ranges of $\mu$ where various phases emerge in the strong coupling limit of the model.}
 \label{mu-wkl-range-st}
 \end{figure}

 With a structure very similar to what is observed in the weak coupling limit, the model admits two or four phases simultaneously under the strong coupling condition, depending strongly on the value of $\mu$. This is depicted in Fig.~\ref{pd strong} and Fig.~\ref{pd strong asymmetric}, where particle-hole symmetry is observed only in Fig.~\ref{pd strong} where symmetric choices of exchange rates, $k_{1}=k_{2}(=0.95)$, are considered. When transitioning between different phases, the density changes continuosly across the phase boundaries, thus implying  second-order phase transitions. Similar transitions were observed in the weak coupling limit. Within a particular range of $\mu$ where all 4 phases exist together in the $\alpha-\beta$ plane, they share a common single point -- the multicritical point. In the subsequent discussion, we provide an in-depth analysis of the phases that manifest within a specific range of $\mu$.

 See the schematic diagram presented in Fig.~\ref{mu-wkl-range-st}. Being the thresholds of $\mu$ for the existence of different phases, the points indicated in that figure are as follows:

 \begin{itemize}
  \item \textbf{A}: $\bigg(\frac{3}{2}+\frac{k_{2}}{k_{1}}-\frac{k_{2}}{2\beta k_{1}}-\frac{1}{2\beta}\bigg)$

  \item \textbf{B}: $\bigg(\frac{1}{2}+\frac{k_{1}}{2\alpha k_{2}}+\frac{1}{2\alpha}\bigg)$

  \item \textbf{C}: $\bigg(\frac{3}{2}+\frac{k_{2}}{k_{1}}\bigg)$

  \item $\mu_\text{max}=\bigg(2+\frac{k_{2}}{k_{1}}\bigg)$
 \end{itemize}
  The maximum value of $\mu$ is subject to the conditions (\ref{ld-upper-mu-st-other}) and (\ref{hd-lower-mu-st-other}). We rewrite these conditions together:
 \begin{eqnarray}
  &&\mu_\text{max} > \bigg(\frac{1}{2}+\frac{k_{1}}{2\alpha k_{2}}+\frac{1}{2\alpha}\bigg), \label{cond-1-st} \\
  &&\mu_\text{max}>\bigg(\frac{1}{2}+\frac{k_{2}}{2\beta k_{1}}+\frac{1}{2\beta}\bigg). \label{cond-2-st}
 \end{eqnarray}
 Clearly, B $\ge$ 1/2 and A $\le$ C for any (positive) values of $\alpha$, $\beta$, $k_{1}$, and $k_{2}$. The positions of points A and B on the $\mu$-axis depend on the values of $\alpha$, $\beta$, $k_{1}$, and $k_{2}$. Specifically, keeping $k_{1}$ and $k_{2}$ fixed and adjusting the values of $\alpha$ and $\beta$, we can slide points A and B along the $\mu$-axis.

 Setting B $\ge$ A, the following condition emerges:

 \begin{equation}
  \label{cond-4-st}
  \mu_\text{max}\le\bigg[1+\frac{1}{2\alpha}\bigg(1+\frac{k_{1}}{k_{2}}\bigg)+\frac{1}{2\beta}\bigg(1+\frac{k_{2}}{k_{1}}\bigg)\bigg],
 \end{equation}
 which is consistent with the conditions (\ref{cond-1-st}) and (\ref{cond-2-st}) for $\mu_\text{max}$, assuming that the control parameters $\alpha$, $\beta$, $k_{1}$, and $k_{2}$ are all positive values. The other possibility B $<$ A, which results into the condition
\begin{equation}
  \label{cond-5-st}
  \mu_\text{max}>\bigg[1+\frac{1}{2\alpha}\bigg(1+\frac{k_{1}}{k_{2}}\bigg)+\frac{1}{2\beta}\bigg(1+\frac{k_{2}}{k_{1}}\bigg)\bigg],
\end{equation}
is discarded as it is not necessarily true along with the conditions (\ref{cond-1-st}) and (\ref{cond-2-st}). This implies that the position of point B can be either to the right of point A or coincide with it, but it cannot be to the left of point A.
 By analysing the expressions for points B and C, the following condition for $\alpha$ is obtained when B is located to the left of C,
\begin{equation}
\label{b-less-than-c-st}
\alpha > \frac{\bigg(1+\frac{k_{1}}{k_{2}}\bigg)}{2\bigg(1+\frac{k_{2}}{k_{1}}\bigg)}.
\end{equation}
For HD-MC phase boundary, given in (\ref{hdmc strong}), to be observed in the phase space, the condition $\mu <$ C must be satisfied, failing to meet which leads to negative values of $\beta$ which is unphysical. Substituing the value of $\alpha$ we got at the LD-MC phase boundary (\ref{ldmc strong}) in (\ref{b-less-than-c-st}) also yields the condition $\mu <$ C. In contrast, when the point B is located to the right side of point C, $\alpha$ takes on the following expression:
\begin{equation}
\label{b-greater-than-c-st}
\alpha < \frac{\bigg(1+\frac{k_{1}}{k_{2}}\bigg)}{2\bigg(1+\frac{k_{2}}{k_{1}}\bigg)}.
\end{equation}
The condition (\ref{b-greater-than-c-st}), when put to the LD-MC phase boundary (\ref{ldmc strong}), corresponds to the condition $\mu >$ C, for which $\beta<0$. Thus B is limited to move between 1/2 and C. Similarly, point A is constrained to move between 1/2 and C.
To summarise, if the exchange rates $k_{1}$ and $k_{2}$ are held constant while $\alpha$, $\beta$ can be varied, all points except A and B remain fixed along the $\mu$-axis. However, points A and B can be adjusted between the values of 1/2 and C by varying the parameters $\alpha$ and $\beta$. It should be noted that point B cannot be located to the left of point A.

Table~\ref{tab2} provides a summary of the range of $\mu$ in which various phases emerge, outlining essential details on the number and nature of phase boundaries, as well as information about the multicritical points.

  \begin{widetext}

\begin{table}[h!]
 \begin{center}
 \begin{tabular} { |p{0.8cm}|p{7.2cm}|p{3cm}|p{2cm}|p{2cm}|p{2cm}| }
  \hline
 \textbf{Row no.} & \textbf{Range of $\mu$} & \textbf{Phases} & \textbf{Domain walls} & \textbf{Phase boundaries} & \textbf{Multicritical points (MCPs)}  \\
 \hline
 1 & $0<\mu<\frac{1}{2}$ & LD and DW  & One LDW & One second-order & None \\
 \hline
 2 & $\frac{1}{2}<\mu<\bigg(\frac{3}{2}+\frac{k_{2}}{k_{1}}-\frac{k_{2}}{2\beta k_{1}}-\frac{1}{2\beta}\bigg)$ & LD, HD, MC, and DW & One LDW & Four second-order & One four-phase MCP\\
 \hline
 3 & $\bigg(\frac{3}{2}+\frac{k_{2}}{k_{1}}-\frac{k_{2}}{2\beta k_{1}}-\frac{1}{2\beta}\bigg)<\mu<\bigg(\frac{1}{2}+\frac{k_{1}}{2\alpha k_{2}}+\frac{1}{2\alpha}\bigg)$ & LD, HD, MC, and DW & One LDW & Four second-order & One four-phase MCP\\
 \hline
 4 & $\bigg(\frac{1}{2}+\frac{k_{1}}{2\alpha k_{2}}+\frac{1}{2\alpha}\bigg)<\mu<\bigg(\frac{3}{2}+\frac{k_{2}}{k_{1}}\bigg)$ & LD, HD, MC, and DW & One LDW & Four second-order & One four-phase MCP\\
 \hline
 5 & $\bigg(\frac{3}{2}+\frac{k_{2}}{k_{1}}\bigg)<\mu<\bigg(2+\frac{k_{2}}{k_{1}}\bigg)$ & HD and DW & One LDW & One second-order & None\\
 \hline
 \end{tabular}
 \caption{Table summarizing the occurence of certain phases and phase boundaries within a range of filling factor $\mu$ according to MFT in the strong coupling limit case. Particle-hole symmetry can be seen when a symmetric choice of exchange rates is considered, i.e., $k_{1}=k_{2}$. } \label{tab2}
 \end{center}

\end{table}

\end{widetext}

\section{Steady state density profiles in the LD, HD, and MC phases for both weak and strong coupling limit}
 \label{den-ld-hd-mc}

\begin{widetext}

 \begin{figure}[ht]
 \begin{minipage}{0.32\linewidth}
\includegraphics[width=\textwidth]{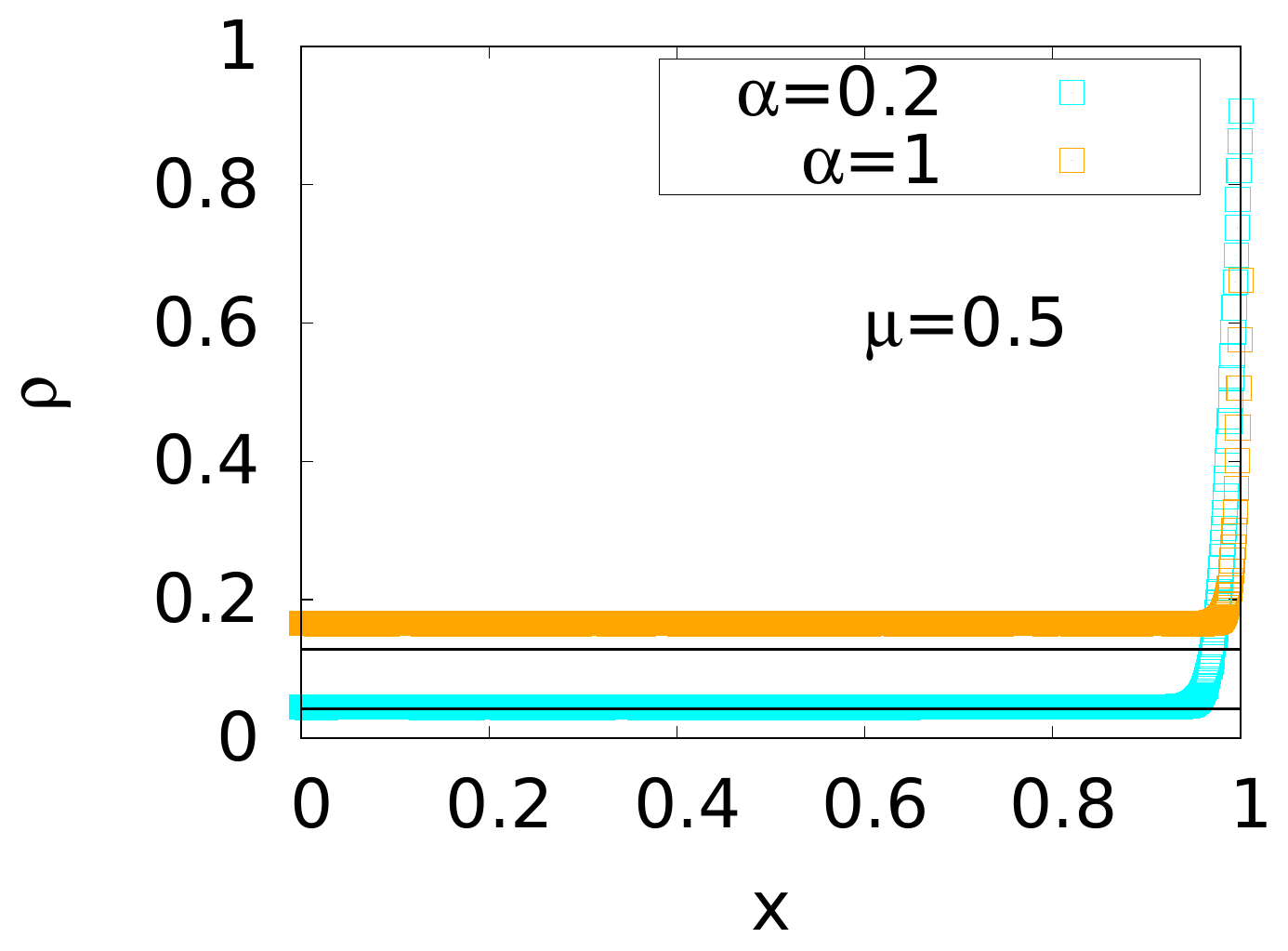}
\label{subfigure d}
\end{minipage}%
\hfill
\begin{minipage}{0.32\linewidth}
\includegraphics[width=\textwidth]{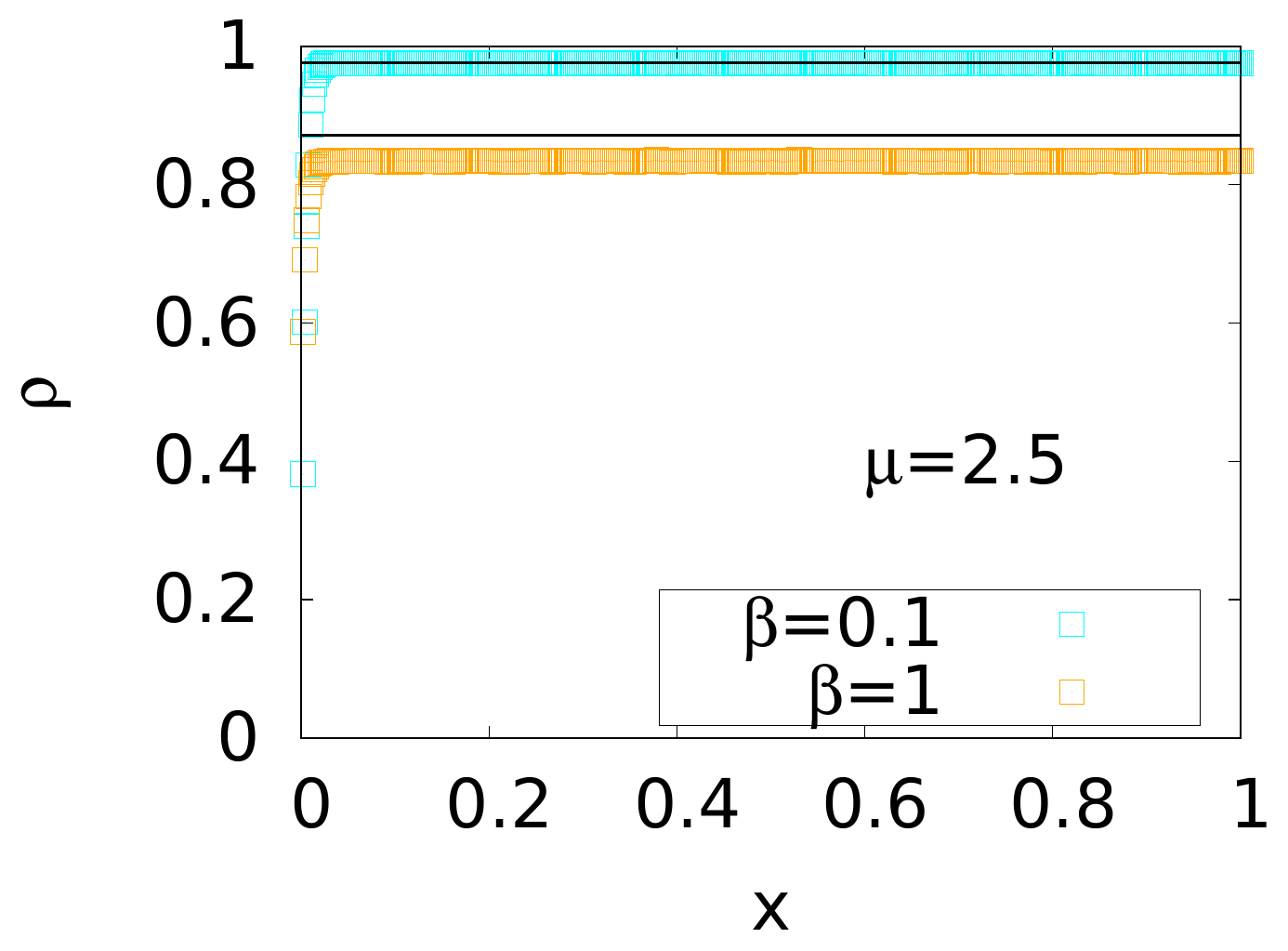}
\label{subfigure e}
\end{minipage}%
\hfill
\begin{minipage}{0.32\linewidth}
\includegraphics[width=\textwidth]{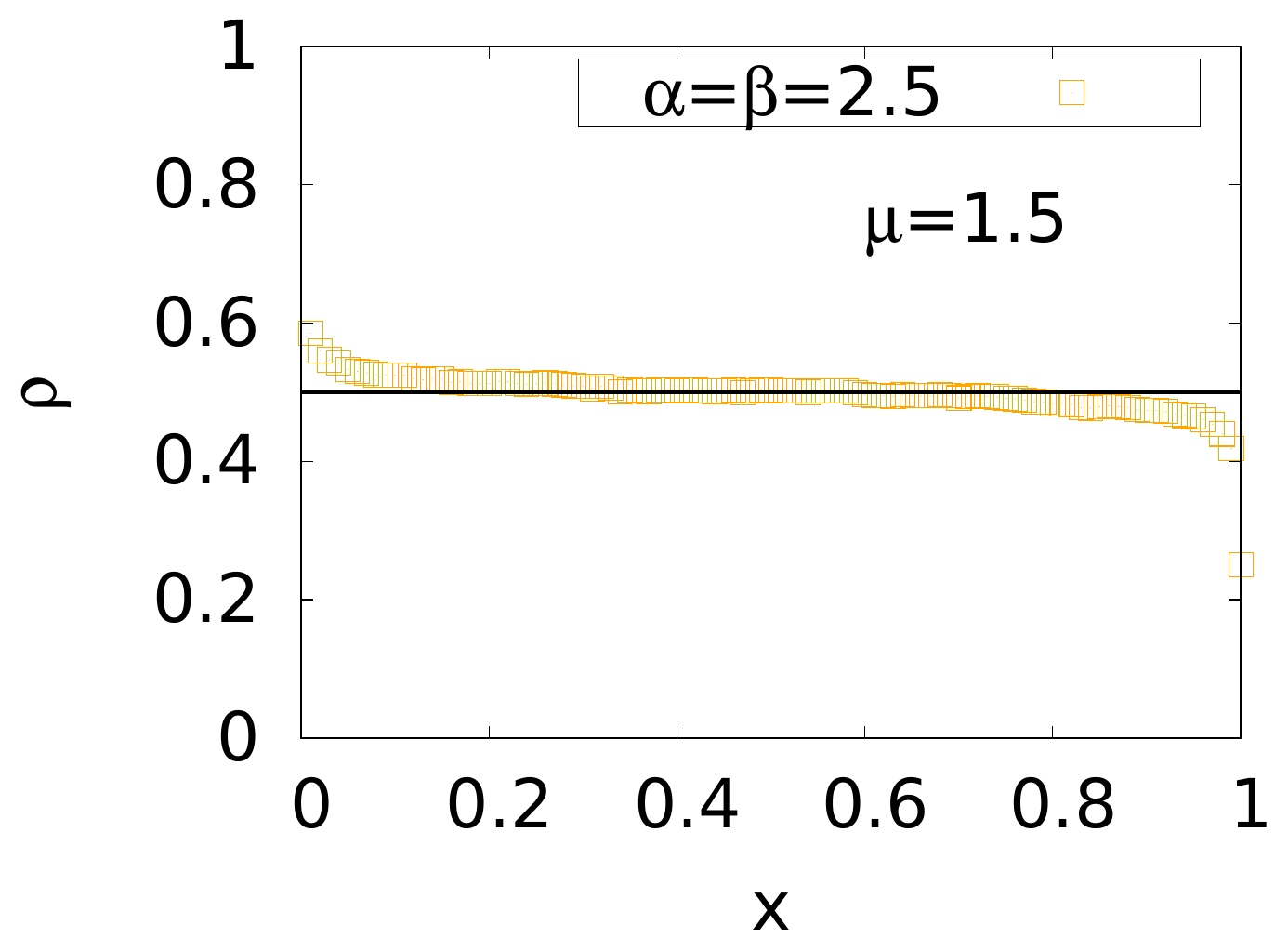}
\label{subfigure f}
\end{minipage}
\begin{minipage}{0.32\linewidth}
\includegraphics[width=\textwidth]{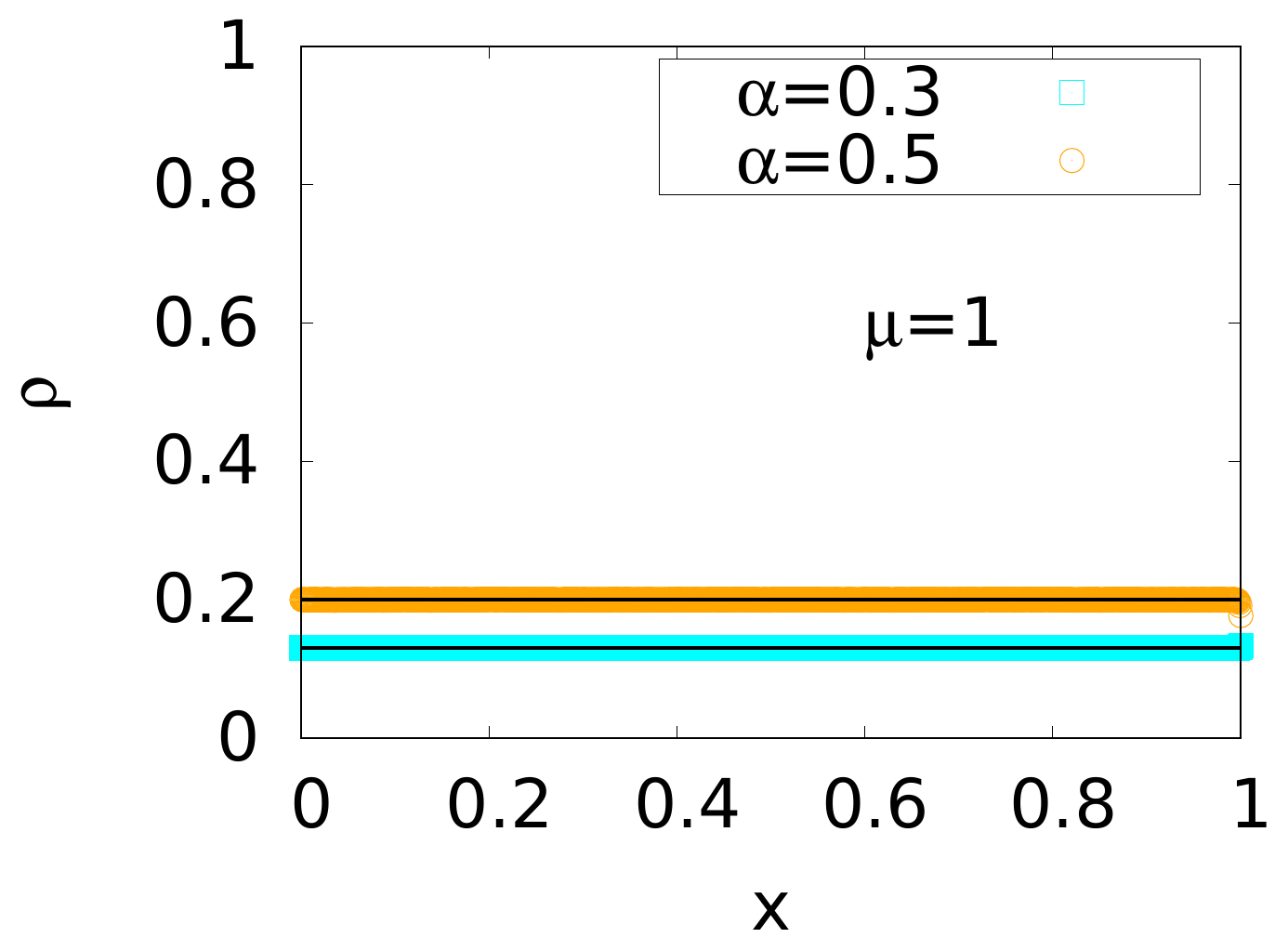}
\label{subfigure a}
\end{minipage}%
\hfill
\begin{minipage}{0.32\linewidth}
\includegraphics[width=\textwidth]{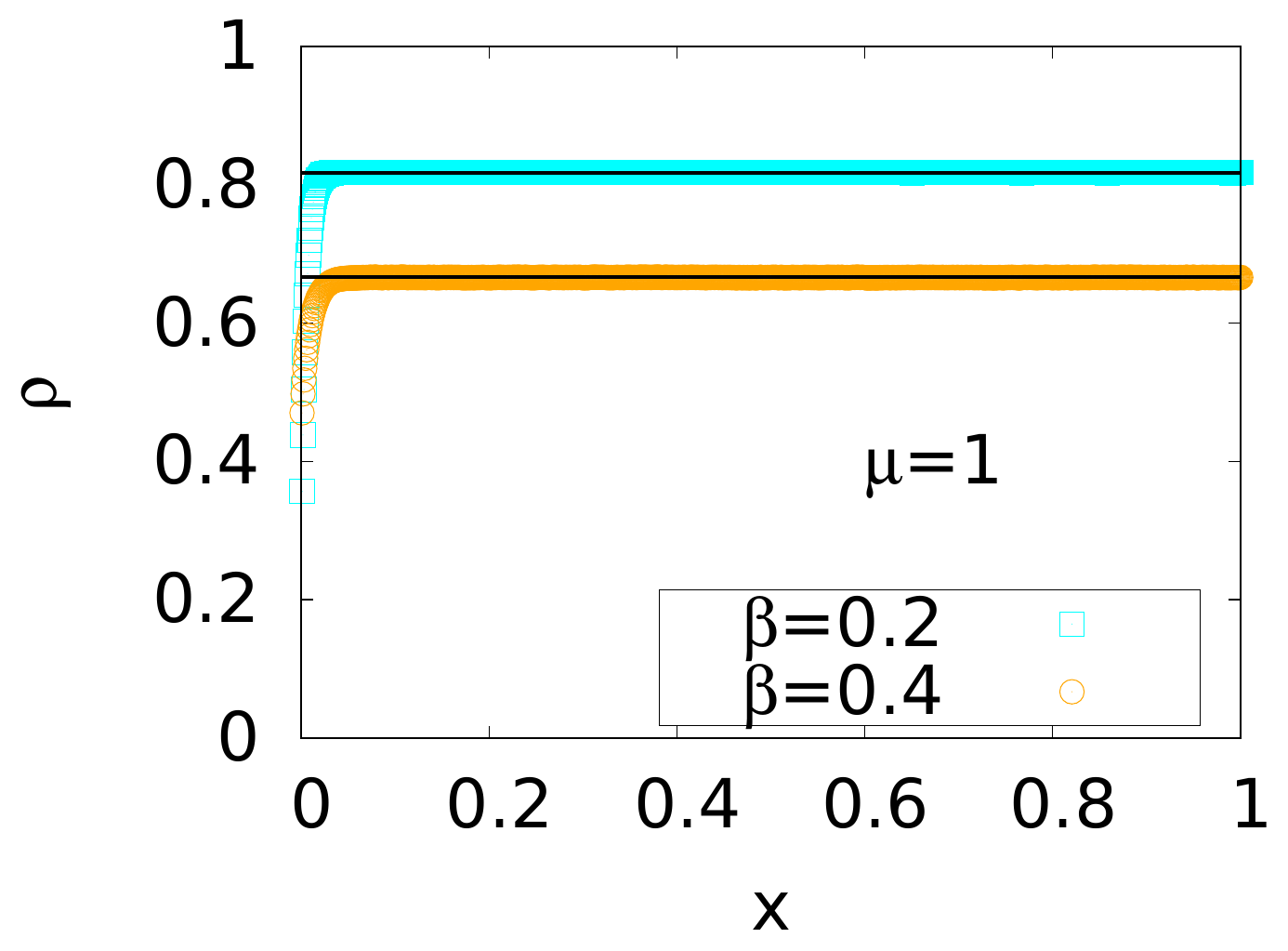}
\label{subfigure b}
\end{minipage}%
\hfill
\begin{minipage}{0.32\linewidth}
\includegraphics[width=\textwidth]{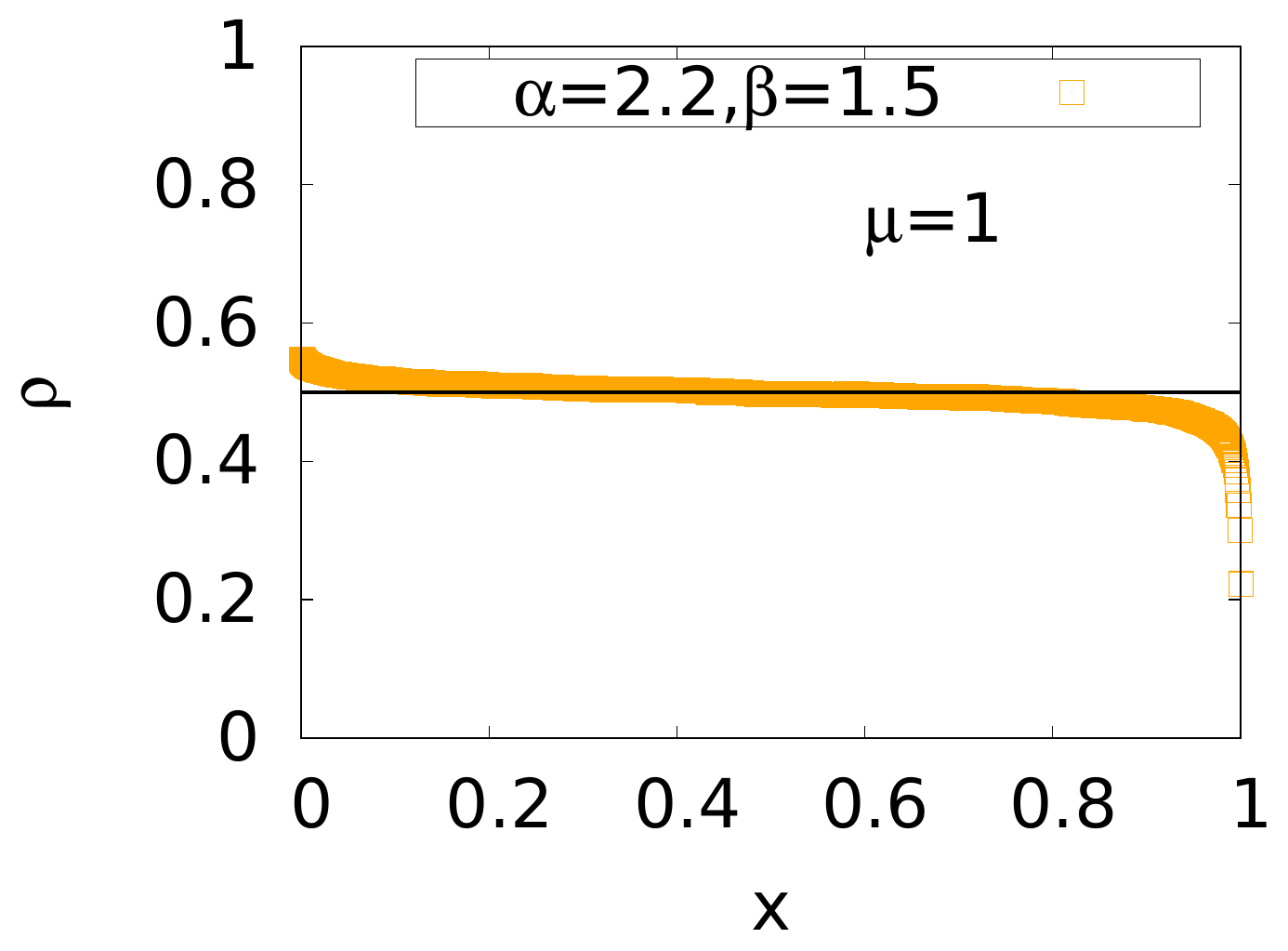}
\label{subfigure c}
\end{minipage}
\caption{These six plots illustrate the steady state density profiles in the LD, HD, and MC phases under both weak and strong coupling limits. The subfigures in the top row represent the weak coupling limit with exchange rates $k_{10}$ and $k_{20}$ set to 0.95, while those in the bottom row represent the strong coupling limit with exchange rates $k_1$ and $k_2$ set to 0.7. The values of the control parameters $\alpha$, $\beta$, and $\mu$ are indicated in each subfigure. {System size $L=1000$ and time-averages over $2 \times 10^{9}$ Monte Carlo steps are performed.} The solid black lines depict the MFT predictions, and the colored points correspond to the results obtained from MCS. In the weak coupling limit, MFT and MCS results closely match for small values of $\alpha$ and $\beta$. However, as $\alpha$ and $\beta$ increase, disagreement between the densities increases. On the other hand, in the strong coupling limit, MFT and MCS results exhibit a remarkable agreement across the entire parameter range. }
\label{ld-hd-mc-st-wk-den}
\end{figure}

\end{widetext}

 In this Section, we present the steady state density profiles in the low-density (LD), high-density (HD), and maximal current (MC) phases of the model for both the weak and strong coupling limit cases; see Fig.~\ref{ld-hd-mc-st-wk-den}.

\section{Simulation algorithm}
\label{sim-algo}

In this study, the mean-field predicted densities and phase diagrams of the model were validated through Monte Carlo simulations using random updates. The simulation rules are as follows: (i) If the first site ($i=1$) of the TASEP lane ($T$) is empty, a particle from reservoir $R_{1}$ enters $T$ with a rate $\alpha_\text{eff}$; (ii) Particles in $T$ can hop with rate 1 to the subsequent site in the bulk of $T$ if that site is empty; (iii) When at the last site ($i=L$) of $T$, a particle exits $T$ with a rate $\beta_\text{eff}$ into reservoir $R_{2}$; (iv) Simultaneously with these movements, particles in the reservoirs ($R_{1}$ and $R_{2}$) can jump directly to the other reservoir with definitive rates: $k_{1}N_{1}$ from $R_{1}$ to $R_{2}$ and $k_{2}N_{2}$ from $R_{2}$ to $R_{1}$; and (v) If, for any iteration, $k_{1}N_{1}$ and/or $k_{2}N_{2}$ are greater than 1, the rates are normalised by dividing each rate by the maximum value between $k_{1}N_{1}$ and $k_{2}N_{2}$. In each iteration of the simulation, the reservoirs or a site from the TASEP lane are randomly chosen for updating. After a sufficient number of iterations to reach steady states, the density profiles are calculated, and temporal averages are performed.



\pagebreak

\end{document}